\documentclass[a4paper,12pt]{article}
\pdfoutput=1 % if your are submitting a pdflatex (i.e. if you have
\usepackage{jheppub} % for details on the use of the package, please
\usepackage{graphicx}
\usepackage{hyperref}
\usepackage{cancel}
\usepackage{amssymb}
\usepackage{textcomp}
\usepackage{amsmath}
\usepackage{bm}
\usepackage{times}
\usepackage{epsfig}
\usepackage{color}
\usepackage{multirow}
\newcommand{\tabincell}[2]{\begin{tabular}{@{}#1@{}}#2\end{tabular}}

\begin{document}
\title{ Unraveling the Scotogenic Model at Muon Collider}
\author{Jiao Liu$^{1,2}$,}
\author{Zhi-Long Han$^1$,}
\emailAdd{sps\_hanzl@ujn.edu.cn}
\author{Yi Jin$^{1,2}$,}
\emailAdd{ss\_jiny@ujn.edu.cn}
\author{Honglei Li$^{1}$}
\affiliation{$^1$School of Physics and Technology, University of Jinan, Jinan, Shandong, 250022, China}
\affiliation{$^2$Guangxi Key Laboratory of Nuclear Physics and Nuclear Technology, Guangxi Normal University, Guilin, Guangxi 541004, China}

\abstract{ The Scotogenic model extends the standard model with three singlet fermion $N_i$ and one inert doublet scalar $\eta$ to address the common origin of tiny neutrino mass and dark matter. For fermion dark matter $N_1$, a hierarchical Yukawa structure $|y_{1e}|\ll|y_{1\mu}|\sim|y_{1\tau}|\sim\mathcal{O}(1)$ is usually favored to satisfy constraints from lepton flavor violation and relic density. Such large $\mu$-related Yukawa coupling would greatly enhance the pair production of charged scalar $\eta^\pm$ at the muon collider. In this paper, we investigate the dilepton and mono-photon signature of the Scotogenic model at a 14 TeV muon collider. For the dimuon signature $\mu^+\mu^-+\cancel{E}_T$, we find that most viable samples can be probed with $200~\text{fb}^{-1}$ data. The ditau signature $\tau^+\tau^-+\cancel{E}_T$ is usually less promising, but it is important to probe the small $|y_{1\mu}|$ region. The mono-photon signature $\gamma+\cancel{E}_T$ could also probe the compressed mass region $M_1\lesssim M_{\eta^\pm}$. Masses of charged scalar $\eta^\pm$ and dark matter $N_1$ can be further  extracted by a binned likelihood fit of the dilepton energy.
}

\maketitle
\flushbottom
%%%%%%%%%%%%%%%%%%%%%%%
\section{Introduction}
%%%%%%%%%%%%%%%%%%%%%%%
There are solid astrophysical and cosmological observations that indicate the existence of dark matter. Meanwhile, the neutrino oscillation experiments have confirmed that neutrinos have nonzero but tiny masses. To explain the origin of dark matter and tiny neutrino masses, we need to seek new physics beyond the standard model. Among various theoretical models to address them, the most appealing pathway is interpreting these two issues within the same framework ~\cite{Krauss:2002px,Asaka:2005an,Ma:2006km,Aoki:2008av,Restrepo:2013aga,Escudero:2016ksa,Cai:2017jrq,Cacciapaglia:2020psm}.

The Scotogenic model \cite{Ma:2006km} is an attractive option, where dark matter acts as the messenger of neutrino mass at the one-loop level. To realize this, an exact $Z_2$ symmetry is imposed, which forbids the tree-level neutrino mass and also stabilizes the dark matter. This model further introduces three singlet fermions $N_i$ and one inert scalar doublet $\eta$. The new particles are odd under the $Z_2$ symmetry. Phenomenological studies of the Scotogenic model have been extensively performed in Refs.~\cite{Kubo:2006yx,AristizabalSierra:2008cnr,Suematsu:2009ww,Kanemura:2011vm,Schmidt:2012yg,Baek:2015mna,Merle:2015gea,Ahriche:2017iar,Hugle:2018qbw,Baumholzer:2018sfb,Ahriche:2018ger,Han:2019lux,Wang:2019byi,Borah:2020wut,Liao:2022cwh}. In principle, either the lightest neutral scalar or the lightest fermion can serve as the dark matter candidate.  For the scalar dark matter scenario, quite a large portion of the parameter space with correct relic density can be probed via direct detection \cite{Dolle:2009fn,Arhrib:2013ela,Belyaev:2016lok}, indirect detection \cite{Queiroz:2015utg,Garcia-Cely:2015khw,Eiteneuer:2017hoh}, and collider experiments \cite{Dolle:2009ft,Miao:2010rg,Kalinowski:2018kdn,Yang:2021hcu,Fan:2022dck}. For instance, the low-mass regime with dark matter mass in the range of $55\sim 75$ GeV can be tested in direct detection experiments \cite{Arhrib:2013ela}. Meanwhile, the high-mass regime with dark matter mass above $500$~GeV is mostly within the reach of the Cherenkov Telescope Array \cite{Queiroz:2015utg,Garcia-Cely:2015khw}.

As for the fermion dark matter scenario, the dark matter-nucleon scattering appears at the one-loop level  \cite{Schmidt:2012yg,Ibarra:2016dlb}. Therefore, the corresponding cross-section is naturally suppressed by the loop factor. However, under certain circumstances, such as the two singlet fermions $N_1$ and $N_2$ nearly degenerating \cite{Schmidt:2012yg} or sufficiently large quartic couplings between inert scalar and Standard Model Higgs \cite{Ibarra:2016dlb}, we still have the chance to probe at upcoming direct detection experiments. On the other hand, the annihilation cross-section of fermion dark matter at present time is $p$-wave suppressed, due to the Majorana nature of $N_1$\cite{Kubo:2006yx}. In this way, the indirect detection experiments are hard to have positive signals. 

One appealing way to probe the fermion dark matter is via the lepton flavor violating  processes \cite{Toma:2013zsa}. The future sensitivities of corresponding observables could exclude dark matter mass below 100~GeV \cite{Vicente:2014wga}. Another promising pathway is the dilepton signatures with missing transverse energy $\cancel{E}_T$ at colliders \cite{Baumholzer:2019twf}. This signature arises from the pair production of  the charged scalar $\eta^\pm$ at colliders with the cascade decay $\eta^\pm\to \ell^\pm N_1$. Searches for similar signatures have been performed in the framework of supersymmetry at LHC \cite{ATLAS:2019lff,ATLAS:2019lng,CMS:2020bfa}. Currently, the 13 TeV LHC has excluded the region with $M_{\eta^\pm}\lesssim 700$~GeV and $M_{N}\lesssim 400$ GeV. For future colliders, the 3 TeV CLIC and the 100 TeV FCC-hh could probe $M_{\eta^\pm}\lesssim 1500$ GeV and $M_{\eta^\pm}\lesssim 2000$ GeV, respectively  \cite{Baumholzer:2019twf}.

The multi-TeV muon collider  has received  growing interest in recent years \cite{Delahaye:2019omf,Long:2020wfp,Han:2020pif,Han:2020uak,Liu:2021jyc,Han:2021udl,AlAli:2021let,Franceschini:2021aqd,Bai:2021ony,Bottaro:2021snn,Chen:2021pqi,Aime:2022flm,Forslund:2022xjq,Yang:2020rjt,Yang:2022fhw,Senol:2022snc,Li:2022kkc,Bottaro:2022one,Chakraborty:2022pcc}. It is an ideal machine to probe potential new physics correlated with muon, e.g., $(g-2)_\mu$ \cite{Capdevilla:2020qel,Buttazzo:2020ibd,Yin:2020afe,Huang:2021nkl,Capdevilla:2021rwo,Li:2021lnz} and $R_{K}$ anomaly \cite{Huang:2021biu,Asadi:2021gah,Qian:2021ihf,Altmannshofer:2022xri,Azatov:2022itm}. Under the tight constrain from $\mu\to e\gamma$, the new Yukawa couplings of the Scotogenic model have a hierarchical structure as $|y_{1e}|\ll |y_{1\mu}|\lesssim |y_{1\tau}|\sim \mathcal{O}(1)$ \cite{Vicente:2014wga}. Therefore, pair production of charged scalar at muon collider could be greatly enhanced due to additional contribution from the $t$-channel exchange of $N_i$ with relatively large Yukawa coupling $|y_{i\mu}|\sim\mathcal{O}(1)$. Then, the muon collider is more competitive than the $e^+e^-$ and $pp$ colliders in searching for the Scotogenic model.

In this paper, we investigate the dilepton signature of the Scotogenic model at a multi-TeV muon collider (MuC). According to our scan of the parameter space in Section \ref{Sec:DM}, the charged scalar mass could be up to about 6~TeV with the Yukawa coupling $|y_{i\alpha}|\lesssim 3$. In order to fully access the charged scalar,  we take the benchmark choice of the collider energy and the corresponding final integrated luminosity as $\sqrt{s}=14$ TeV with  $\mathcal{L}=20~\text{ab}^{-1}$ \cite{Han:2020uak}. As will be shown later, a 14 TeV muon collider can easily cover a significant part of the parameter space of fermion dark matter. Furthermore, the mass of charged scalar $\eta^\pm$ and dark matter $N_1$ are able to be extracted from the kinematic edges of the lepton energy distribution \cite{Battaglia:2013bha,Homiller:2022iax}. 

The rest of the paper is organized as follows. In Section \ref{Sec:MD} , we review the key structure of the Scotogenic model. Constraints from lepton favor violating processes are discussed in Section \ref{Sec:LFV} . Phenomenology of dark matter is considered in Section \ref{Sec:DM} . Viable parameter space is obtained under constraints from lepton flavor violation and relic density in this section. Dilepton signature, mono-photon signature and mass measurement at muon collider are studied in Section~\ref{Sec:MC}~. The conclusion is in Section \ref{Sec:CL} . 

\section{The Scotogenic Model} \label{Sec:MD}

The Scotogenic model is originally proposed in Ref.~\cite{Ma:2006km} to explain the common origin of tiny neutrino mass and dark matter. This model has three singlet fermions $N_i$ and one inert scalar doublet $\eta$, which are odd under a discrete $Z_2$ symmetry. In this paper, we consider the lightest fermion $N_1$ as the dark matter candidate  and produced via the freeze-out mechanism.  The scalar potential under the exact $Z_2$ symmetry is
\begin{eqnarray}
	\mathcal{V}&=&m_\phi^2 \phi^\dag\phi + m_\eta^2 \eta^\dag \eta + \frac{\lambda_1}{2}(\phi^\dag\phi)^2 + \frac{\lambda_2}{2}(\eta^\dag \eta)^2 + \lambda_3 (\phi^\dag\phi)(\eta^\dag\eta) \\ \nonumber
	&& + \lambda_4 (\phi^\dag \eta)(\eta^\dag \phi) + \frac{\lambda_5}{2}\left[(\phi^\dag\eta)^2+(\eta^\dag\phi)^2\right],
\end{eqnarray}
where $m_\eta^2>0$ is required to avoid the broken of the $Z_2$ symmetry \cite{Alvarez:2021otp}.  Notably, the $\lambda_5$-term is the only source of lepton number violation, thus is naturally small. All the parameters in the scalar potential are taken to be real \cite{Ginzburg:2010wa}. Constrained by the vacuum stability condition, the couplings in the scalar potential should satisfy \cite{Branco:2011iw}
\begin{equation}\label{Eq:Vac}
	\lambda_1>0,\lambda_2>0, \lambda_3+\sqrt{\lambda_1\lambda_2}>0, \lambda_3+\lambda_4-|\lambda_5|+\sqrt{\lambda_1\lambda_2}>0.
\end{equation}

After the electroweak symmetry breaking, masses of the Standard Model Higgs boson and inert scalars are
\begin{eqnarray}
	M_h^2 &=& \lambda_1 v^2,\\
	M_{\eta^\pm}^2&=& m_\eta^2+ \frac{1}{2}\lambda_3v^2,\\
	M_{R}^2&=& m_\eta^2 + \frac{1}{2}(\lambda_3+\lambda_4+\lambda_5)v^2,\\
	M_{I}^2&=& m_\eta^2 + \frac{1}{2}(\lambda_3+\lambda_4-\lambda_5)v^2,
\end{eqnarray}
with $M_R(M_I)$ denoting the mass of the real(imaginary) part of $\eta^0$. As discussed in Section~\ref{Sec:DM}, we fix $\lambda_3=\lambda_4=0.01$ to escape the tight constrain from dark matter direct detection. Meanwhile, the coupling $\lambda_5$ is also very small to obtain tiny neutrino mass with sizable Yukawa coupling. Therefore, the mass spectrum of inert scalars are highly degenerate, i.e., $M_{\eta^\pm}\simeq M_R\simeq M_I$. The contributions of inert scalars to the electroweak precision observables $S$ and $T$ are quite small with degenerate masses \cite{Barbieri:2006dq}.

\begin{figure} 
	\centering
	\includegraphics[width=0.5\textwidth]{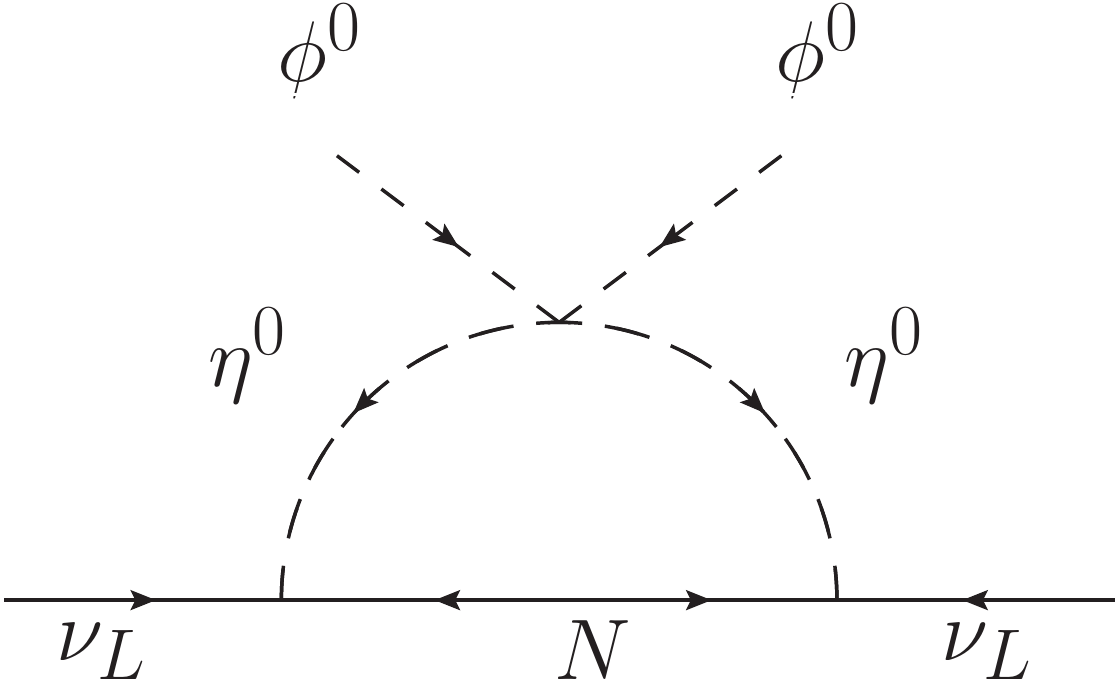}
	\caption{The one-loop neutrino mass generation in Scotogenic model.
		\label{FIG:mv}} 	
\end{figure}

The new Yukawa interactions for neutrino masses generation are given by 
\begin{equation}
	-\mathcal{L}_Y \supset y_{i\alpha} \overline{N_i} \tilde{\eta}^\dag  L_\alpha + \frac{1}{2} M_i \overline{N^c_i} N_i + h.c.,
\end{equation}
where  $\tilde{\eta}= i \sigma_2 \eta^*$. We also assume the mass matrix of singlet fermion $M$ to be diagonal for simplicity. As shown in Figure~\ref{FIG:mv}, the  neutrino mass is generated at one-loop level
\begin{eqnarray}
	m_\nu^{\alpha\beta} 
	&=& \sum_{i=1}^3  \frac{y_{i\alpha}y_{i\beta}}{32\pi^2}M_i \left[\frac{M_R^2}{M_R^2 -  M_i^2}\log \left(\frac{M_R^2}{M_i^2}\right) - \frac{M_I^2}{M_I^2 -  M_i^2}\log \left(\frac{M_I^2}{M_i^2}\right)\right]\\
	&\equiv&  \sum_{i=1}^3 y_{i\alpha} y_{i\beta} \Lambda_i
\end{eqnarray}

Using the Casas-Ibarra parameterization \cite{Casas:2001sr}, the Yukawa coupling $y$ can be determined by the neutrino oscillation parameters 
\begin{equation}
	y=\sqrt{\Lambda}^{-1}R\sqrt{\hat{m}_\nu}U^\dag_\text{PMNS},
\end{equation}
where $\Lambda=\text{diag}(\Lambda_1,\Lambda_2,\Lambda_3)$, and $R$ is a complex orthogonal matrix. In this paper, we assume $R$ is real for simplicity. $\hat{m}_\nu=\text{diag}(m_1,m_2,m_3)$ is the diagonalized neutrino mass matrix. $U_\text{PMNS}$ is the Pontecorvo-Maki-Nakagawa-Sakata (PMNS) matrix for neutrino mixing. In the following study, all the neutrino oscillation parameters are varied in the $3\sigma$ range of the global fit result \cite{deSalas:2020pgw}.

As will shown in Figure \ref{FIG:DM}, sizable Yukawa coupling $y\sim\mathcal{O}(1)$ is preferred by dark matter relic density. Such large Yukawa coupling  has a dominant contribution to the evolution of parameter $m_\eta^2$ \cite{Merle:2015gea,Alvarez:2021otp}
	\begin{equation}
		Q\frac{d m_\eta^2}{dQ}\simeq\frac{1}{16\pi^2}\left[2\text{Tr}(y^\dag y)m_\eta^2 - 4 \sum_{i=1}^3 M_i^2 (y y^\dag)_{ii}\right],
	\end{equation}
where Q is the renormalization scale. The second term seems can drive $m_{\eta}^2$ to negative, thus breaking the $Z_2$ symmetry spontaneously. However, as shown in Ref. \cite{Alvarez:2021otp}, no spontaneous $Z_2$ breaking occurs when considering the thermal effects. Meanwhile, the large Yukawa coupling can greatly affect the vacuum stability conditions in Equation \eqref{Eq:Vac}. The most dangerous one is the evolution of parameter $\lambda_2$ \cite{Merle:2015gea}
	\begin{equation}
		Q \frac{d \lambda_2}{dQ}\simeq \frac{1}{16\pi^2}\left[12\lambda_2^2-4\text{Tr}(y^\dag y y^\dag y)\right].
	\end{equation}
In the limit $\lambda_2^2\ll\text{Tr}(y^\dag y y^\dag y)$ and assuming constant Yukawa coupling, we can obtain an approximate solution as $\lambda_2(Q)\simeq\lambda_2(Q_0)-\text{Tr}(y^\dag y y^\dag y)\times\log(Q/Q_0)/4\pi^2$. By setting $\lambda_2(Q)=0$, we then have $\log(Q/Q_0)=4\pi^2\lambda_2(Q_0)/\text{Tr}(y^\dag y y^\dag y)$. Typically for $\lambda_2(Q_0=1~\text{TeV})=0.1$ and $y\sim\mathcal{O}(1)$, we have $Q/Q_0\sim50$, which indicates that Equation \eqref{Eq:Vac} is only valid up to about 50 TeV. To avoid the $Z_2$ symmetry breaking and vacuum unstable at high scale, additional scalar that couples strongly to the inert doublet $\eta$ might be further introduced.

\section{Lepton Flavor Violation} \label{Sec:LFV}
With TeV-scale new particles in the Scotogeinc model, the Yukawa interaction $y_{i\alpha} \overline{N_i} \tilde{\eta}^\dag  L_\alpha$ would also induce observable lepton flavor violation (LFV) processes. Here, we briefly summarize the results of experimental limits and theoretical prediction. A detailed discussion has been performed in Ref.~\cite{Toma:2013zsa}. One well-studied process is the radiative decay $\ell_\alpha\to \ell_\beta\gamma$. Currently, the experimental limits on such processes are
BR$(\mu\to e\gamma)<4.2\times10^{-13}$ \cite{MEG:2016leq}, BR$(\tau\to e\gamma)<3.3\times10^{-8}$ \cite{BaBar:2009hkt}, and BR$(\tau\to \mu\gamma)<4.4\times10^{-8}$ \cite{BaBar:2009hkt}.
The corresponding branching ratios are calculated as \cite{Toma:2013zsa}
\begin{equation}
	\text{BR}(\ell_\alpha\to \ell_\beta\gamma) = \frac{3(4\pi)^3\alpha}{4G_F^2} |A_D|^2 \text{BR}(\ell_\alpha\to \ell_\beta \nu_\alpha \bar{\nu}_\beta),
\end{equation}
with the dipole form factor
\begin{equation}
	A_D=\sum_{i=1}^3 \frac{y_{i\beta}^* y_{i\alpha}}{2(4\pi)^2 M^2_{\eta^\pm}}
	F_2\left(\xi_i\right).
\end{equation}
Here $\xi_i=M_i^2/M_{\eta^+}^2$, and the loop function $F_2(x)$ is given by
\begin{equation}
	F_2(x)=\frac{1-6x+3x^2+2x^3-6x^2\log x}{6(1-x)^4}.
\end{equation}

The three-body decay $\ell_\alpha\to 3\ell_\beta$ is also a competitive process. The present limits are 
BR$(\mu\to 3e)<1.2\times10^{-12}$ \cite{SINDRUM:1987nra}, BR$(\tau\to 3e)<2.7\times10^{-8}$ \cite{Hayasaka:2010np}, and BR$(\tau\to 3\mu)<2.1\times10^{-8}$ \cite{Hayasaka:2010np}. In the Scotogenic model, the dominant contributions are $\gamma$-penguins and box diagrams \cite{Guo:2020qin}. Meanwhile, the $Z$-penguins and Higgs-penguins contributions are suppressed.
The branching ratios are given by
\begin{eqnarray}
	\text{BR}(\ell_\alpha\to 3 \ell_\beta) &=&
	\frac{3 (4\pi)^2 \alpha^2}{8G_F^2}\bigg[ |A_{ND}|^2 +|A_D|^2\left(\frac{16}{3}\log\left(\frac{m_\alpha}{m_\beta}\right)
	-\frac{22}{3}\right) +\frac{1}{6}|B|^2  \\ \nonumber
	&+& \big(-2 A_{ND} A_D^*+\frac{1}{3}A_{ND} B^*-\frac{2}{3}A_D B^*+\text{h.c.}\big)\bigg] \text{BR}(\ell_\alpha\to \ell_\beta \nu_\alpha \bar{\nu}_\beta),
\end{eqnarray}
with the non-dipole form factor
\begin{equation}
	A_{ND}=\sum_{i=1}^3 \frac{y_{i\beta}^* y_{i\alpha}}{6(4\pi)^2 M^2_{\eta^\pm}}
	G_2\left(\xi_i\right),
\end{equation}
and the loop function $G_2(x)$
\begin{equation}
	G_2(x)=\frac{2-9x+18x^2-11x^3+6x^3\log x}{6(1-x)^4}.
\end{equation}
The contribution of box diagram $B$ is
\begin{equation}
	e^2 B = \frac{1}{(4\pi)^2 M_{\eta^+}^2} \sum_{i,j=1}^3 \left[\frac{1}{2}D_1(\xi_i,\xi_j)y_{j\beta}^*y_{j\beta}y_{j\beta}^* y_{i\alpha}+\sqrt{\xi_i\xi_j}D_2(\xi_i,\xi_j)y_{j\beta}^*y_{j\beta}^*y_{j\beta} y_{i\alpha}\right],
\end{equation}
with the loop function $D_1(x,y),D_2(x,y)$ of the form
\begin{eqnarray}
	D_1(x,y) & = & -\frac{1}{(1-x)(1-y)}-\frac{x^2 \log x}{(1-x)^2(x-y)}-\frac{y^2\log y}{(1-y)^2(y-x)},\\
	D_2(x,y) & = & -\frac{1}{(1-x)(1-y)}-\frac{x \log x}{(1-x)^2(x-y)}-\frac{y\log y}{(1-y)^2(y-x)}.
\end{eqnarray}

Finally, we consider $\mu-e$ conversion in nuclei. The most stringent limits  are
CR$(\mu,\text{Ti}\to e,\text{Ti})<4.3\times 10^{-12}$ \cite{SINDRUMII:1993gxf} and CR$(\mu,\text{Au}\to e,\text{Au})<7\times 10^{-13}$ \cite{SINDRUMII:2006dvw}. The $\mu-e$ conversion rates, normalized to the muon capture rate $\Gamma_\text{capt}$, are expressed as
\begin{eqnarray}
	\text{CR}(\mu-e,\text{Nucleus})&\simeq&\frac{p_e E_e m_\mu^3 G_F^2 \alpha^3 Z_\text{eff}^4 F_p^2}{8\pi^2 Z \Gamma_\text{capt}} \left|(Z+N)g_{LV}^{(+)}+(Z-N)g_{LV}^{(-)}\right|^2.~
\end{eqnarray}
Here, $p_e\simeq E_e\simeq m_\mu$ are the momentum and energy of the electron. $Z_\text{eff}$ is the effective atomic charge, and $F_p$ is the nuclear matrix element \cite{Kitano:2002mt}. $Z$ and $N$ are the number of protons and neutrons in the nucleus. In the above approximation, we only take into account the left-hand vector effective operators, because contributions from scalar and right-hand vector effective operators are suppressed in the Scotogenic model \cite{Toma:2013zsa,Vicente:2014wga}. The factors $g_{LV}^{(\pm)}$ are calculated as
\begin{equation}
	g_{LV}^{(\pm)}=\frac{1}{2}\sum_{q=u,d,s} \left(g_{LV(q)}G_V^{(q,p)}\pm g_{LV(q)}G_V^{(q,n)}\right).
\end{equation}
The numerical values of the $G_V$ coefficients are $G_V^{u,p}=G_V^{(d,n)}=2,G_V^{d,p}=G_V^{(u,n)}=1$ \cite{Arganda:2007jw}. The effective coupling $g_{LV}(q)$ is dominant by the $\gamma$-penguins
\begin{equation}
	g_{LV(q)}\approx g_{LV(q)}^\gamma =\frac{\sqrt{2}}{G_F} e^2 Q_q (A_{ND}-A_D),
\end{equation}
with $Q_q$ the electric charge of the corresponding quark.

With new source of CP-violation, the Yukawa interaction $y \overline{N} \tilde{\eta}^\dag  L$ contributes to the electric dipole moment (EDM) of charged leptons $d_\ell$ at two-loop level \cite{Borah:2016zbd}. An order of magnitude estimation of $d_\ell$ gives \cite{Borah:2016zbd}
\begin{equation}
	\frac{d_\ell}{e} \sim \frac{m_\ell \lambda_3~ \text{Im}(y^2 ) }{(16\pi^2)^2 M_{\eta^\pm}^2}.
\end{equation}
Provided $\lambda_3=0.01,M_{\eta^\pm}=3$ TeV and $\text{Im}(y)\sim\mathcal{O}(1)$, we have $d_e/e\sim4.5\times10^{-31}$ cm. Such result is much smaller than current experimental bound $d_e/e<8.7\times10^{-29}$ cm \cite{ACME:2013pal}. So the EDM limit is easy to satisfy.

\section{Dark Matter} \label{Sec:DM}

\begin{figure} 
	\centering
	\includegraphics[width=0.45\textwidth]{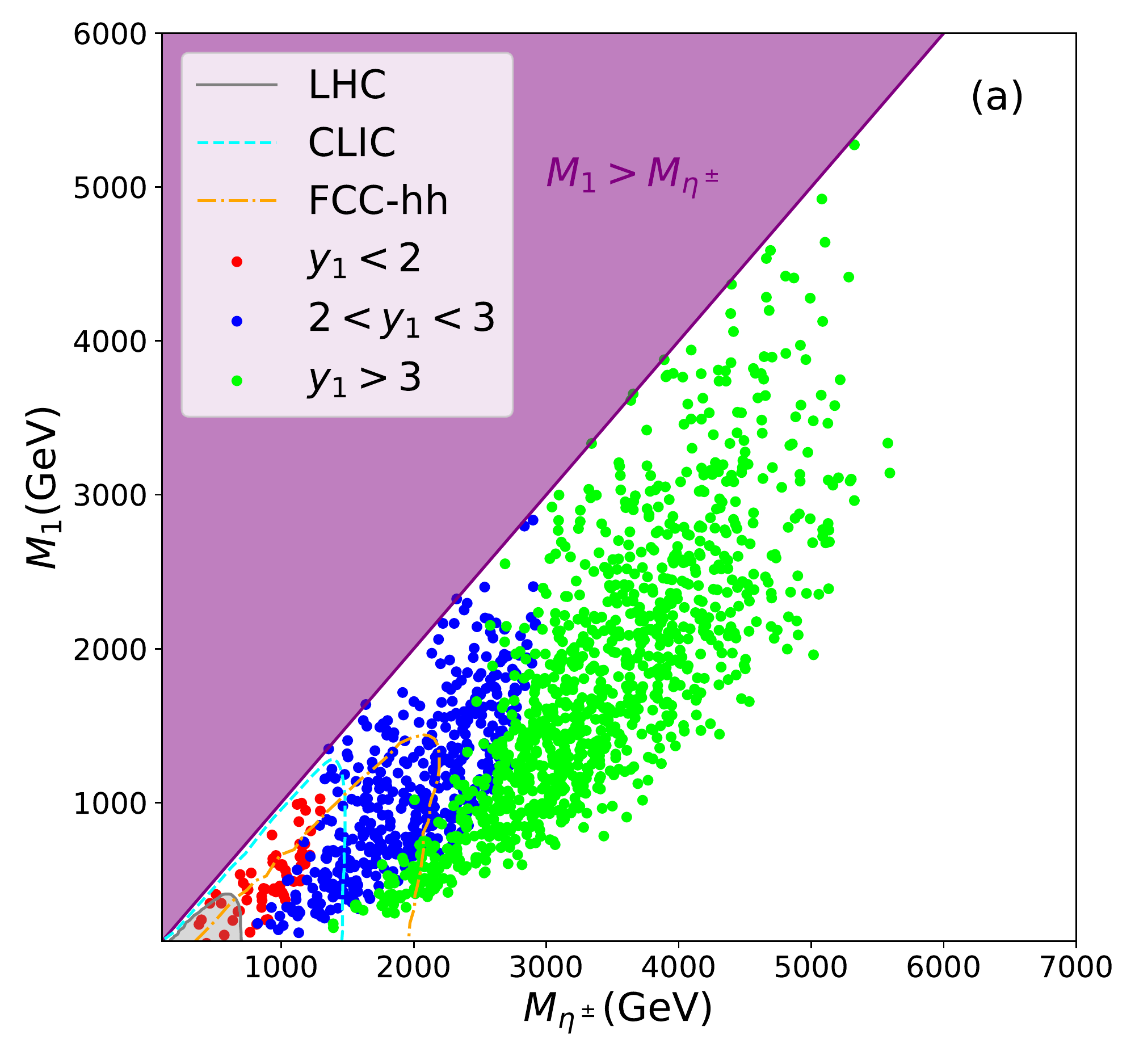}
	\includegraphics[width=0.45\textwidth]{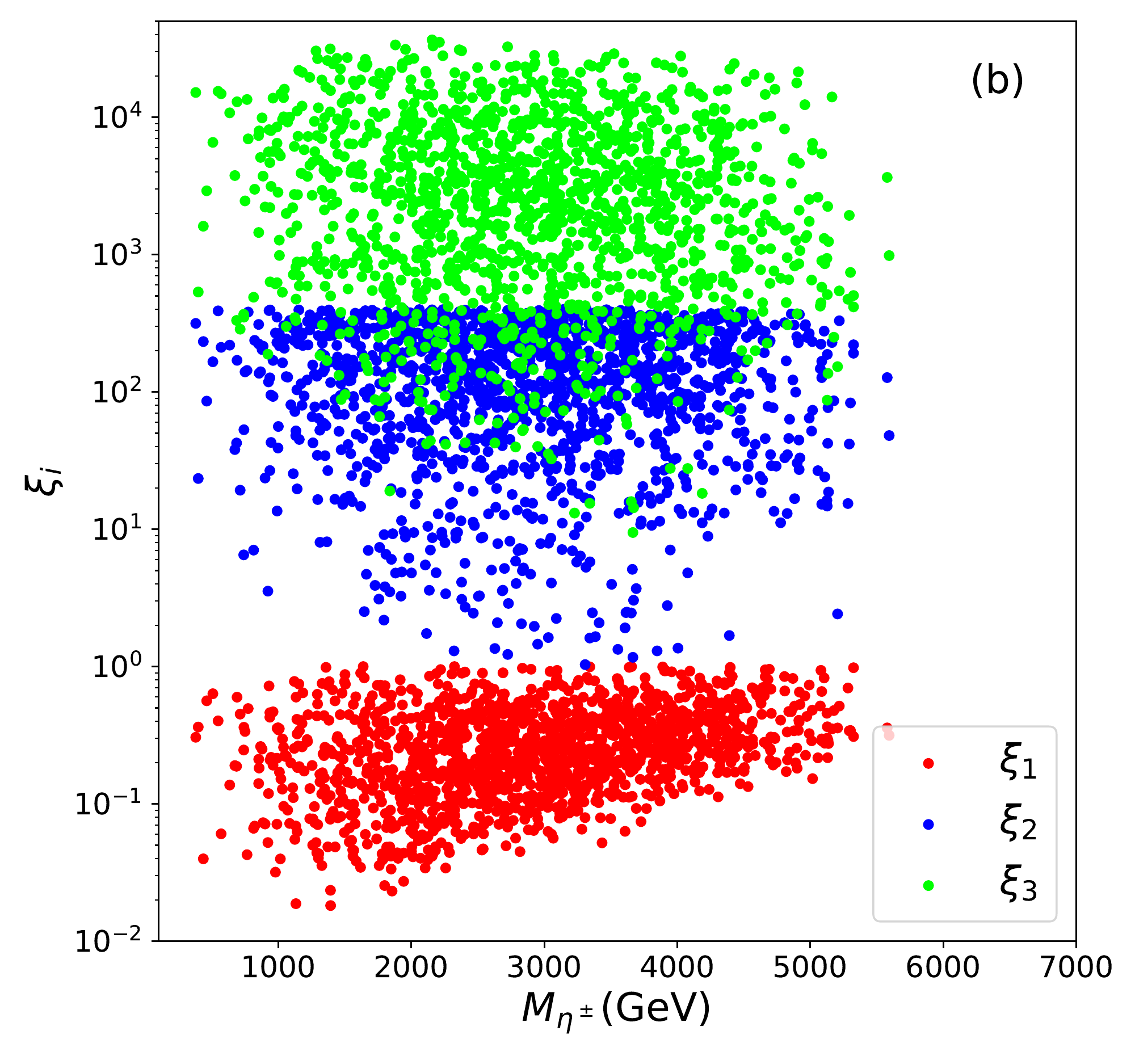}
	\includegraphics[width=0.45\textwidth]{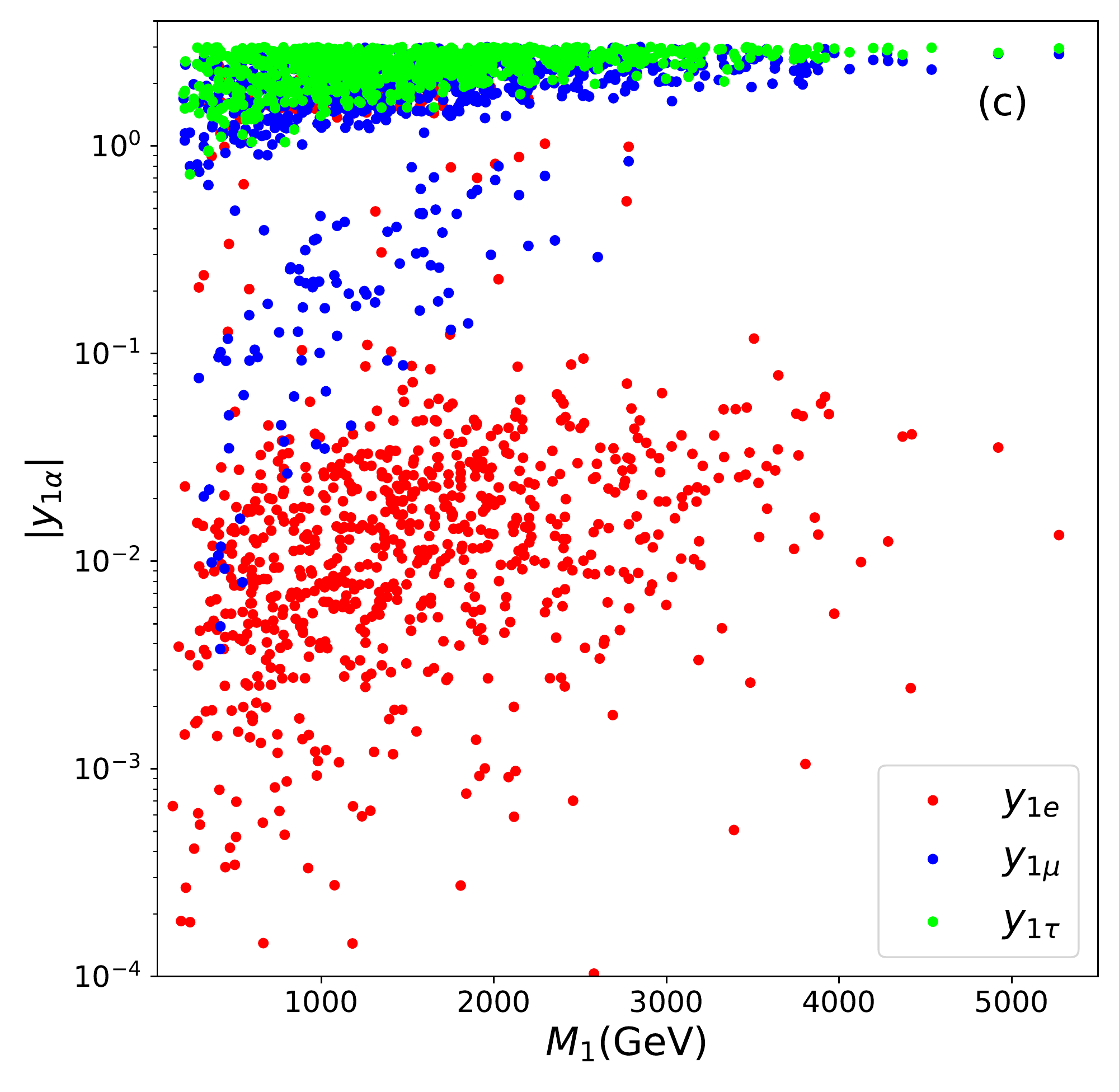}
	\includegraphics[width=0.45\textwidth]{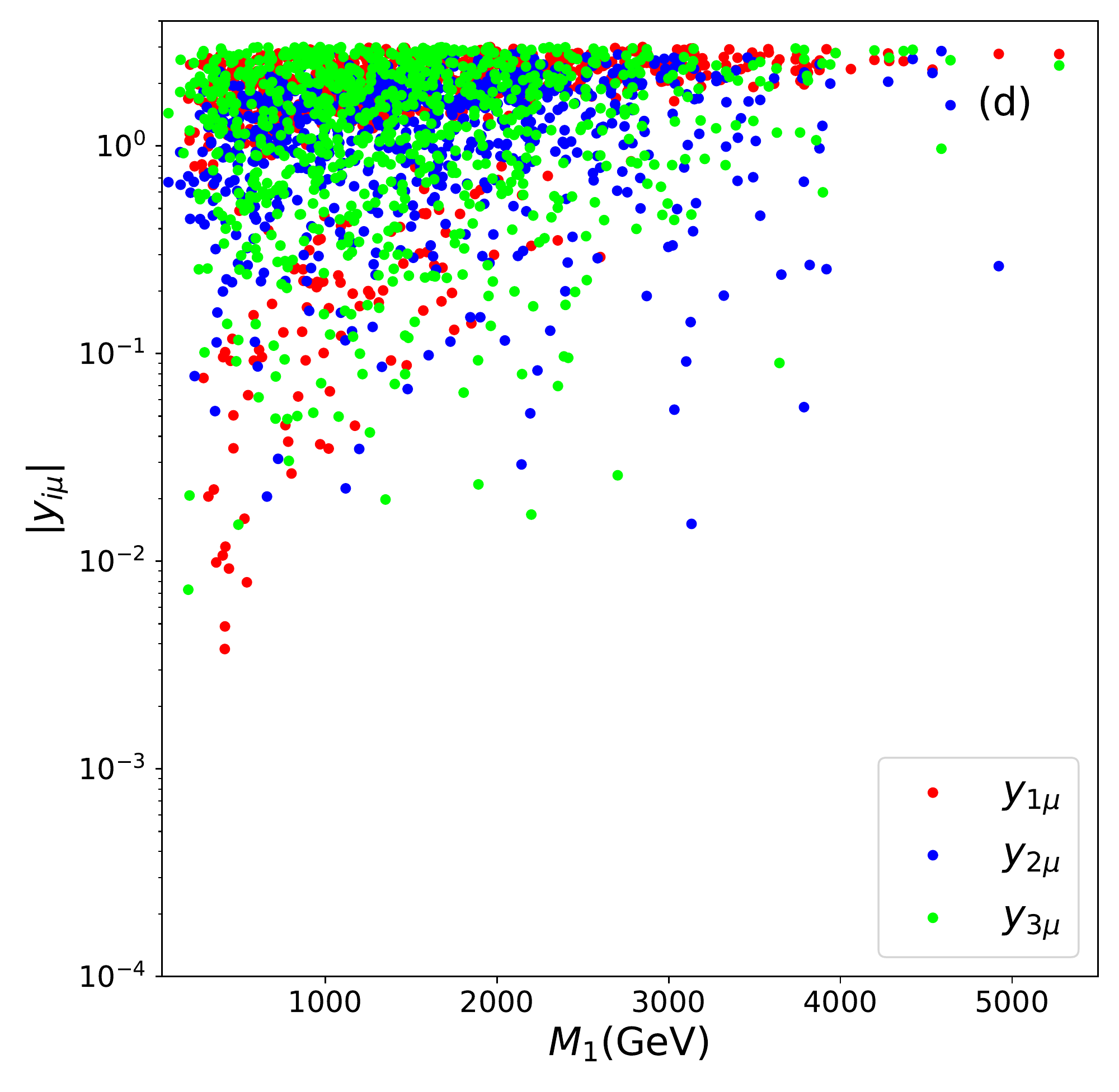}
	\caption{The allowed samples under constraints from lepton flavor violation and dark matter relic density. The gray region in panel (a) is excluded by LHC \cite{ATLAS:2019lff}. The cyan and orange lines correspond to the future limits by CLIC and FCC-hh \cite{Baumholzer:2019twf}.
		\label{FIG:DM}} 	
\end{figure}

In this paper, we assume $N_1$ as the dark matter candidate in the freeze-out scenario. We only consider the pair annihilation channels $N_1 N_1 \to \ell^+\ell^-, \bar{\nu}\nu$. For a compressed mass spectrum $M_1\sim M_{\eta^\pm}$, the $N_1\eta$ coannihilation channels are also possible and may become dominant \cite{Vicente:2014wga}. Provided vanishing lepton masses, the pair annihilation cross-section  in powers of the
relative dark matter velocity ($v_r$) is \cite{Kubo:2006yx}
\begin{equation}
	\sigma v_r = 0+ \frac{y_1^4 r_1^2(1-2r_1+2r_1^4)}{24\pi M_1^2} v_r^2 \equiv a+b v_r^2,
\end{equation}
where $y_1^4=\sum_{\alpha\beta}|y_{1\alpha} y_{1\beta}^*|^2 $, $r_1=M_1^2/(M_{\eta^\pm}^2+M_1^2)$. For simplicity, we assume masses of inert scalars are degenerate $M_{\eta^\pm}\simeq M_{R}\simeq M_{I}$. The thermally averaged cross-section can be calculated as $\langle \sigma v_r \rangle=a+6b/x_f$. Then we can obtain the freeze-out parameter $x_f$ by numerically solving 
\begin{equation}
	x_f=\log \frac{0.0764c(2+c) M_\text{Pl} M_1 6 b/x_f}{\sqrt{g_* x_f}},
\end{equation}
where $c\simeq1/2$, $M_\text{Pl}=1.22\times 10^{19} $ GeV, and $g_*$ is the number of relativistic degrees of freedom. The final dark matter relic density is calculated as
\begin{equation}
	\Omega h^2 = \frac{1.07\times 10^9 \text{GeV}^{-1}}{M_\text{Pl}} \frac{x_f}{\sqrt{g_*}}\frac{1}{a+3b/x_f}.
\end{equation}

\begin{figure} 
	\centering
	\includegraphics[width=0.5\textwidth]{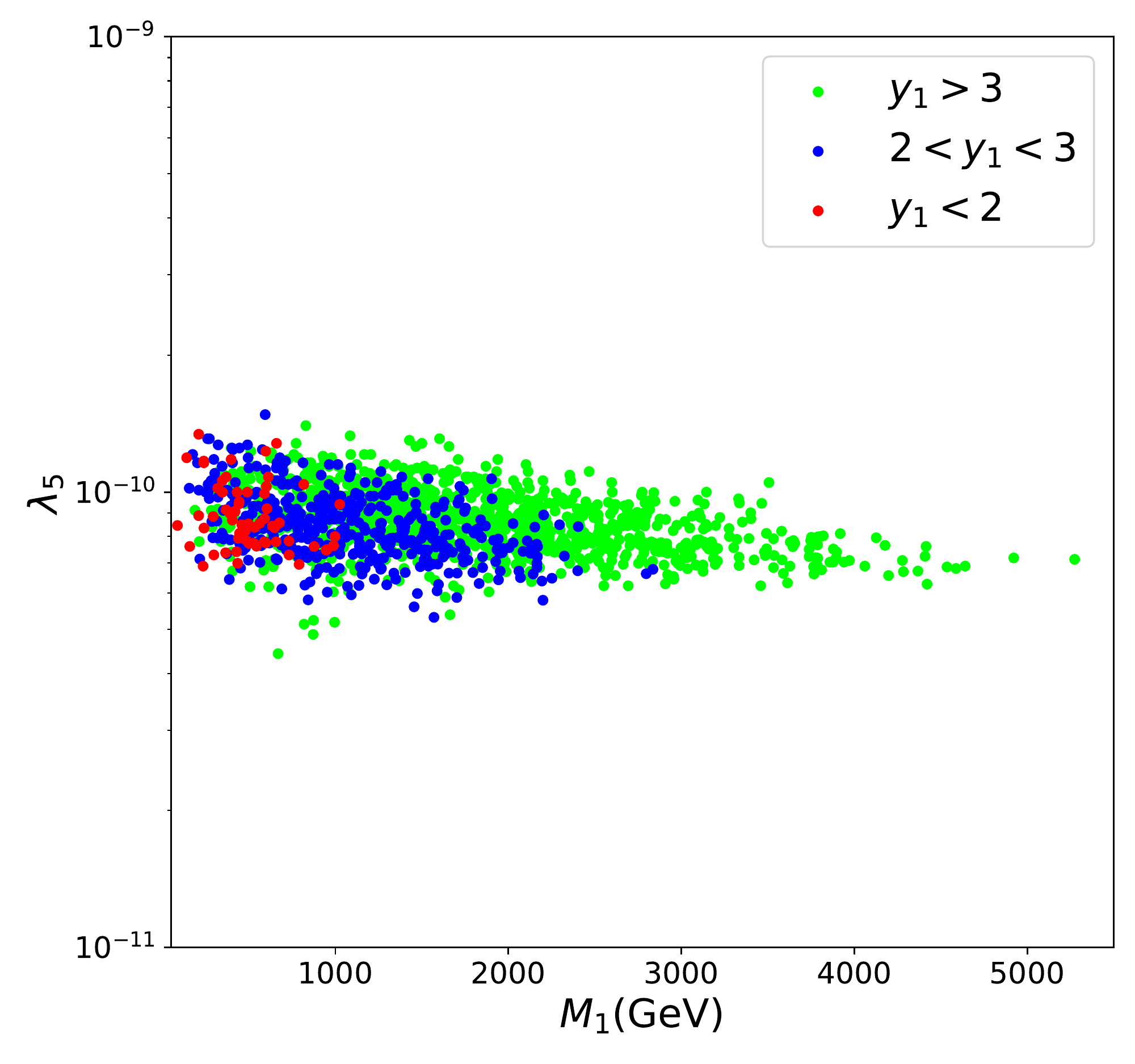}
	\caption{Same as Figure \ref{FIG:DM}, but in the $M_1-\lambda_5$ plane.
		\label{FIG:lam5}} 	
\end{figure}

For the fermion dark matter $N_1$, the interactions with nucleons appear at the one-loop level. As already discussed in Ref.~\cite{Ibarra:2016dlb}, the spin-independent cross-section via  the Higgs exchange is
 \begin{equation}
 	\sigma_\text{SI}=\frac{4}{\pi} \frac{M_1^2 m_p^2}{(M_1+m_p)^2} m_p^2 \left(\frac{\Lambda_q}{m_q}\right)^2 f_p^2,
 \end{equation}
where $m_p$ is the proton mass, $f_p\approx 0.3$ is the scalar form factor, and the effective scalar coupling $\Lambda_q$ is
\begin{equation}
	\Lambda_q=-\frac{y_1^2}{16\pi^2 M_h^2 M_1}\left[\lambda_3 G_1\left(\frac{M_1^2}{M_{\eta^\pm}^2}\right)+\frac{\lambda_3+\lambda_4}{2}G_1\left(\frac{M_1^2}{M^2_{R,I}}\right)\right]m_q,
\end{equation}
with the loop function $G_1(x)$ defined as
\begin{equation}
	G_1(x)=\frac{x+(1-x)\log(1-x) }{x}.
\end{equation}
To make sure the explored parameter space is safe under the tight constraints from direct detection \cite{XENON:2018voc,PandaX-4T:2021bab}, we assume the couplings $\lambda_{3,4}= 0.01$ in the following study \cite{Ibarra:2016dlb}.

Since the neutrino mass, lepton flavor violation, and dark matter relic density involve the same Yukawa interactions, a combined analysis is required to explore the parameter space. Based on the strategy in Ref.~\cite{Vicente:2014wga}, we perform a random scan over the following parameter space:
\begin{eqnarray}
	M_1\in[100,10000]~\text{GeV}, M_{\eta^\pm}\in [100,10000]~\text{GeV}, \\ \nonumber
	M_2\in[M_1,20M_{\eta^\pm}], M_3\in[M_2,200 M_{\eta^\pm}],\lambda_5 \in [10^{-11},10^{-9}].
\end{eqnarray}
During the scan, we assume flat prior distributions for all the parameters.

For $M_1<100$~GeV, the parameter space has been mostly excluded by lepton flavor violation \cite{Vicente:2014wga} and direct search at LHC \cite{ATLAS:2019lff,ATLAS:2019lng,CMS:2020bfa}, thus this region is not considered in this work. Slightly different from the previous study in Ref.~\cite{Vicente:2014wga}, the upper bounds on the heavier singlet fermions $N_2$ and $N_3$ are much higher in our studies. The viable samples are required to satisfy current LFV bounds and predict the correct dark matter relic density as well \cite{Planck:2018vyg}. During the scan, we also impose a perturbativity limit on the Yukawa couplings $|y_{i\alpha}|<3$. 
\begin{table}
		\renewcommand\arraystretch{1.25}
		\begin{center}	
	\begin{tabular}{|c|c|c|c|c|c|c|c|c|c|}
		\hline
		& $M_1$  & $M_2$   & $M_3$   &  $M_{\eta^\pm}$   & $|y_{1e}|$ & $|y_{1\mu}|$ & $|y_{1\tau}|$ &  $\Omega h^2$
		\\\hline 
		BP-1 & 648.7 & $5.06\!\times\!10^3$ & $3.39\!\times\!10^4$ & 1141 & $2.62\!\times\!10^{-2}$ & 1.17 & 1.49  & 0.122
		\\ \hline
		BP-2 & 1579  & $1.82\!\times\!10^4$ & $1.22\!\times\!10^5$ & 2647 & $9.05\!\times\!10^{-4}$ & 1.99 & 2.08  & 0.123 
		\\ \hline
		BP-3 & 2521 & $6.30\!\times\!10^4$  & $5.52\!\times\!10^5$ & 4371 & $2.17\!\times\!10^{-2}$ & 2.31 & 2.95  & 0.121
		\\ \hline
	\end{tabular}
	\end{center}
	\caption{Benchmark points for MuC studies. Here, all the messes are in the unit of GeV.}
	\label{TB:BP}
\end{table}

The allowed samples and possible collider limits are shown in Figure~\ref{FIG:DM} and \ref{FIG:lam5}. From the distribution in the $M_{\eta^\pm}-M_1$ plane, it is clear that the allowed region satisfies $M_{\eta^\pm}\lesssim5.5$ TeV and $M_1\lesssim 5$ TeV. In Figure~\ref{FIG:DM} (a), the excluded regions by current LHC, future CLIC and FCC-hh are also shown. The current LHC exclusion region only covers a small portion of the parameter space. The CLIC has the potential to cover the whole region with $y_1<2$. Meanwhile, FCC-hh could probe the region $M_{\eta^\pm}\lesssim 2$ TeV with most samples satisfying $y_1\lesssim 3$. For samples with $y_1\gtrsim 3$, most of them are beyond the reach of LHC, CLIC and FCC-hh. However, as will be shown later, such samples can be easily probed at a multi-TeV MuC. The parameter $\xi_i=M_i^2/M_{\eta^\pm}^2$ is shown in Figure~\ref{FIG:DM}~(b). We notice that the upper bounds on $\xi_i$ shown in Figure 4 of Ref.~\cite{Vicente:2014wga} actually correspond to the requirements $M_{2(3)}<5(10)$ TeV. Therefore, by allowing much larger values of $M_{2,3}$, we obtain a wider parameter space in the $M_{\eta^\pm}-M_1$ plane. Our results imply that $M_1<M_{\eta^\pm}<M_2<M_3$ is the favored mass spectrum for TeV-scale $M_1$. In Figure~\ref{FIG:DM}~(c), the distributions of $|y_{1\alpha}|$, which is correlated with LFV and relic density, are shown. The TeV scale dark matter usually requires a hierarchical structure $|y_{1e}|\ll|y_{1\mu}|\sim|y_{1\tau}|$ with $|y_{1e}|\lesssim0.1, 1\lesssim |y_{1\mu}|\lesssim |y_{1\tau}|$. Hence, the dark matter $N_1$ and inert charged scalar $\eta^\pm$ behave as $\mu\tau$-philic particles. As a complementary, the $\mu$-related Yukawa couplings $|y_{i\mu}|$ are also depicted in Figure~\ref{FIG:DM}~(d), which are involved in the production of $\eta^+\eta^-$ at MuC. Most samples predict $|y_{i\mu}|\gtrsim0.1$ without any special hierarchical structure. With sizable Yukawa couplings, correct neutrino mass can still be obtained by tiny $\lambda_5$. In Figure~\ref{FIG:lam5}, the allowed values of $\lambda_5$ are shown, which indicates that $\lambda_5\sim10^{-10}$ should be satisfied. 

Based on the above discussion, we have selected three benchmark points (BP) for the following MuC studies. Detailed information of the BPs are listed in Table~\ref{TB:BP}. BP-1 is also within the reach of CLIC and FCC-hh, while BP-2 and BP-3 are out reach of them. Notably, $|y_{1\tau}|=2.954$ in BP-3 is close to the upper bound of the perturbativity limit $|y_{i\alpha}|<3$.

\section{Signatures at Muon Collider} \label{Sec:MC}

\begin{figure}
	\centering
	\includegraphics[width=0.9\textwidth]{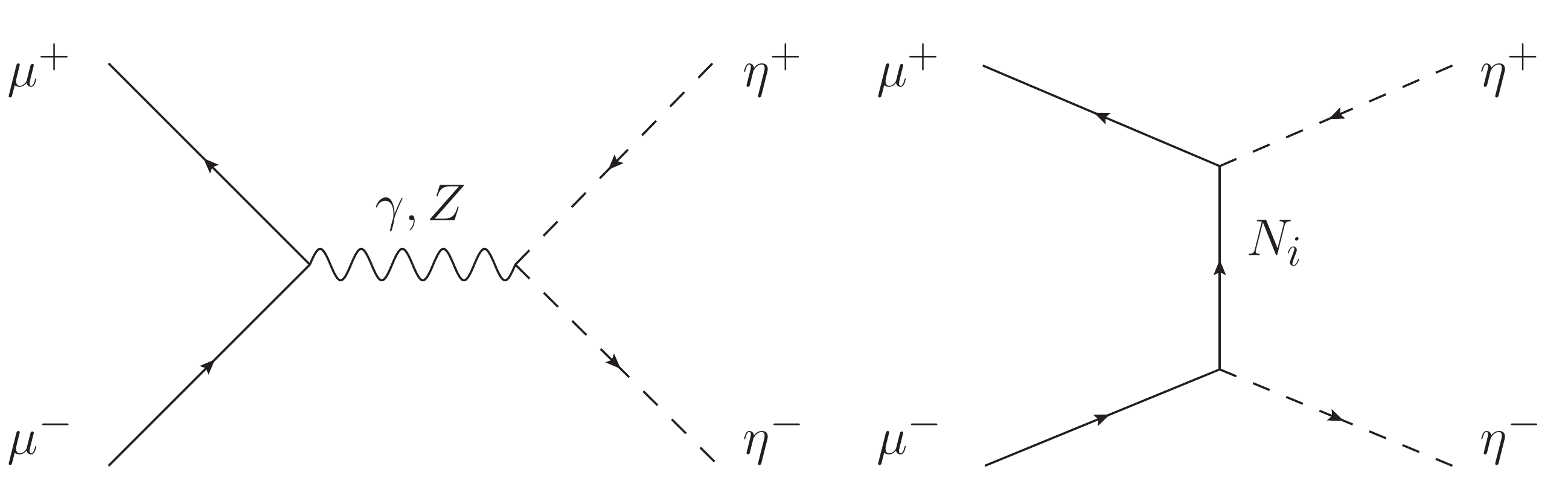}
	\caption{Charged scalar pair production at the muon collider.}
	\label{FIG:MuC}
\end{figure}

In this section, we investigate the signatures of the Scotogenic model at a multi-TeV MuC, which is well motivated due to large $\mu$-related Yukawa coupling.
As shown in Figure~\ref{FIG:MuC}, the charged scalar $\eta^\pm$ can be pair produced via the $s$-channel exchange of $\gamma/Z$ boson, and via the $t$-channel exchange of $N_i$.
The corresponding production cross-section is then calculated as \cite{Djouadi:2001yk}
\begin{eqnarray}\notag
	\sigma(\mu^+\mu^-\!\!\to\! \eta^+\eta^-)\!\! &=&\! \frac{\pi\alpha^2}{s}\Bigg\{\frac{1}{3}\beta^3\!\!\left[1\!+\! g_L(g_L\!+\! g_R) \frac{s}{s-M_Z^2}\!+\! g_L^2(g_L^2\!+\! g_R^2) \frac{s^2}{(s-M_Z^2)^2}\right] \\ 
	&&\sum_{i=1}^3\sum_{j=1}^3\frac{|y_{i \mu}|^2 |y_{j \mu}|^2}{64\pi^2\alpha^2} H_{ij}+
	 \sum_{i=1}^3 \frac{|y_{i \mu}|^2}{8\pi \alpha} \left[1+g_L^2 \frac{s}{s-M_Z^2}\right] F_i\Bigg\},
\end{eqnarray}
where the left- and right-chiral $Z$ couplings are 
\begin{equation}
	g_L=\frac{-1+2s_W^2}{2s_W c_W}, g_R=\frac{s_W}{c_W},
\end{equation}
with $s_W=\sin\theta_W,c_W=\cos\theta_W$, and $\theta_W$ is the weak mixing angle. The kinematical functions $H_{ij}$ and $F_i$ read
\begin{eqnarray}
		H_{ij} &=& \left\{
		\begin{array}{lc}
			-2\beta+\Delta_i \ln \frac{\Delta_i+\beta}{\Delta_i-\beta} & \quad i=j \\
			\frac{F_j-F_i}{\Delta_i-\Delta_j} & \quad i\neq j
		\end{array}\right.,\\
	F_i &=&  \Delta_i \beta - \frac{\Delta_i^2-\beta^2}{2}\ln \frac{\Delta_i+\beta}{\Delta_i-\beta},
\end{eqnarray}
where the parameters $\Delta_i$ and $\beta$ are given by
\begin{equation}
	\Delta_i=\frac{2}{s}(M_{\eta^\pm}^2-M_i^2)-1, \beta = \sqrt{1-4M_{\eta^\pm}^2/s}.
\end{equation}

\begin{figure}
	\centering
	\includegraphics[width=0.5\textwidth]{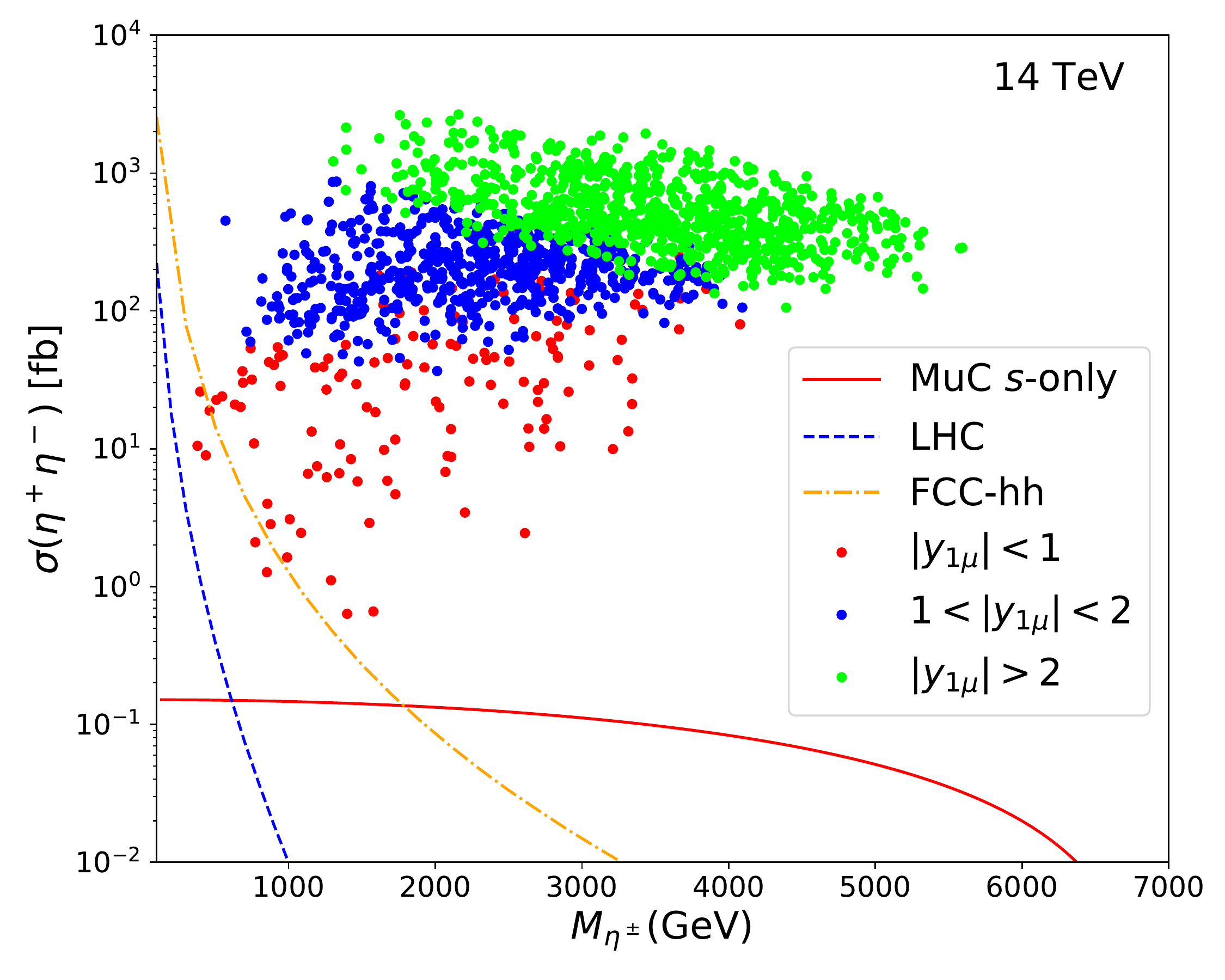}
	\caption{The production cross-section of $\mu^+\mu^-\to \eta^+\eta^-$ at 14 TeV MuC. The red solid line is the cross-section at MuC with $s$-channel contribution only. The blue dashed (yellow dash-dotted) line corresponds to cross-section at 14 TeV LHC (100 TeV FCC-hh).}
	\label{FIG:PR}
\end{figure}

The theoretically predicted production cross-section $\sigma(\eta^+\eta^-)$ at 14 TeV MuC is shown in Figure~\ref{FIG:PR}. It is obvious that the cross-section can be greatly enhanced by the $t$-channel contribution. The strong correlation between $\sigma(\eta^+\eta^-)$ and $|y_{1\mu}|$ indicates that the $t$-channel exchange of $N_1$ is the dominant contribution. Compared with the $s$-channel, the enhancement factor due to the $t$-channel might be naively estimated as $|y_{1\mu}|^4/(16\pi^2\alpha^2)$, e.g., $|y_{1\mu}|=2$ leading to an enhancement factor at the order of $\sim$2000. For a fixed value of $|y_{1\mu}|$, $\sigma(\eta^+\eta^-)$ tends to decrease as $M_{\eta^\pm}$ increases. Meanwhile, as the lower limits on $|y_{1\mu}|$ becomes higher when $\eta^\pm$ is heavier, the resulting predicted lower bounds of $\sigma(\eta^+\eta^-)$  will increase. The $\sigma(\eta^+\eta^-)$ could reach a maximum value of about 3000~fb with $M_{\eta^\pm}\sim$ 2~TeV. For charged scalar $\eta^\pm$ heavier than 4 TeV, the $\sigma(\eta^+\eta^-)$ is roughly in the range of [100,1000]~fb with $|y_{1\mu}|>2$. When $|y_{1\mu}|>1$, the $\sigma(\eta^+\eta^-)\gtrsim 100$ fb is always expected. Samples with $|y_{1\mu}|<1$ usually predict a cross-section smaller than 100~fb, where contributions from $N_{2,3}$ may become important if $|y_{1\mu}|$ is too small. All the samples lead to a larger cross-section at MuC than at LHC. Only a few samples with $M_{\eta^\pm}\lesssim 1000$ GeV and $|y_{1\mu}|<1$ have a cross-section at MuC smaller than at FCC-hh. Actually, for $M_{\eta^\pm}\gtrsim1800$~GeV, the cross-section at MuC with only $s$-channel contribution is larger than at FCC-hh.

\begin{figure} 
	\centering
	\includegraphics[width=0.45\textwidth]{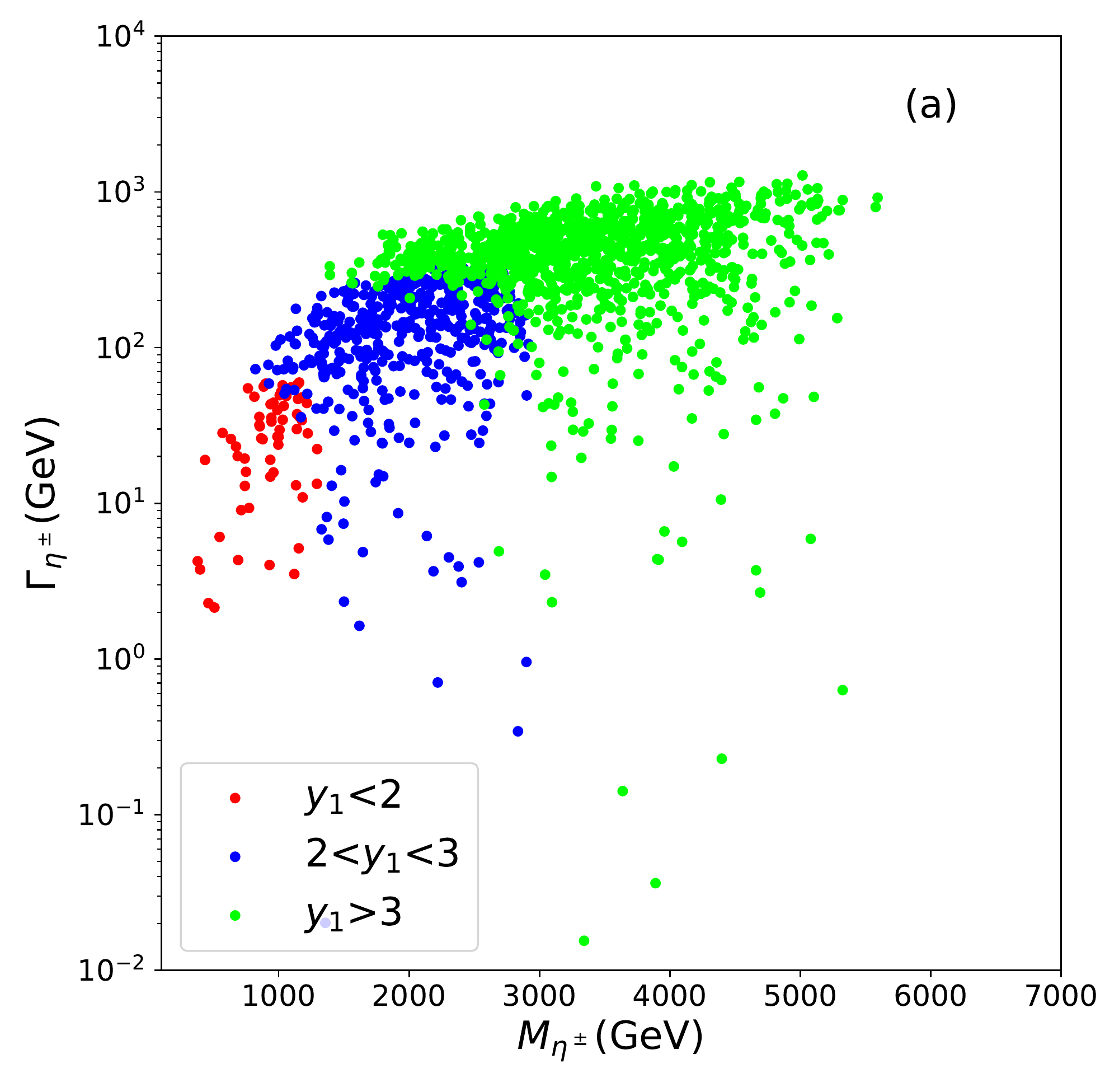}
	\includegraphics[width=0.45\textwidth]{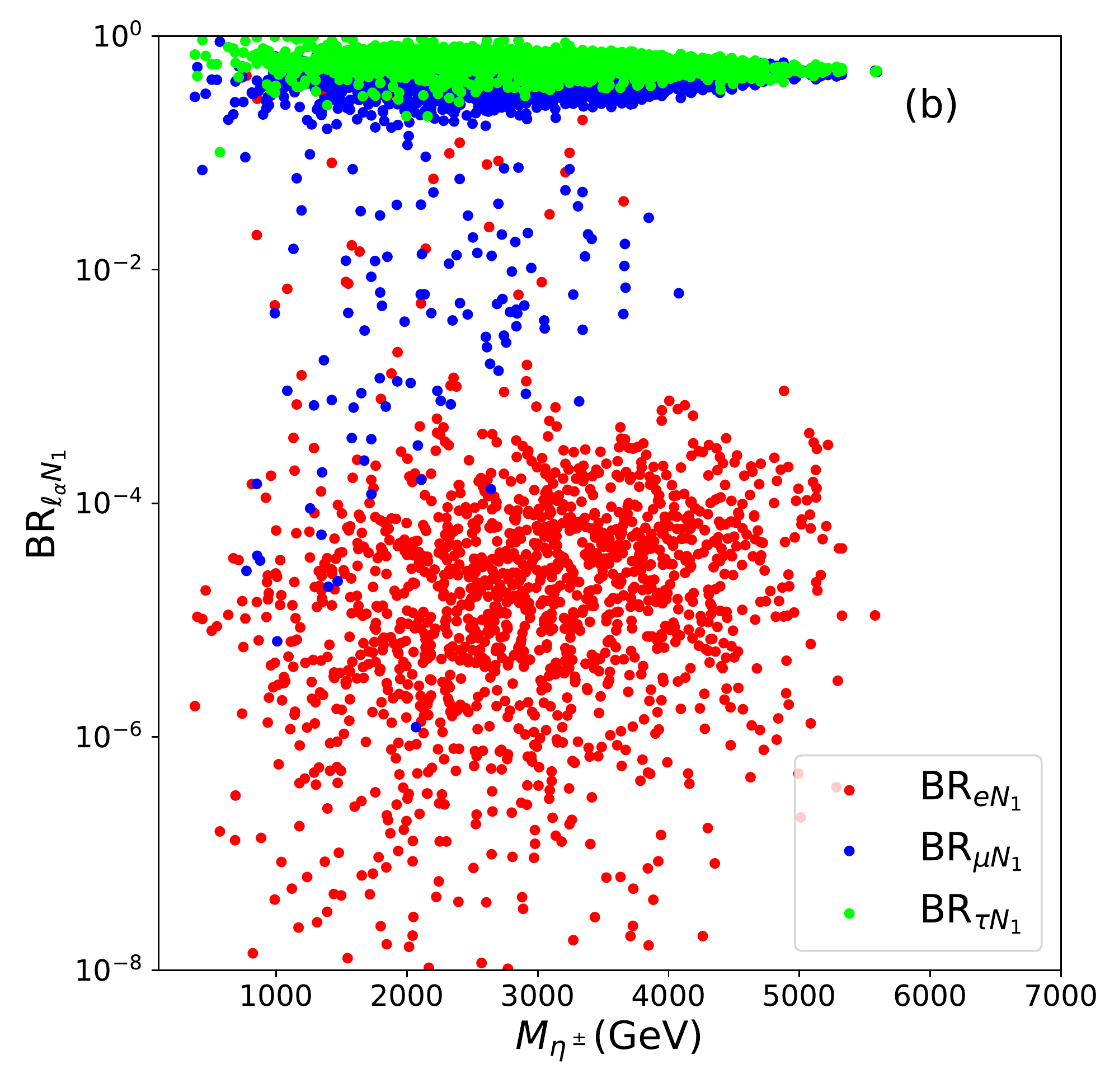}
	\caption{Left: The total decay width of charged scalar $\eta^\pm$. Right: The branching ratios of charged scalar $\eta^\pm$.
		\label{FIG:BR}} 	
\end{figure}
Now we consider the decays of charged scalar $\eta^\pm$. The corresponding decay width is
\begin{equation}
	\Gamma(\eta^\pm\to \ell^\pm_\alpha N_i)=\frac{|y_{i\alpha}|^2}{16\pi}M_{\eta^\pm}\left(1-\frac{M_i^2}{M_{\eta^\pm}^2}\right)^2.
\end{equation}
In principle, $\eta^\pm$ can decay into all three generations of $N_i$. However, our scanned samples favor the spectrum $M_1<M_{\eta^\pm}<M_2<M_3$, which means $\eta^\pm$ can only decay into $N_1$ via $\eta^\pm \to \ell^\pm_\alpha N_1$.
The total decay width $\Gamma_{\eta^\pm}$ and branching ratios are shown in Figure~\ref{FIG:BR}. The total decay width $\Gamma_{\eta^\pm}$ can be as large as 1 TeV with $y_1>3$. Meanwhile for $y_1<2$, the total decay width is less than 100 GeV. Suppression of $\Gamma_{\eta^\pm}$ is also possible with compressed spectrum $M_1\sim M_{\eta^\pm}$. With hierarchical Yukawa couplings $|y_{1e}|\ll |y_{1\mu}|\sim |y_{1\tau}|$, the corresponding branching ratios are also hierarchical, i.e., BR$_{e N_1}\ll$ BR$_{\mu N_1}\lesssim$ BR$_{\tau N_1}$. The $\tau$ final state has a branching ratio always larger than 0.2. Most samples predict BR$_{\mu N_1}\gtrsim 10^{-2}$ and BR$_{eN_1}\lesssim10^{-4}$.

\subsection{$\mu^+\mu^-+\cancel{E}_T$ Signature}

The usual dilepton signature considers both electron and muon final states. However, in the Scotogenic model, the BR$_{eN_1}$ is suppressed by tiny $|y_{1e}|$. So we only consider the muon channel. The explicit opposite-sign dimuon signature at MuC is
\begin{equation}
	\mu^+\mu^-\to \eta^+\eta^-\to \mu^+N_1 + \mu^- N_1 \to \mu^+\mu^-+\cancel{E}_T.
\end{equation}
The corresponding backgrounds are
\begin{eqnarray}
	\mu^+\mu^-\to \mu^+\mu^-, \mu^+ \mu^- \bar{\nu}_\ell\nu_\ell, W^+W^-\bar{\nu}_\ell \nu_\ell,
\end{eqnarray}
with the leptonic decay $W^\pm \to \mu^\pm \nu_\mu$ and $\nu_\ell=\nu_{e,\mu,\tau}$ including all flavors. Here, the $\mu^+\mu^-\to W^+W^- \bar{\nu}_\ell \nu_\ell$ is dominant by the VBF process \cite{Costantini:2020stv,Han:2020uid,Ruiz:2021tdt}.
The dimuon signature and corresponding backgrounds are simulated with {\bf MadGraph5\_aMC@NLO}~\cite{Alwall:2014hca}. The detector effects are illustrated by using {\bf Delphes}~\cite{deFavereau:2013fsa} with the parameters in the delphes\_card\_MuonColliderDet.tcl card. In this paper, we perform a simple cut-based analysis of the signature. The cut flow for the dimuon signature and backgrounds are summarized in Table~\ref{Tab:MuMu}.

\begin{figure}
	\begin{center}
		\includegraphics[width=0.45\linewidth]{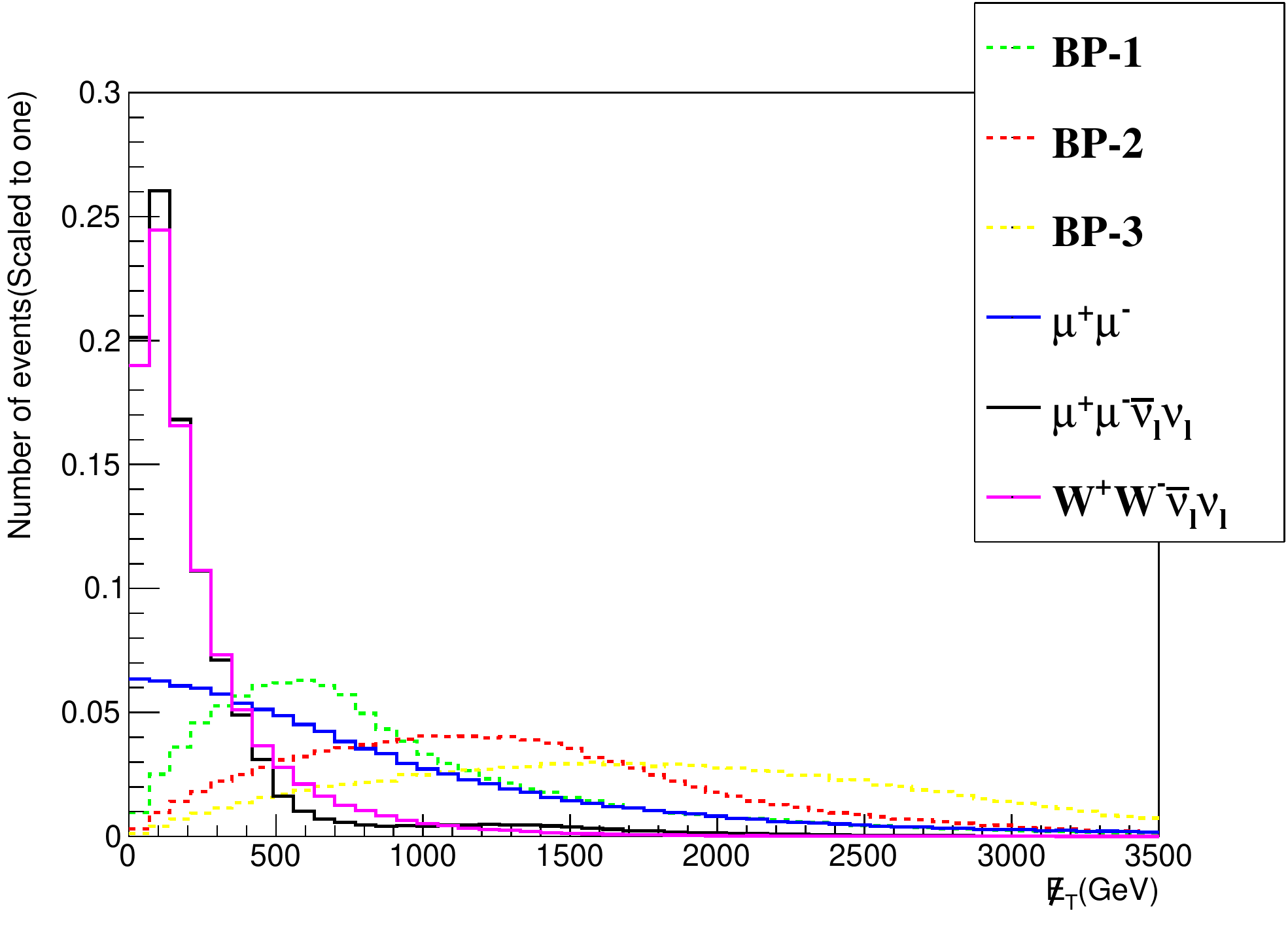}
		\includegraphics[width=0.45\linewidth]{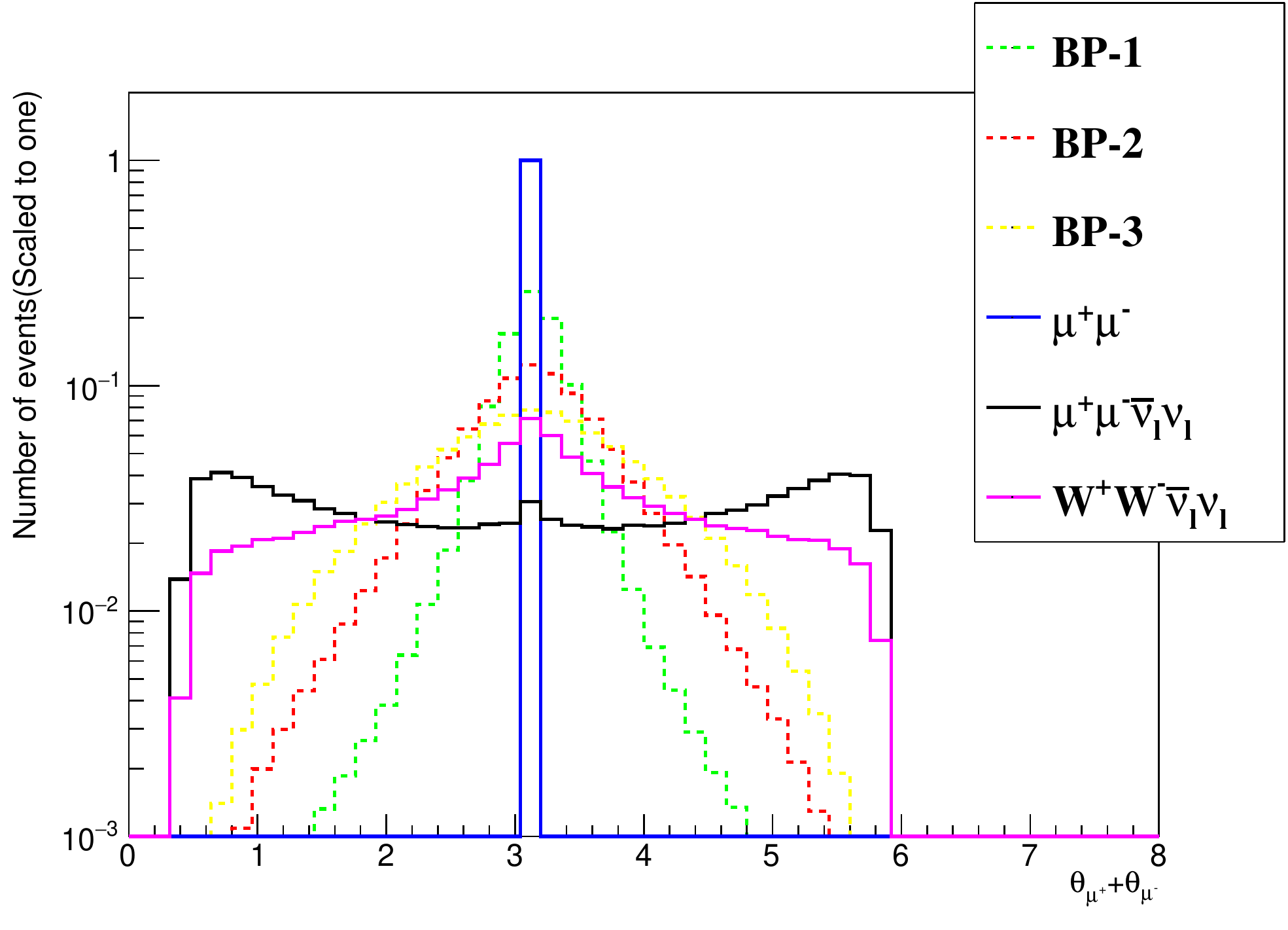}
		\includegraphics[width=0.45\linewidth]{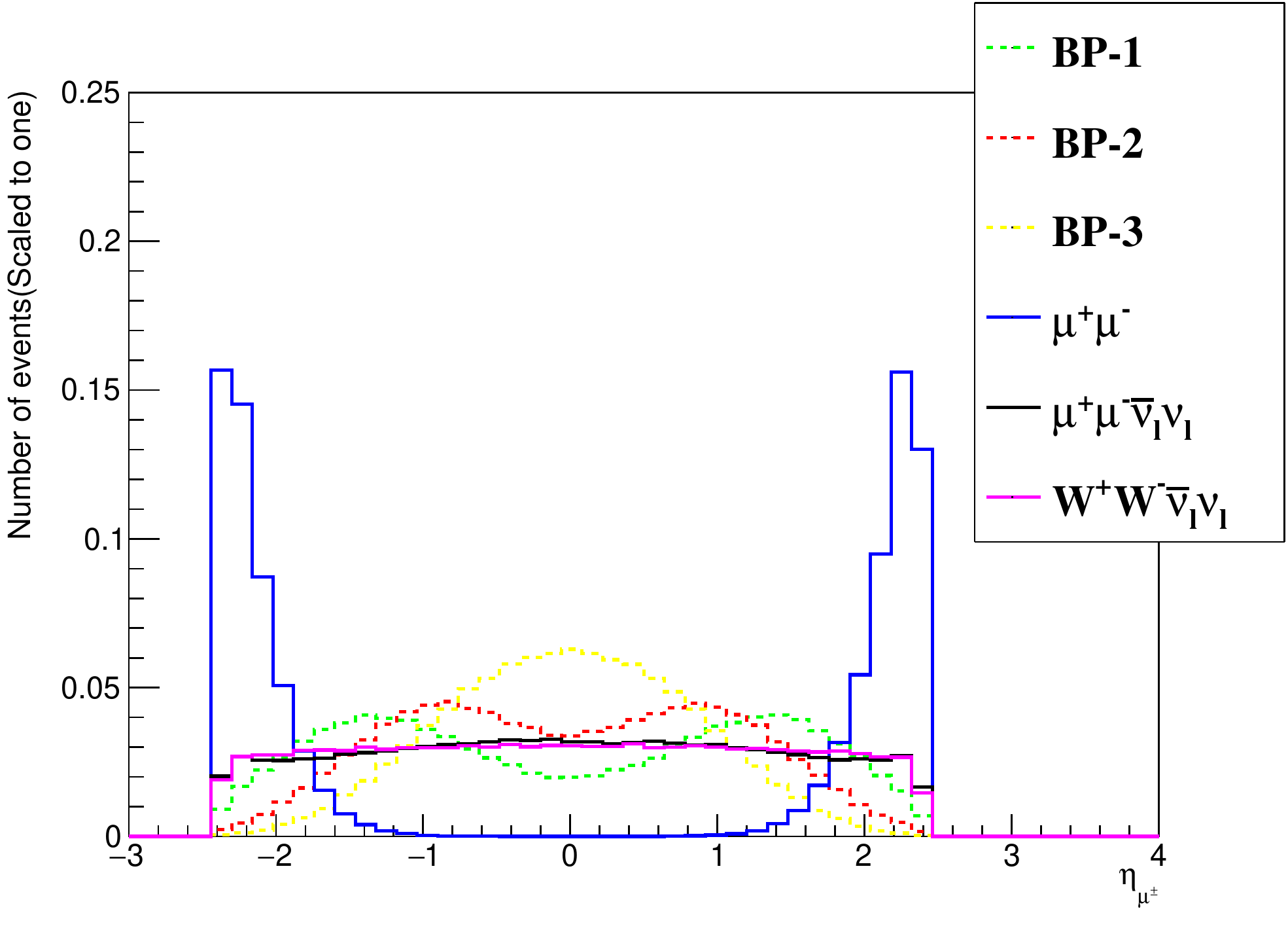}
		\includegraphics[width=0.45\linewidth]{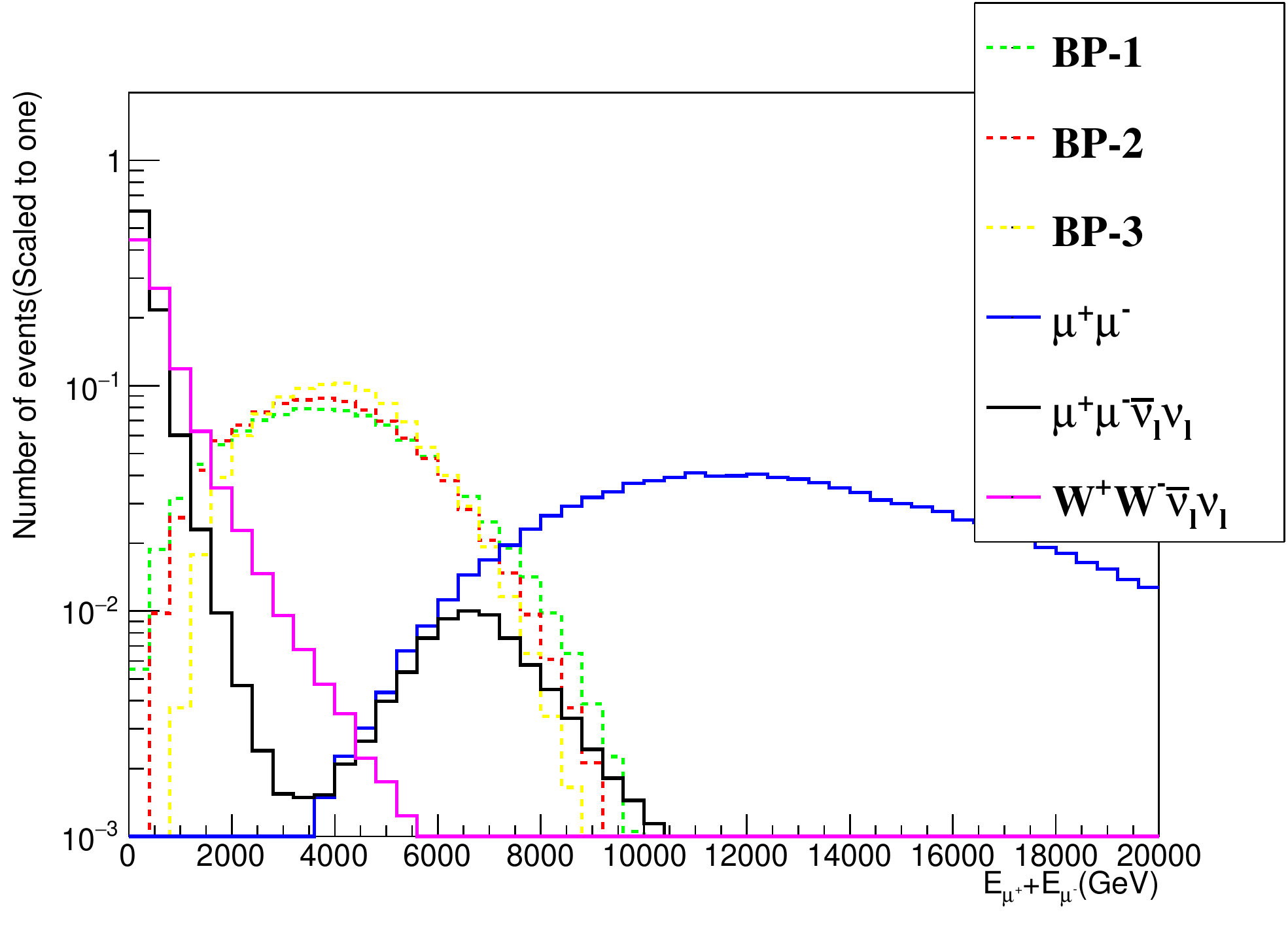}
		\includegraphics[width=0.45\linewidth]{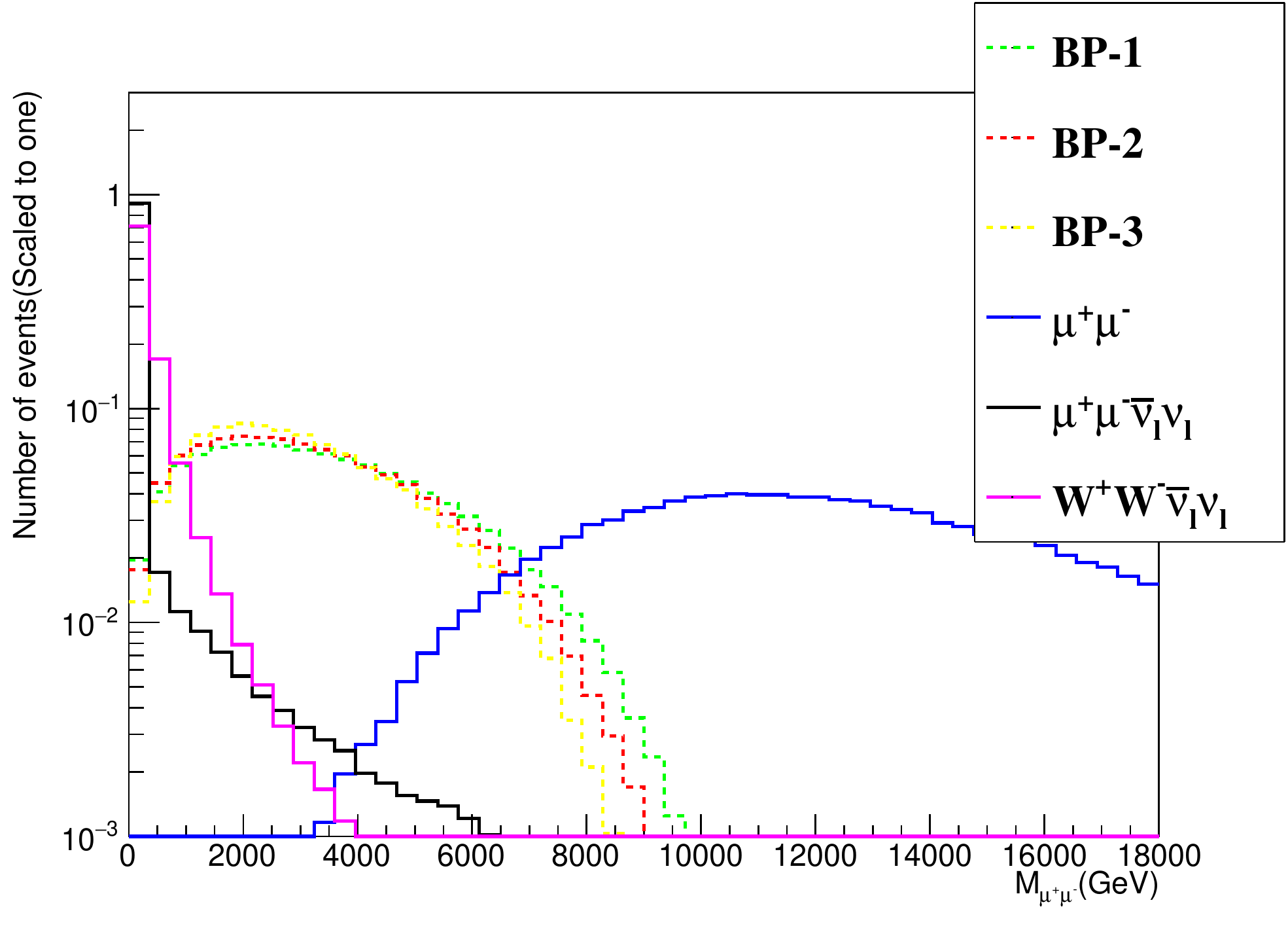}
		\includegraphics[width=0.45\linewidth]{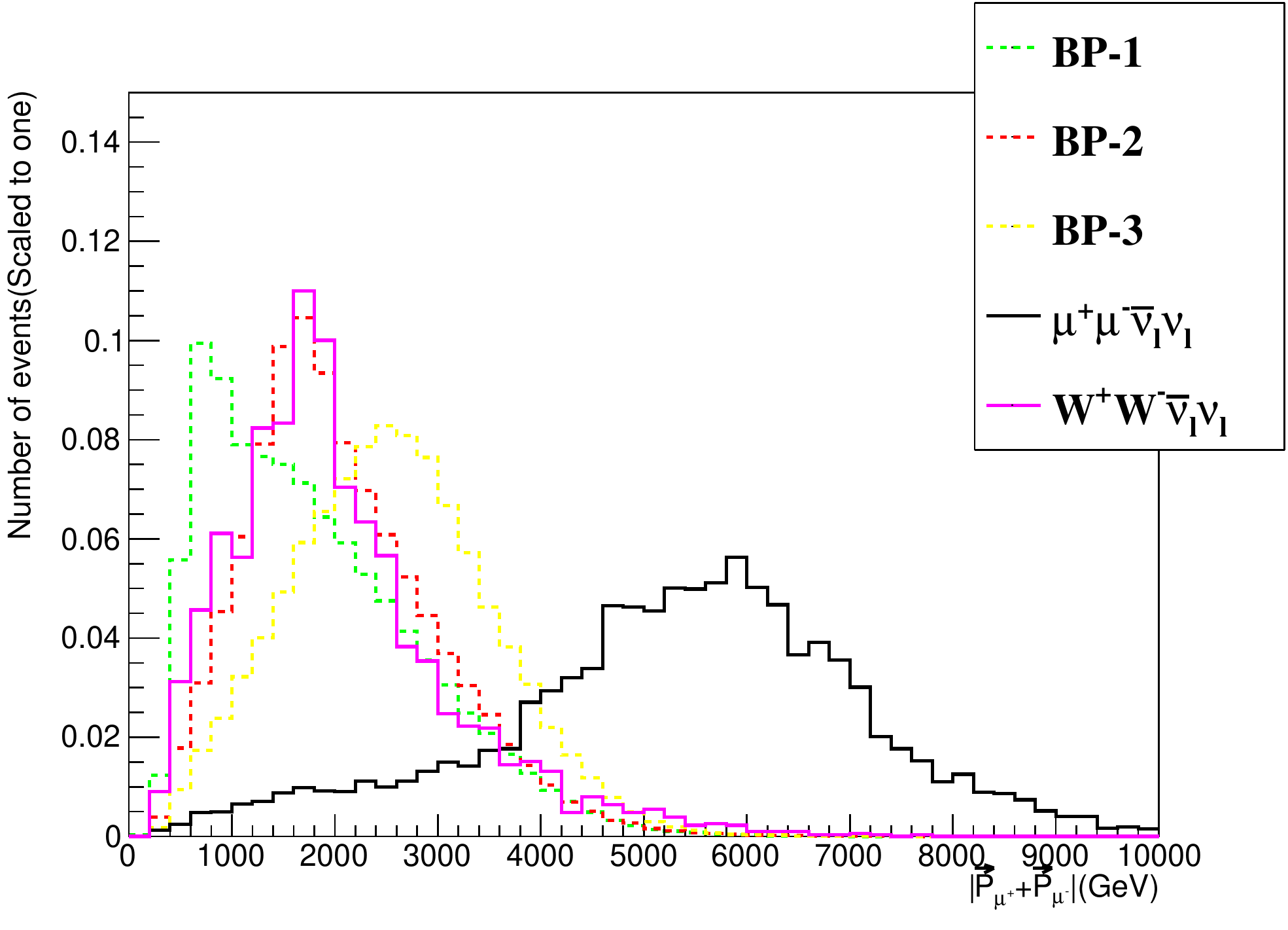}
	\end{center}
	\caption{Normalized distribution of missing transverse energy $\cancel{E}_T$ (up-left panel), dimuon acollinearity $\theta_{\mu^+}+\theta_{\mu^-}$ (up-right panel),  pseudorapidity $\eta_{\mu^\pm}$ (middle-left panel), dimuon energy $E_{\mu^+}+E_{\mu^-}$ (middle-right panel), dimuon invariant mass $M_{\mu^+\mu^-}$ (down-left panel), vector sum momentum of dimuon $|\vec{P}_{\mu^+}+\vec{P}_{\mu^-}|$ (down-right panel) for benchmark points (dashed lines) and corresponding backgrounds (solid lines) at the $s=\sqrt{14}$ TeV muon collider. Variable $|\vec{P}_{\mu^+}+\vec{P}_{\mu^-}|$ is shown after applying cuts from Equation \eqref{Eq:Nmu} to \eqref{Eq:Mmu}, while the other five variables are shown with only the cuts in Equation \eqref{Eq:Nmu}.}
	\label{FIG:DISMu}
\end{figure}

First, we select events with exact two opposite-sign muons via the following cuts 
\begin{equation}\label{Eq:Nmu}
	N_{\mu^\pm}=1, P_T^{\mu^\pm}>10~\text{GeV}.
\end{equation}
Because the dark matter $N_1$ is invisible at detectors, the missing transverse energy $\cancel{E}_T$ is the characteristic feature of this signal. Distributions of $\cancel{E}_T$ for the signals and backgrounds are shown in the up-left panel of Figure~\ref{FIG:DISMu}. The signals tend to have larger $\cancel{E}_T$ than the backgrounds. In our analysis, we require a relatively large $\cancel{E}_T$ to suppress the backgrounds
\begin{equation}
	\cancel{E}_T>300~\text{GeV}.
\end{equation}

\begin{table}
	\renewcommand\arraystretch{1.25}
	\begin{center}
		\begin{tabular}{c| c c c | c c c} 
			\hline
			\hline
			$\sigma$(fb) & \textbf{BP-1} & \textbf{BP-2} & \textbf{BP-3} & $\mu^+\mu^-$ & $\mu^+ \mu^-\bar{\nu}_\ell\nu_\ell$ & $W^+W^-\bar{\nu}_\ell \nu_\ell$ \\
			\hline
			Preselection & 19.7  & 93.8  & 35.1  & 220 & 93.4  & 6.41 \\
			\hline
			\tabincell{c}{$N_{\mu^\pm}=1$\\ $P_T^{\mu^\pm}>10$ GeV}
			  & 8.10  &  40.6  &  12.9  &  69.7 &  69.3 & 2.25 \\
			\hline
			$\cancel{E}_T>300$ GeV & 7.08 & 38.7 & 12.5 & 52.3 & 16.5 & 0.63 \\
			\hline
			\tabincell{c}{$|\pi-\theta_{\mu^+}-\theta_{\mu^-}|$ \\ $ >0.0016$}
			 & 7.03 & 38.6 & 12.5 & 0.00 & 16.5 & 0.63 \\
			\hline
		    $|\eta_{\mu^\pm}|<2$& 5.86 & 34.5 & 11.7 & 0.00 & 11.1 & 0.52 \\
			\hline
			$E_{\mu^+}+E_{\mu^-}>2$ TeV & 4.90 & 29.6 & 11.0 & 0.00 & 1.95 & 0.10 \\
			\hline
			$M_{\mu^+ \mu^-}>1$ TeV & 4.77 & 28.4 & 10.2 & 0.00 & 1.23 & 0.06 \\
			\hline
			 $|\vec{P}_{\mu^+}\!+\!\vec{P}_{\mu^-}|<3.5$ TeV 
			 & 4.51 & 26.2 & 8.50 & 0.00 & 0.15 & 0.06 \\ 
			\hline
			\hline
			Significance &  29.4  &  72.1   &   40.7 & \multicolumn{2}{c|}{\multirow{2}{*}{Total Background}} & \multirow{2}{*}{0.21}\\
			\cline{1-4}
			$5\sigma$ Luminosity (fb$^{-1}$)&  5.80 &  0.96 &  3.01  & \multicolumn{2}{c|}{ } \\
			\hline
			\hline
		\end{tabular}
	\end{center}
	\caption{Cut flow table for the dimuon signature $\mu^+\mu^-+\cancel{E}_T$ from BP-1, BP-2, BP-3, and various background processes at the $\sqrt{s}=14$ TeV MuC. The significance $ S/\sqrt{S+B} $ is calculated by assuming an integrated luminosity $\mathcal{L} = 200~\text{fb}^{-1}$. 
		\label{Tab:MuMu}}
\end{table} 

At this level, the cross-section of the $\mu^+\mu^-\to \mu^+\mu^-$ channel is still quite large. To dig signal events from background events, further cuts should be applied. The normalized distribution of relevant parameters are also shown in Figure~\ref{FIG:DISMu}.  From the distribution of dimuon acollinearity $\theta_{\mu^+}+\theta_{\mu^-}$ in the up-right panel, it is clear that the corresponding out-going muons in the $\mu^+\mu^-$ channel are back-to-back. Therefore, this background can be easily suppressed to a negligible level by the requirement
\begin{equation}\label{Eq:theta}
	|\pi-\theta_{\mu^+}-\theta_{\mu^-}|>0.0016.
\end{equation} 

After suppressing $\mu^+\mu^-$, the $\mu^+\mu^-\bar{\nu}_\ell \nu_\ell$ channel with a cross-section of 15.0 fb becomes the dominant one, which is already smaller than the cross-section of BP-2. In order to determine the masses of charged-scalar $\eta^\pm$ and dark matter $N_1$ more precisely, we try to eliminate the backgrounds as small as possible. In the distribution of pseudorapidity $\eta_{\mu^\pm}$ in the middle-left panel of Figure~\ref{FIG:DISMu}, the background $\mu^+\mu^-\bar{\nu}_\ell \nu_\ell$ has a nearly flat distribution, so we tighten the events with the cut 
\begin{equation}
	|\eta_{\mu^\pm}|<2.
\end{equation}
As shown in the middle-right panel of Figure~\ref{FIG:DISMu}, the dimuon energy $E_{\mu^+}+E_{\mu^-}$ of the backgrounds $\mu^+\mu^-\bar{\nu}_\ell \nu_\ell$ and $W^+W^-\bar{\nu}_\ell \nu_\ell$ are typically less than 2 TeV,  while the signals are in the range of $1\sim 9$ TeV. So we require 
\begin{equation}
	E_{\mu^+}+E_{\mu^-}> 2~\text{TeV}.
\end{equation}

Notably, there are a few parts of the $\mu^+\mu^-\bar{\nu}_\ell \nu_\ell$ samples leading to $E_{\mu^+}+E_{\mu^-}$ in $4\sim 9$~TeV, which is overlap with the signals.
We do not apply an upper bound on $E_{\mu^+}+E_{\mu^-}$, because the energetic $\mu^+\mu^-$ background has already been suppressed to zero by the cut on the dimuon acollinearity in Equation~\eqref{Eq:theta}. Another distinguishable viable is the dimuon invariant mass $M_{\mu^+\mu^-}$ as shown in the down-left panel of Figure~\ref{FIG:DISMu}. About 90\% of the $\mu^+\mu^-\bar{\nu}_\ell \nu_\ell$ samples have an invariant mass $M_{\mu^+\mu^-}$ less than 100 GeV, which is due to such samples produced from the on-shell decay of $Z$ boson. For the $W^+W^-\bar{\nu}_\ell \nu_\ell$ background, it usually leads to $M_{\mu^+\mu^-}< 1$ TeV. And most of the signal samples predict $M_{\mu^+\mu^-}\in [0,8]$~TeV. Based on this distribution, we then apply the cut
\begin{equation}\label{Eq:Mmu}
	M_{\mu^+\mu^-} > 1~\text{TeV}.
\end{equation}

In the down-right panel of Figure \ref{FIG:DISMu}, we illustrate the distribution of vector sum momentum of dimuon $|\vec{P}_{\mu^+}+\vec{P}_{\mu^-}|$ after applying the cuts from Equation \eqref{Eq:Nmu} to \eqref{Eq:Mmu}. The distribution of survived $\mu^+\mu^-\bar{\nu}_\ell \nu_\ell$ samples has a peak value around 5 to 6 TeV, meanwhile the peak values of signals and $W^+W^-\bar{\nu}_\ell \nu_\ell$ background are less than 3 TeV. We adopt the cut 
\begin{equation}\label{Eq:Pmu}
	|\vec{P}_{\mu^+}+\vec{P}_{\mu^-}| < 3.5~\text{TeV}
\end{equation}
to further suppress the $\mu^+\mu^-\bar{\nu}_\ell \nu_\ell$ background.

The final cross-sections of signals are 4.51 fb, 26.2 fb, and 8.50 fb for BP-1, BP-2, and BP-3, respectively. The final cross-section of total backgrounds is 0.21 fb, with 0.15~fb from $\mu^+\mu^-\bar{\nu}_\ell \nu_\ell$ channel and 0.06 fb from $W^+W^-\bar{\nu}_\ell \nu_\ell$ channel. Provided an integrated luminosity of 200 fb$^{-1}$, the expected significance will reach 29.4 for BP-1, 72.1 for BP-2, and 40.7 for BP-3. To reach the $5\sigma$ discovery limit, BP-2 is the most promising one, which only requires about 1 fb$^{-1}$ data. For BP-1 and BP-3, 10 fb$^{-1}$ data will make sure to discover them.

\begin{figure} 
	\centering
	\includegraphics[width=0.45\textwidth]{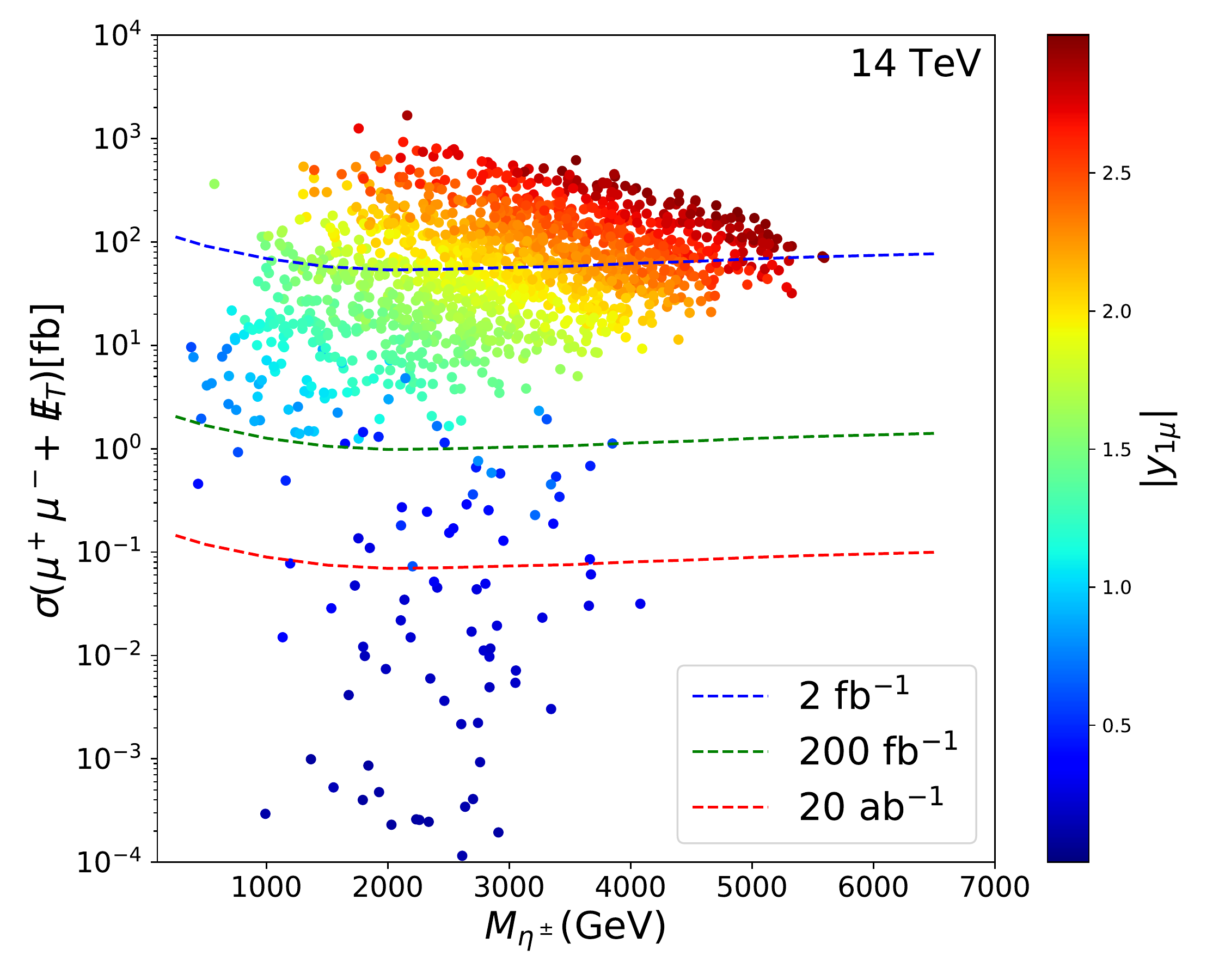}
	\includegraphics[width=0.45\textwidth]{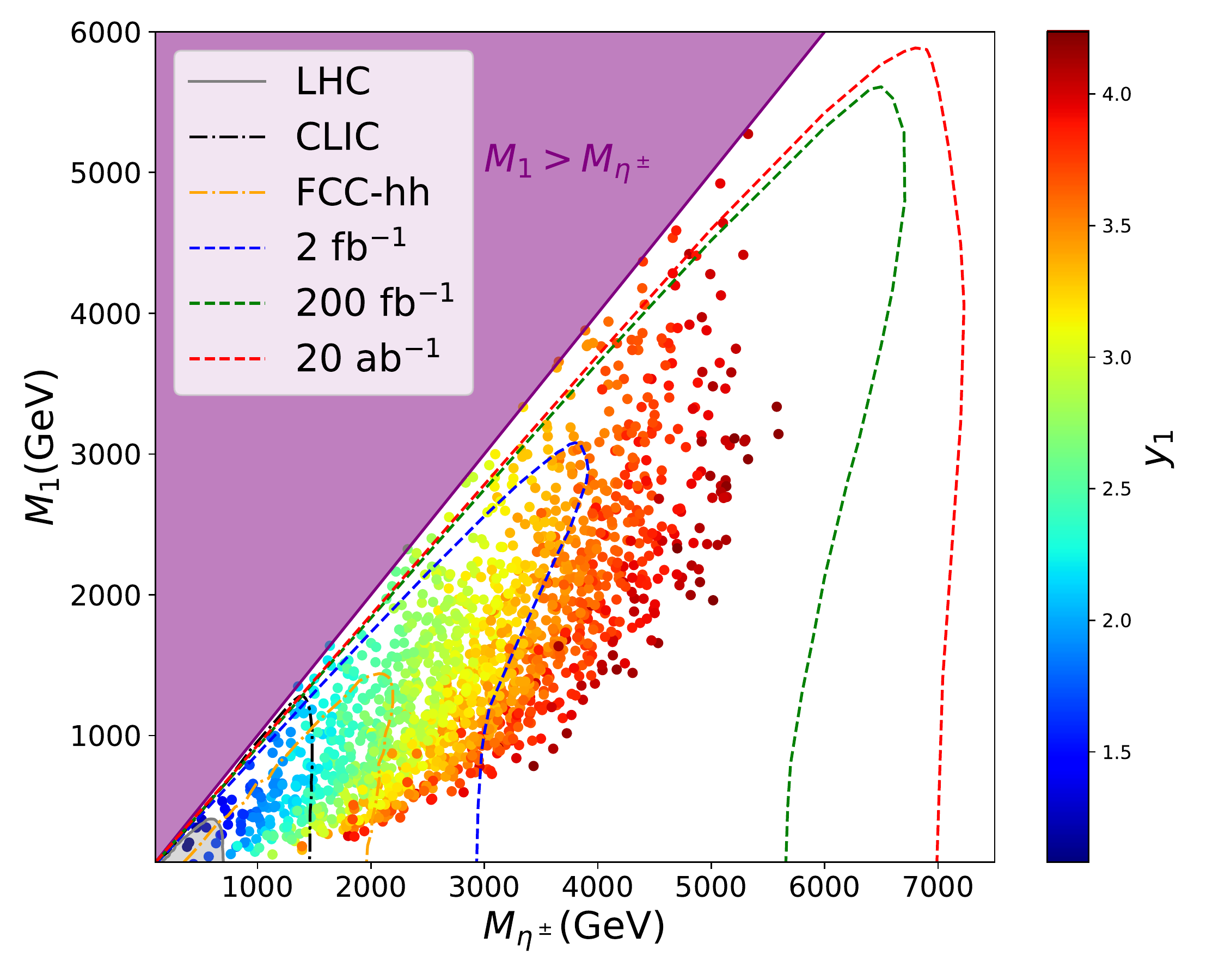}
	\caption{ The $5\sigma$ discovery reaches at the 14 TeV MuC for different luminosity. In the left panel, the discovery limits are obtained by assuming $M_1=M_{\eta^\pm}/2$. In the right panel, we have fixed  $y_{1e}=0.02,y_{1\mu}=y_{1\tau}=2$ and neglected the contribution of $N_{2,3}$ to derive the discovery limits. The gray region is excluded by LHC \cite{ATLAS:2019lff}. The black and orange lines are the future limits by CLIC and FCC-hh \cite{Baumholzer:2019twf}.
		\label{FIG:SG1}} 	
\end{figure}

Based on the above cuts in Equation \eqref{Eq:Nmu} - \eqref{Eq:Pmu}, we explore the $5\sigma$ discovery reach at the 14 TeV MuC. For the Scotogenic model at the MuC, both the Yukawa couplings and the mass spectrum will affect the significance.  A full simulation and scan over the whole parameter space are beyond the scope of this work. For simplicity, we consider the scenarios with fixed mass relation or fixed Yukawa couplings to qualitatively obtain the discovery reach. The results are shown in Figure~\ref{FIG:SG1}. First, we assume the mass relation $M_1=M_{\eta^\pm}/2$. The total cut efficiency is about 0.2 for the signal. In the left panel of Figure~\ref{FIG:SG1}, we show the $5\sigma$ discovery reach on the theoretical signal cross-section, which is calculated as
\begin{equation}
	\sigma(\mu^+\mu^-+\cancel{E}_T)=\sigma(\eta^+\eta^-)\times
	\text{BR}^2_{\mu N_1}.
\end{equation}
With $2~\text{fb}^{-1}$ data, the 14 TeV MuC is able to probe $\sigma(\mu^+\mu^-+\cancel{E}_T)\gtrsim 100$ fb, covering the most region with $|y_{1\mu}|\gtrsim 2$. The discovery limit is down to about 1 fb, when the integrated luminosity reaches $200~\text{fb}^{-1}$. This will unravel the whole region with $|y_{1\mu}|\gtrsim 1$. With $20~\text{ab}^{-1}$ data, $\sigma(\mu^+\mu^-+\cancel{E}_T)\gtrsim 0.1$ fb can be discovered. However, the production cross-section $\sigma(\eta^+\eta^-)$ and the decay branching ratio BR$_{\mu N_1}$ are both suppressed for $|y_{1\mu}|<1$. The theoretical cross-section can be much smaller than 0.1 fb, thus beyond the reach of MuC. 

In the right panel of Figure~\ref{FIG:SG1}, we fix $y_{1e}=0.02,y_{1\mu}=y_{1\tau}=2$ and 
consider the effect of mass spectrum. It is obvious that the discovery reach of 14 TeV MuC for the Scotogenic model can easily exceed the 100 TeV FCC-hh. Most samples with $M_{\eta^\pm}\lesssim$ 3.9~TeV and $M_1\lesssim$ 2.9~TeV are within the reach of $2~\text{fb}^{-1}$ data. With $200~\text{fb}^{-1}$ data, it is able to probe the region with $M_{\eta^\pm}\lesssim$ 6.7~TeV and $M_1\lesssim$ 5.6~TeV, which almost cover the whole region with correct relic density and satisfying lepton flavor violation. For the compressed mass region $M_1\simeq M_{\eta^\pm}$, the final states muons are relatively soft and become hard to pass through the energetic cut as $E_{\mu^+}+E_{\mu^-}>2$~TeV. Therefore, only increasing the integrated luminosity to $20~\text{ab}^{-1}$ will not help too much to probe such a region. Because of the large decay width of $\eta^\pm$,  the $\eta^\pm$ can still be pair produced via the off-shell process even when $M_{\eta^\pm}>7$ TeV. We find that the 14 TeV MuC could probe $M_{\eta^\pm}\lesssim 7.2$ TeV with $20~\text{ab}^{-1}$ data, although such large $M_{\eta^\pm}$ can not lead to correct relic density.

\subsection{$\tau^+\tau^-+\cancel{E}_T$ Signature}

Besides the opposite-sign dimuon signature $\mu^+\mu^-+\cancel{E}_T$, the opposite-sign ditau signature $\tau^+\tau^-+\cancel{E}_T$ are usually less promising due to the lower tau-tagging efficiency \cite{Baumholzer:2019twf}.  However, with a relatively large branching ratio BR$_{\tau N_1}\gtrsim0.2$, the ditau signature is expected more promising than the dimuon channel when the latter is suppressed by the branding ratio. In this paper, we consider the hadronic decay of $\tau$, and assume the $\tau$-tagging efficiency to be 0.4 \cite{CMS:2018jrd}.
The opposite-sign ditau signature at MuC is
\begin{equation}
	\mu^+\mu^-\to \eta^+\eta^-\to \tau^+N_1 + \tau^- N_1 \to \tau^+\tau^-+\cancel{E}_T.
\end{equation}
The corresponding backgrounds are from
\begin{eqnarray}
	\mu^+\mu^-\to \tau^+\tau^- , \tau^+ \tau^- \bar{\nu}_\ell\nu_\ell, W^+W^-\bar{\nu}_\ell \nu_\ell,
\end{eqnarray}
followed by $W^\pm \to \tau^\pm \nu_\tau$. Different from the $\mu^+\mu^-\to \mu^+\mu^-$ process, the  $\mu^+\mu^-\to \tau^+\tau^-$ only has $s$-channel contribution, so $\sigma(\tau^+\tau^-)$ is relatively small.

\begin{table}
	\renewcommand\arraystretch{1.25}
	\begin{center}
		\begin{tabular}{c| c c c | c c c} 
			\hline
			\hline
			$\sigma$(fb) & \textbf{BP-1} & \textbf{BP-2} & \textbf{BP-3} & $\tau^+\tau^-$ & $\tau^+ \tau^-\bar{\nu}_\ell\nu_\ell$ & $W^+W^-\bar{\nu}_\ell \nu_\ell$ \\
			\hline
			Preselection & 51.9   &  111  & 93.9   & 0.531  & 102.4 & 6.405  \\
			\hline
		    \tabincell{c}{$N_{\tau^\pm}=1$\\ $P_T^{\tau^\pm}>20$ GeV} 
		     & 3.38  & 7.52   & 6.34   & 0.075  & 8.349  & 0.278 \\
			\hline
			$\cancel{E}_T>100$ GeV & 3.24 & 7.43 & 6.30 & 0.072 & 4.202 & 0.162 \\
			\hline
			\tabincell{c}{$|\pi-\theta_{\tau^+}-\theta_{\tau^-}|$ \\ $ >0.04$}
			 & 2.74 & 6.93 & 6.02 & 0.003 & 4.149 & 0.155 \\
			\hline
			$|\eta_{\tau^\pm}|<2.5$ & 2.52 & 6.67 & 5.89 & 0.003 & 4.139 & 0.153 \\ 
			\hline
			$E_{\tau^+}+E_{\tau^-}>1$ TeV & 2.26 & 6.04 & 5.56 & 0.003  &  0.246  &  0.039 \\
			\hline
			$M_{\tau^+ \tau^-}>0.5$ TeV & 2.22 & 5.87  & 5.36  & 0.003  & 0.034   & 0.025  \\
			\hline
			\hline
			Significance & 20.8     &  34.1   &   32.6  & \multicolumn{2}{c|}{\multirow{2}{*}{Total Background}} & \multirow{2}{*}{0.062 }\\
			\cline{1-4}
			$5\sigma$ Luminosity (fb$^{-1}$)&  11.6  &  4.30  &   4.72  & \multicolumn{2}{c|}{ } \\
			\hline
			\hline
		\end{tabular}
	\end{center}
	\caption{Same as Table.~\ref{Tab:MuMu}, but for the $\tau^+\tau^-+\cancel{E}_T$ signature. In this channel, we do not apply cut on $|\vec{P}_{\tau^+}+\vec{P}_{\tau^-}| $.
		\label{Tab:TaTa}}
\end{table} 

\begin{figure}
	\begin{center}
		\includegraphics[width=0.45\linewidth]{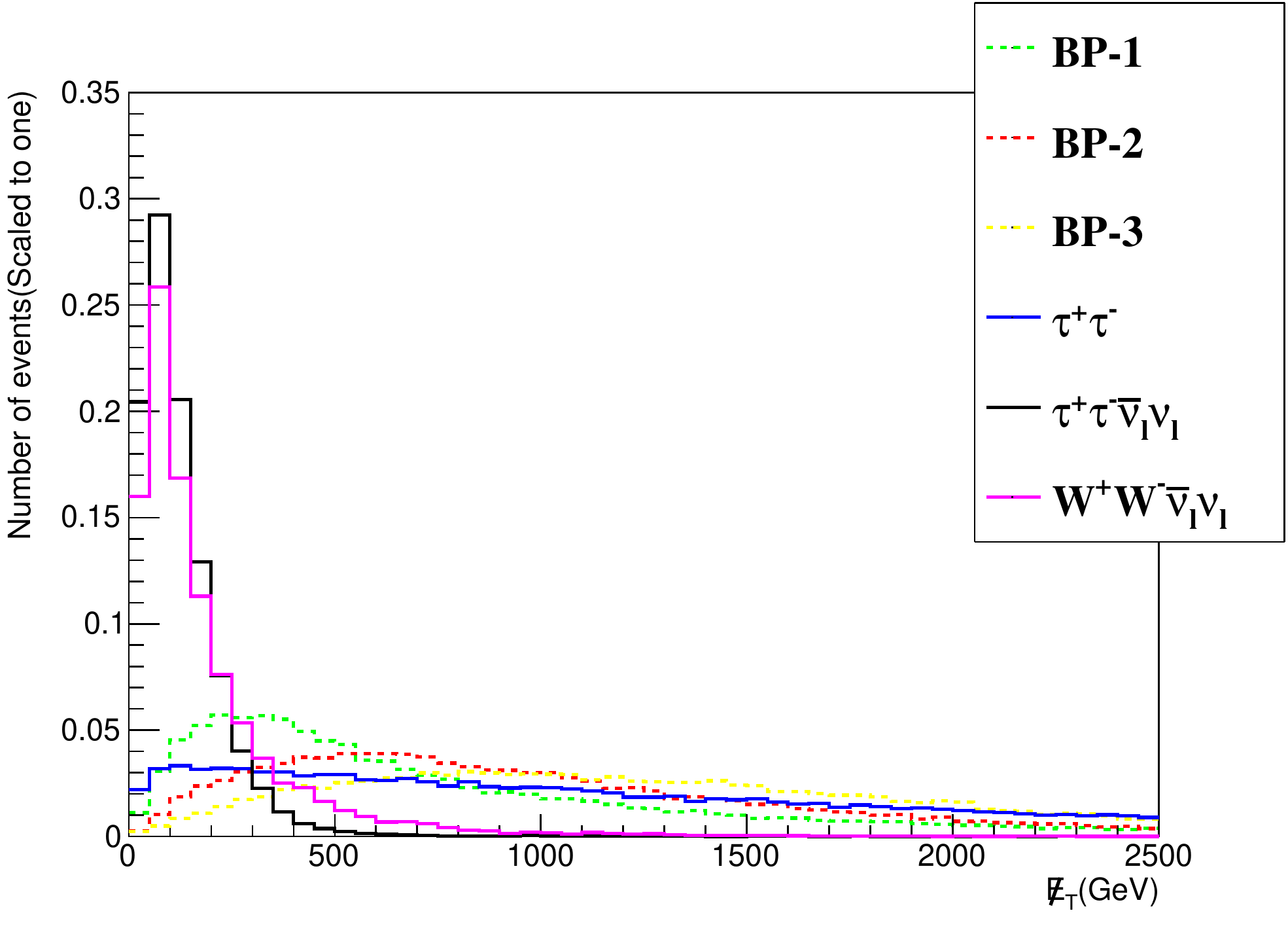}
		\includegraphics[width=0.45\linewidth]{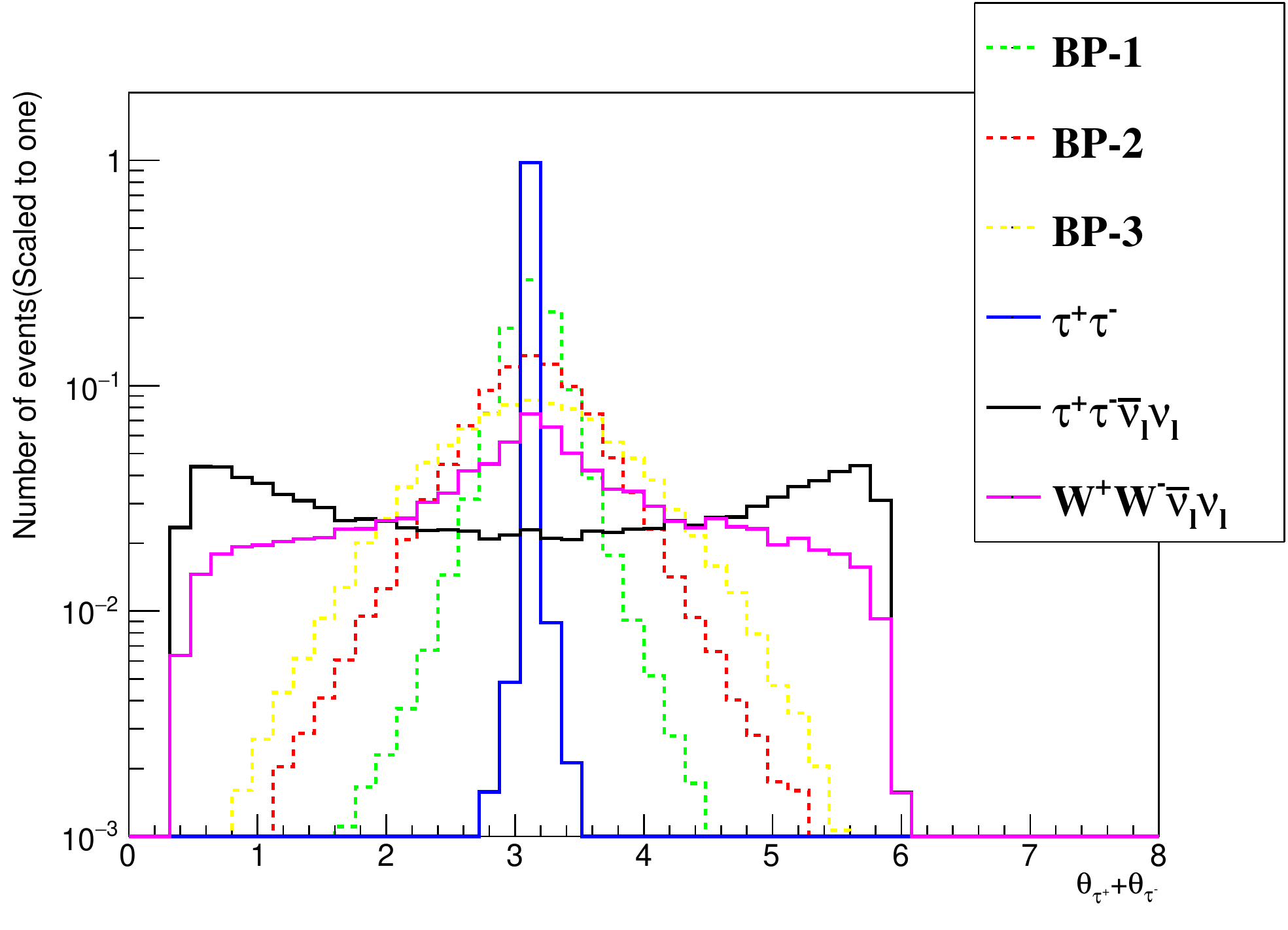}
		\includegraphics[width=0.45\linewidth]{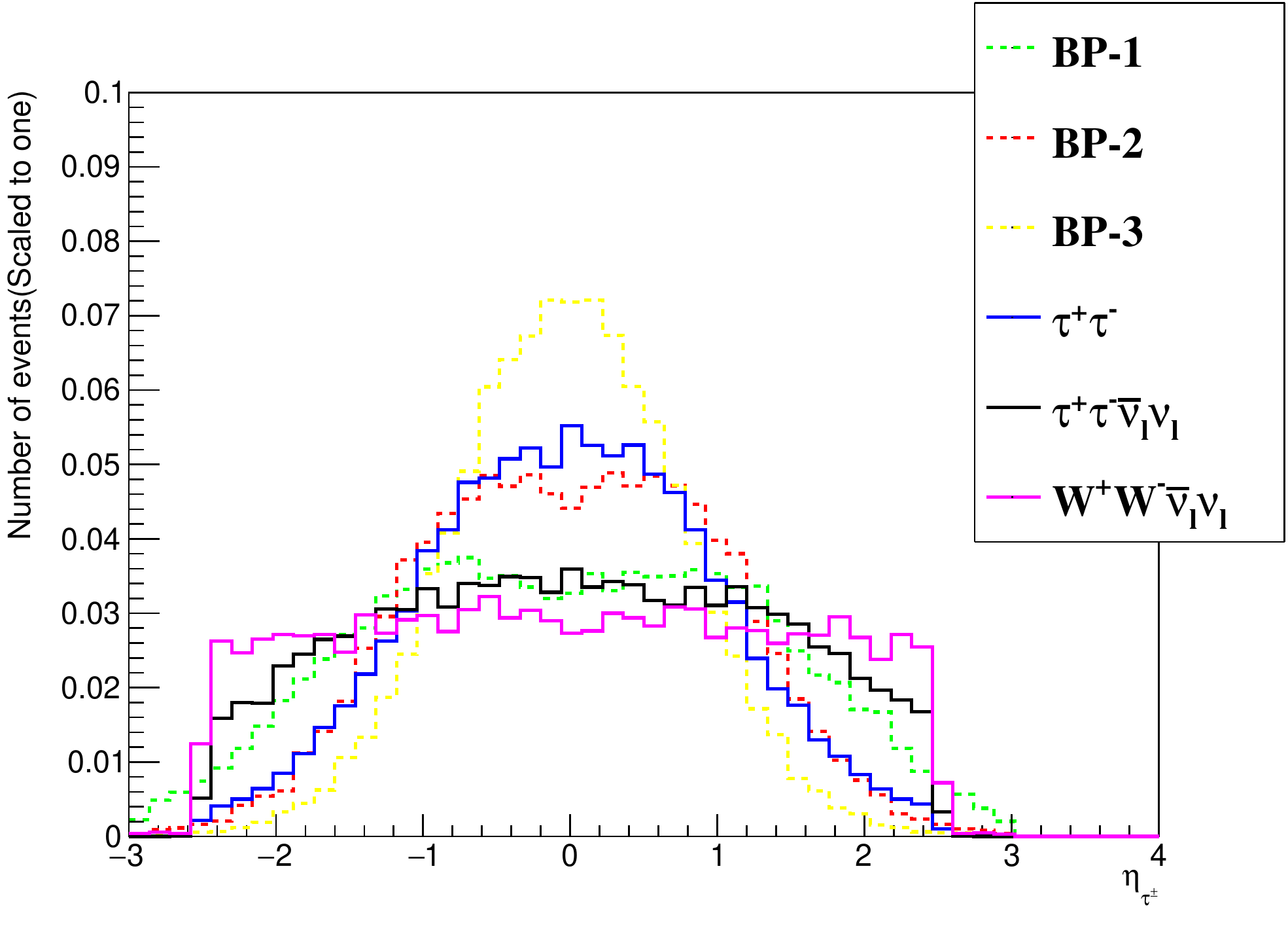}
		\includegraphics[width=0.45\linewidth]{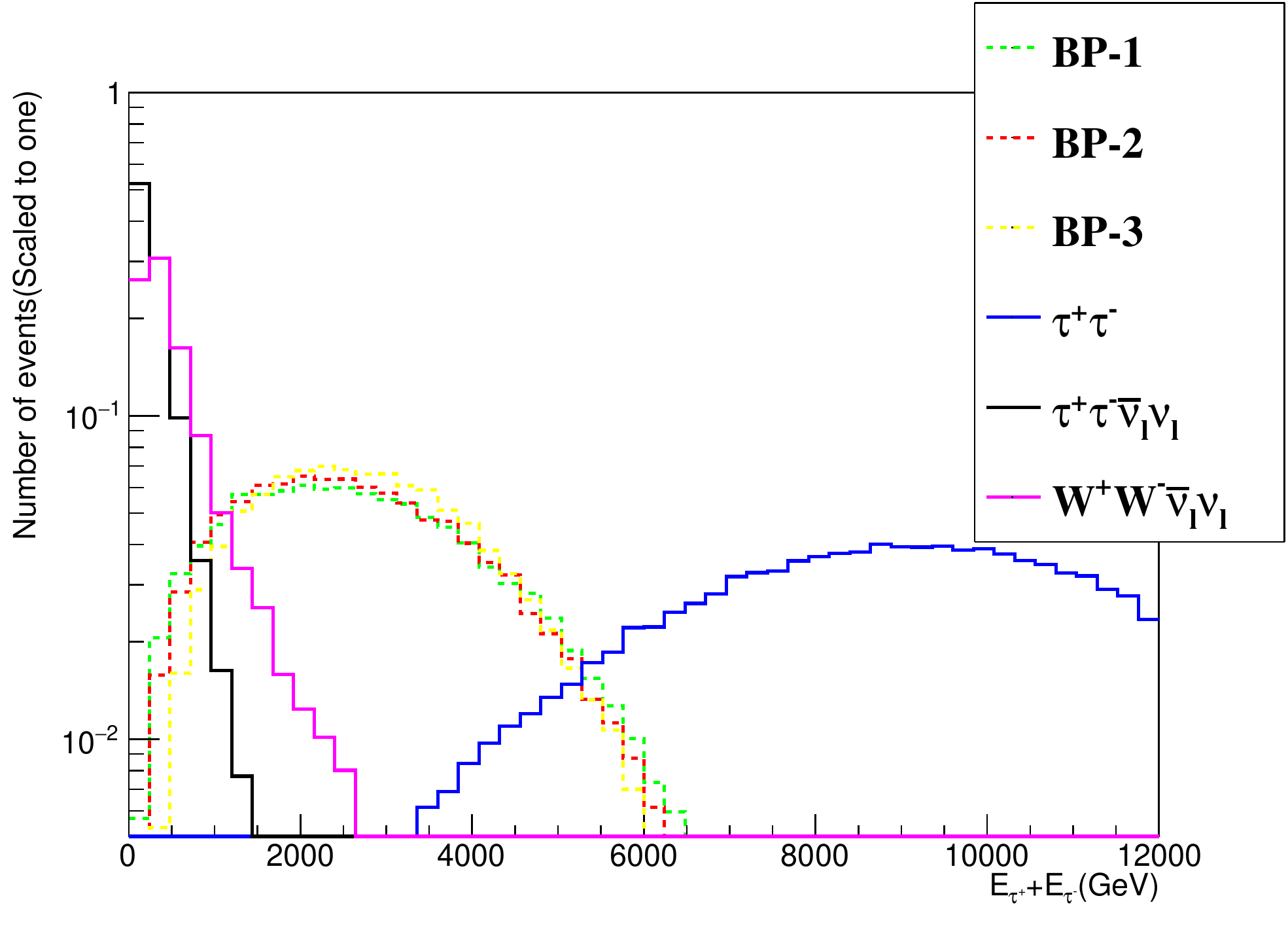}
		\includegraphics[width=0.45\linewidth]{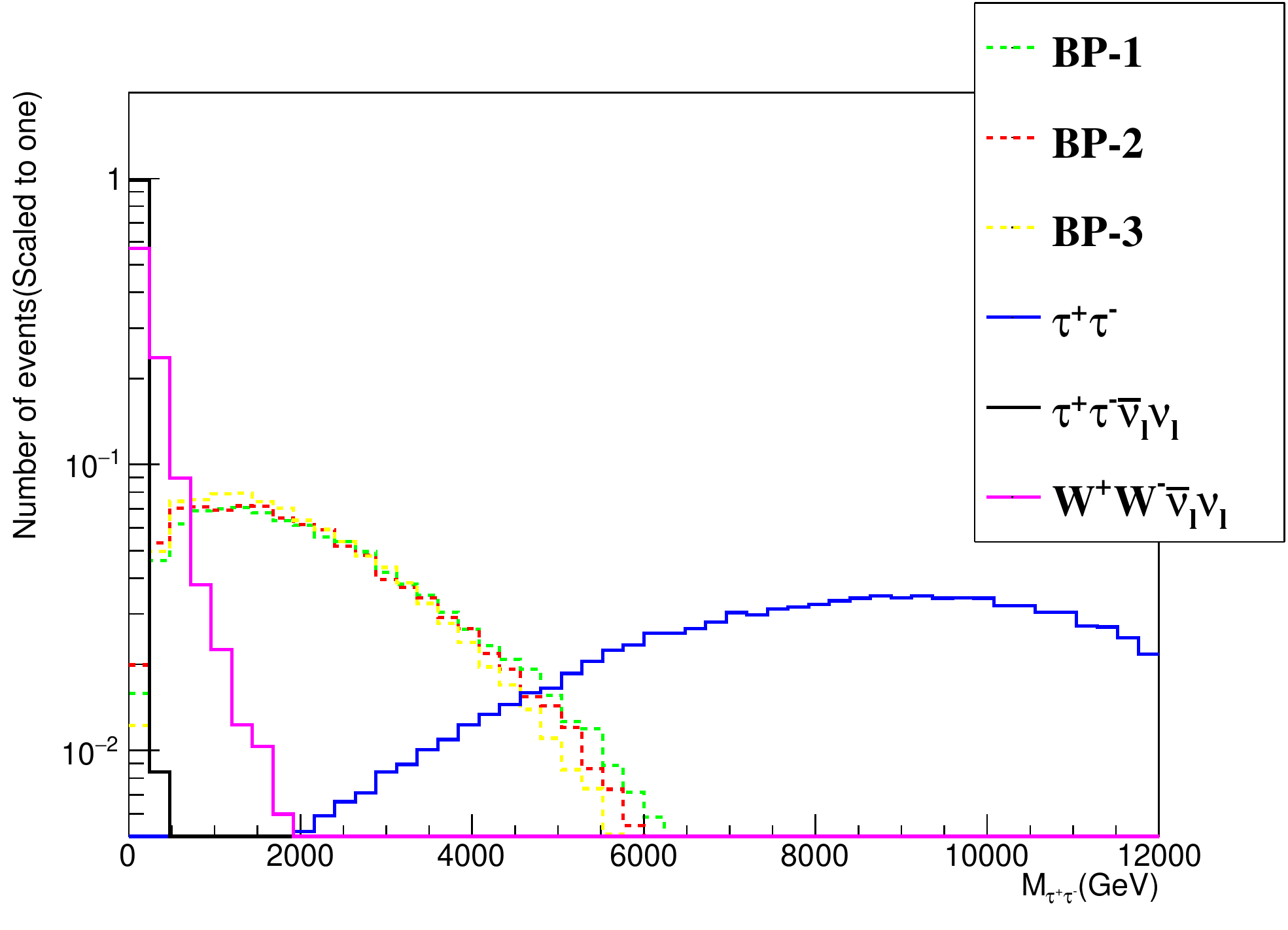}
		\includegraphics[width=0.45\linewidth]{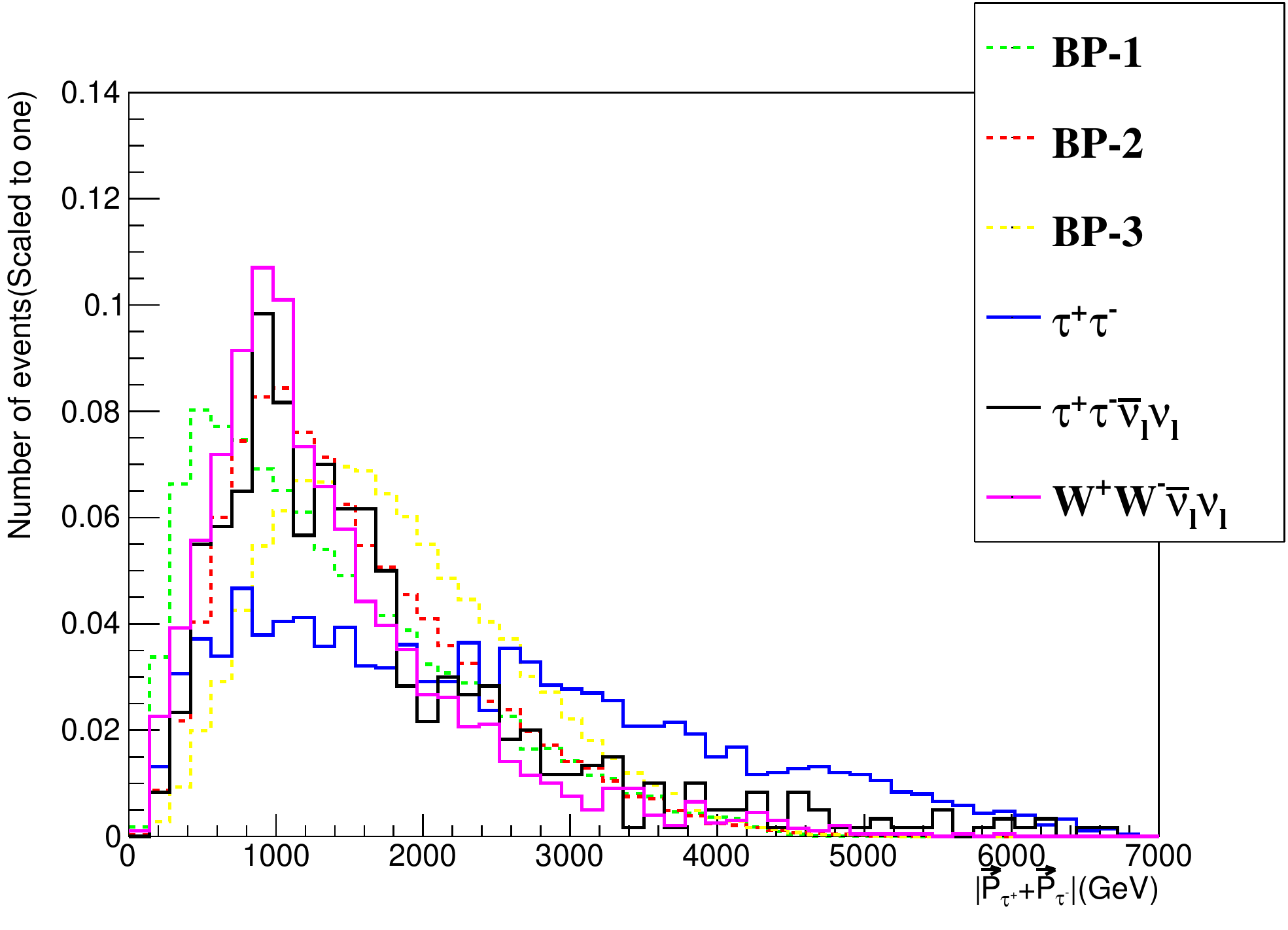}
	\end{center}
	\caption{ Same as Figure \ref{FIG:DISMu}, but for the ditau signature.}
	\label{FIG:DISTau}
\end{figure}

 First, events with two opposite-sign taus are selected by
\begin{equation}\label{Eq:Ntau}
	N_{\tau^\pm}=1, P_T^{\tau^\pm}>20~\text{GeV}.
\end{equation} 
Since hadronic decay of $\tau$ is considered, we require $P_T^{\tau^\pm}>20$ GeV to pass the trigger cut. At this level, the dominant background is $\tau^+\tau^-\bar{\nu}_\ell\nu_\ell$, which is slightly larger than the signal. In Figure~\ref{FIG:DISTau}, the normalized distribution of relevant parameters are shown. Generally speaking, distributions of these variables for the ditau signature are similar to the dimuon signature. The ditau signature thus is analyzed with similar cuts as the dimuon signature. The cut flow for the ditau signature and backgrounds are summarized in Table~\ref{Tab:TaTa}. However, in order to keep as much as the signal events, we lose some cuts compared with the dimuon signal.
For instance, we apply the cut
\begin{equation}
	\cancel{E}_T>100~\text{GeV}
\end{equation}
to select samples with missing transverse energy. After this cut, the total background is already smaller than the cross-section of BP-2 and BP-3.

An efficient cut to suppress the $\tau^+\tau^-$ channel is 
\begin{equation}
	|\pi-\theta_{\tau^+}-\theta_{\tau^-}|>0.04.
\end{equation}
Because the angle resolution of $\tau$ is worse than that of $\mu$, the rejected acollinearity region of the ditau signature is much larger than the dimuon signature region. And the direct ditau events can not be suppressed to a negligible level by the above acollinearity cut. The following cuts in Equation \eqref{Eq:EtaTau} and \eqref{Eq:Etau} are also hard to reject all the $\tau^+\tau^-$ samples.  We then require 
\begin{equation}\label{Eq:EtaTau}
	|\eta_{\tau^\pm}|<2.5.
\end{equation}

To further suppress the $\tau^+\tau^-\bar{\nu}_\ell \nu_\ell$ and $W^+W^-\bar{\nu}_\ell \nu_\ell$ background, the cuts on ditau energy and invariant mass are adopted
\begin{equation}\label{Eq:Etau}
	E_{\tau^+}+E_{\tau^-}>1~\text{TeV}, M_{\tau^+\tau^-}>0.5~\text{TeV}.
\end{equation}
These two cuts are able to suppress the total background less than 0.1 fb. In the down-right panel of  Figure~\ref{FIG:DISTau}, we show the distribution of $|\vec{P}_{\tau^+}+\vec{P}_{\tau^-}| $ after applying all cuts from Equation \eqref{Eq:Ntau} - \eqref{Eq:Etau}. Comparing with the distribution of $|\vec{P}_{\mu^+}+\vec{P}_{\mu^-}| $ in Figure~\ref{FIG:DISMu}, the $\tau^+\tau^-\bar{\nu}_\ell\nu_\ell$ channel does not have any peak structure around 5 TeV. This is because the ditau energy $E_{\tau^+}+E_{\tau^-}$ is always smaller than 1.5 TeV for the $\tau^+\tau^-\bar{\nu}_\ell\nu_\ell$ channel as shown in the middle-right panel of Figure \ref{FIG:DISTau}. Except for the tiny $\tau^+\tau^-$ channel, both the $\tau^+\tau^-\bar{\nu}_\ell\nu_\ell$ and $W^+W^-\bar{\nu}_\ell\nu_\ell$ channel have quite a similar $|\vec{P}_{\tau^+}+\vec{P}_{\tau^-}| $ distribution as the signal. Therefore, we do not apply any cut on $|\vec{P}_{\tau^+}+\vec{P}_{\tau^-}| $. 

\begin{figure} 
	\centering
	\includegraphics[width=0.45\textwidth]{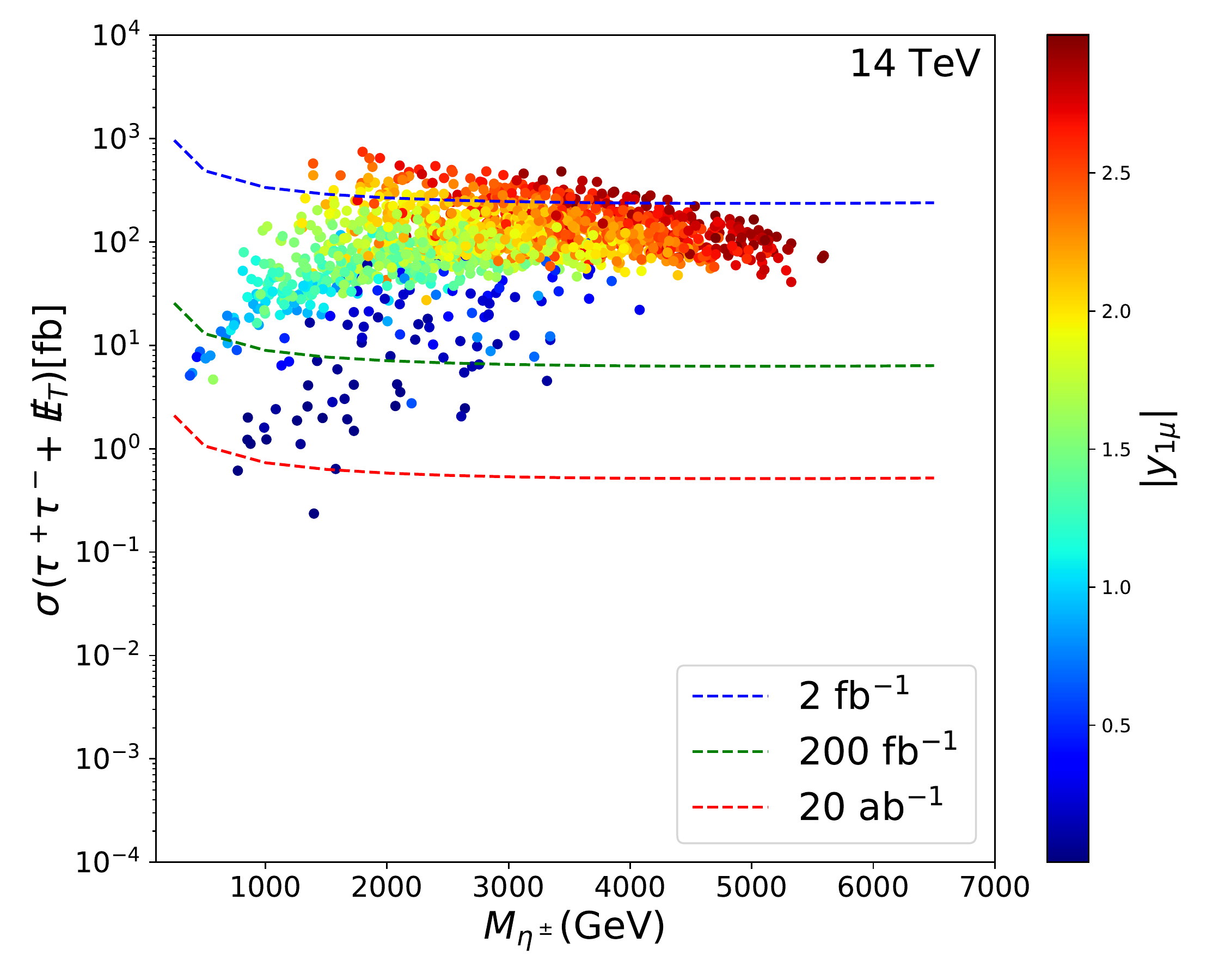}
	\includegraphics[width=0.45\textwidth]{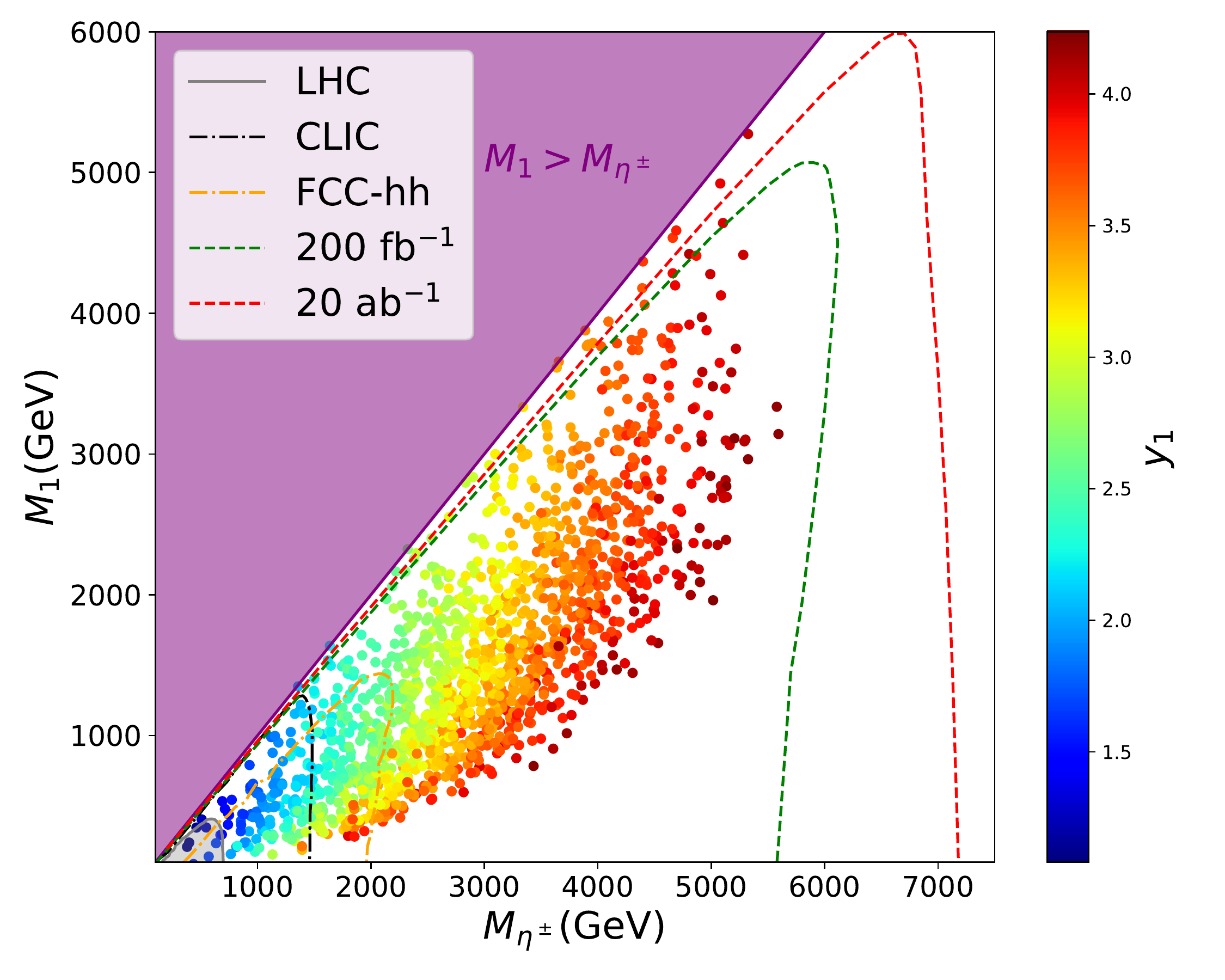}
	\caption{Same as Fig.~\ref{FIG:SG1}, but for the ditau signature. 
		\label{FIG:SG2}} 	
\end{figure}

After the above cuts on the ditau signature, we finally have 2.22 fb, 5.87 fb and 5.36 fb for BP-1 to BP-3. These values are slightly smaller than the corresponding dimuon channel. The total background is 0.062 fb after all cuts. The expected significance would be 20.8 for BP-1, 34.1 for BP-2, and 32.6 for BP-3 with $200~\text{fb}^{-1}$ luminosity. For BP-2 and BP-3, 5 fb$^{-1}$ data is able to reach the $5\sigma$ discovery limit. Meanwhile, we need 11.6~fb$^{-1}$ data to discover BP-1.

In Figure~\ref{FIG:SG2}, the $5\sigma$ discovery reaches of the two specific scenarios for the ditau signature are shown. In the left panel, the theoretical ditau cross-section is calculated as
\begin{equation}
	\sigma(\tau^+\tau^-+\cancel{E}_T)=\sigma(\eta^+\eta^-)\times
	\text{BR}^2_{\tau N_1}.
\end{equation}
Different from the dimuon signature, the branching ratio of $\tau$ final state is never suppressed. So the ditau cross-section is always larger than 0.1 fb. Provided $M_1=M_{\eta^\pm}/2$, we then obtain the total cut efficiency is approximately 0.05 for the signal. For the most promising case, the 14 TeV MuC is able to probe $\sigma(\tau^+\tau^-+\cancel{E}_T)\gtrsim$ 250 fb in the range of $M_{\eta^\pm}\in[1.5,4]$ TeV with only $2~\text{fb}^{-1}$ data. Samples with $\sigma(\tau^+\tau^-+\cancel{E}_T)\gtrsim$ 6.5 fb are within the reach of $200~\text{fb}^{-1}$. This covers most samples with $|y_{1\mu}|\gtrsim1$. By increasing the integrated luminosity to $20~\text{ab}^{-1}$, the $5\sigma$ discovery limit is down to about 0.5 fb, which unravels almost all samples. 

In the right panel of Figure~\ref{FIG:SG2}, the $5\sigma$ discovery limits on the $M_{\eta^\pm}-M_1$ plane with $y_{1e}=0.02,y_{1\mu}=y_{1\tau}=2$ are shown. With lower tagging efficiency for the ditau system, we find that the significance of this channel at best could reach $4.9\sigma$ with $2~\text{fb}^{-1}$ data, so there is no corresponding $5\sigma$ discovery reach for $2~\text{fb}^{-1}$ in the plot. With $200~\text{fb}^{-1}$ data, the ditau channel could discover the region with $M_{\eta^\pm}\lesssim$ 6.1~TeV and $M_1\lesssim$ 5.1~TeV. Although this region is smaller than the dimuon channel with same luminosity, it is enough to cover most allowed samples. Similarly, the compressed mass region can be hardly probe by the ditau channel even with $20~\text{ab}^{-1}$ data either.

\subsection{Mono-$\gamma$ Signature}
\begin{table}
	\renewcommand\arraystretch{1.25}
	\begin{center}
		\begin{tabular}{c| r r r | r } 
			\hline
			\hline
			$\sigma$(fb) & ~~\textbf{BP-1} & ~~\textbf{BP-2} & ~~\textbf{BP-3} & ~~$\gamma \bar{\nu}_\ell \nu_\ell$ \\
			\hline
			Preselection &  2.76   &  15.99   &  18.91  &  3277   \\
			\hline
			$0.70<\theta_\gamma<2.44$ & 1.19  & 7.09  &  8.47 & 1399 \\
			\hline
			$E_\gamma>500$ GeV & 0.55  & 2.86  & 3.00  &  54.19  \\ 
			\hline
			$\cancel{E}_T>1$ TeV & 0.39  & 1.82  & 1.72  &  10.96   \\
			\hline
			$P_T^\gamma>2$ TeV & 0.27  &  1.07  & 0.84   &   2.23  \\
			\hline
			\hline
			Significance &  2.41   &  8.33    &  6.78   &\\
			\cline{1-4}
			$5\sigma$ Luminosity (fb$^{-1}$)&  857   &   72  &  109    &  \\
			\hline
			\hline
		\end{tabular}
	\end{center}
	\caption{Cut flow table for the mono-photon signature $\gamma+\cancel{E}_T$ from BP-1, BP-2, BP-3 and the dominant background $\gamma \bar{\nu}_\ell \nu_\ell$.
		\label{Tab:MonoPhoton}}
\end{table} 

	Another interesting channel is the mono-photon signature  $\mu^+\mu^-\to N_1 N_1 \gamma\to \gamma +\cancel{E}_T$ \cite{Casarsa:2021rud}, with the photon from initial state radiation \cite{Habermehl:2020njb}. This signature is mediated by the $t$-channel exchange of $\eta^\pm$, thus the cross-section is also enhanced by relatively large $|y_{1\mu}|$. The dominant background is from $\gamma \bar{\nu}_\ell\nu_\ell$ \cite{Black:2022qlg}.

\begin{figure} 
	\centering
	\includegraphics[width=0.45\textwidth]{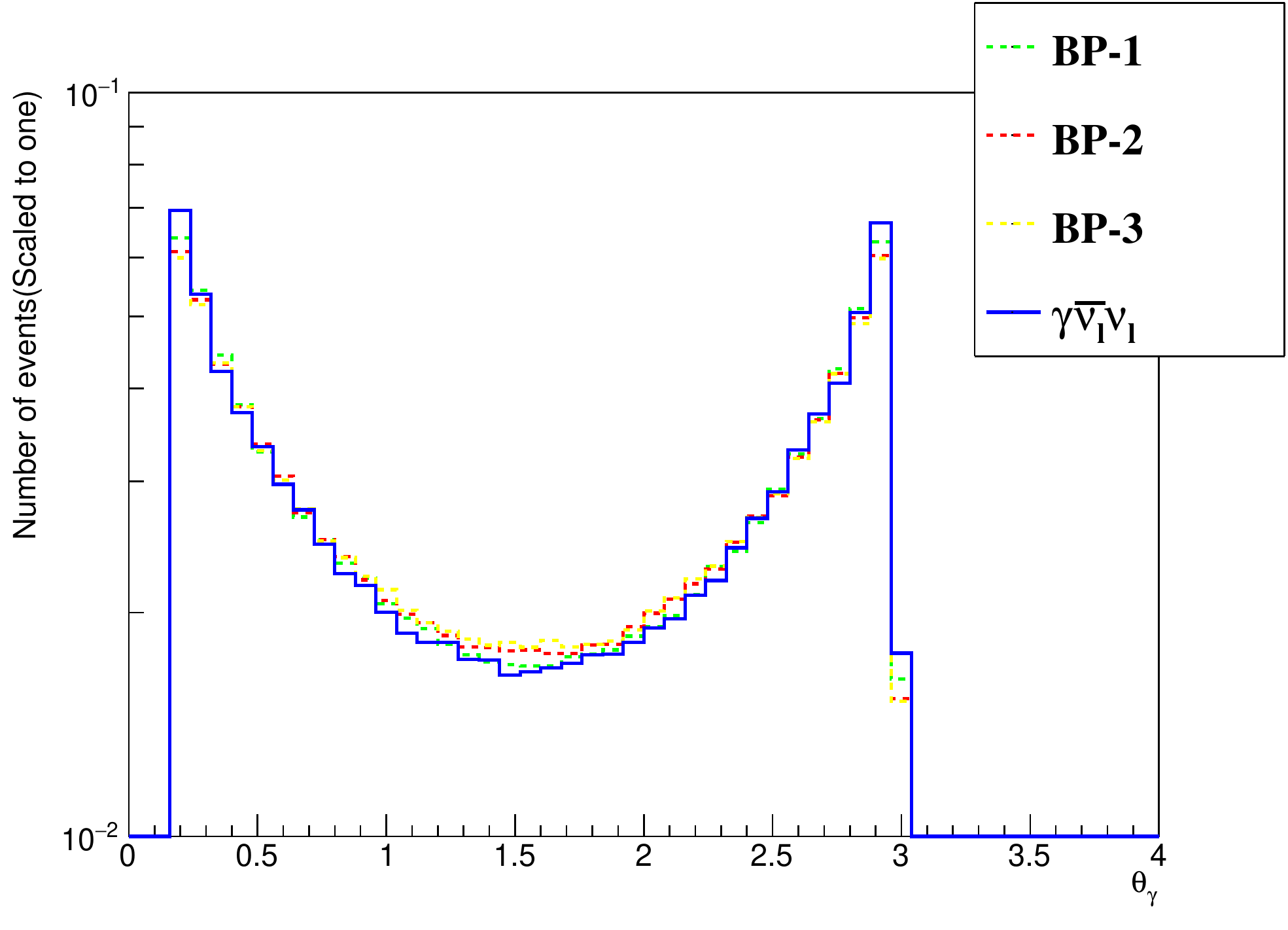}
	\includegraphics[width=0.45\textwidth]{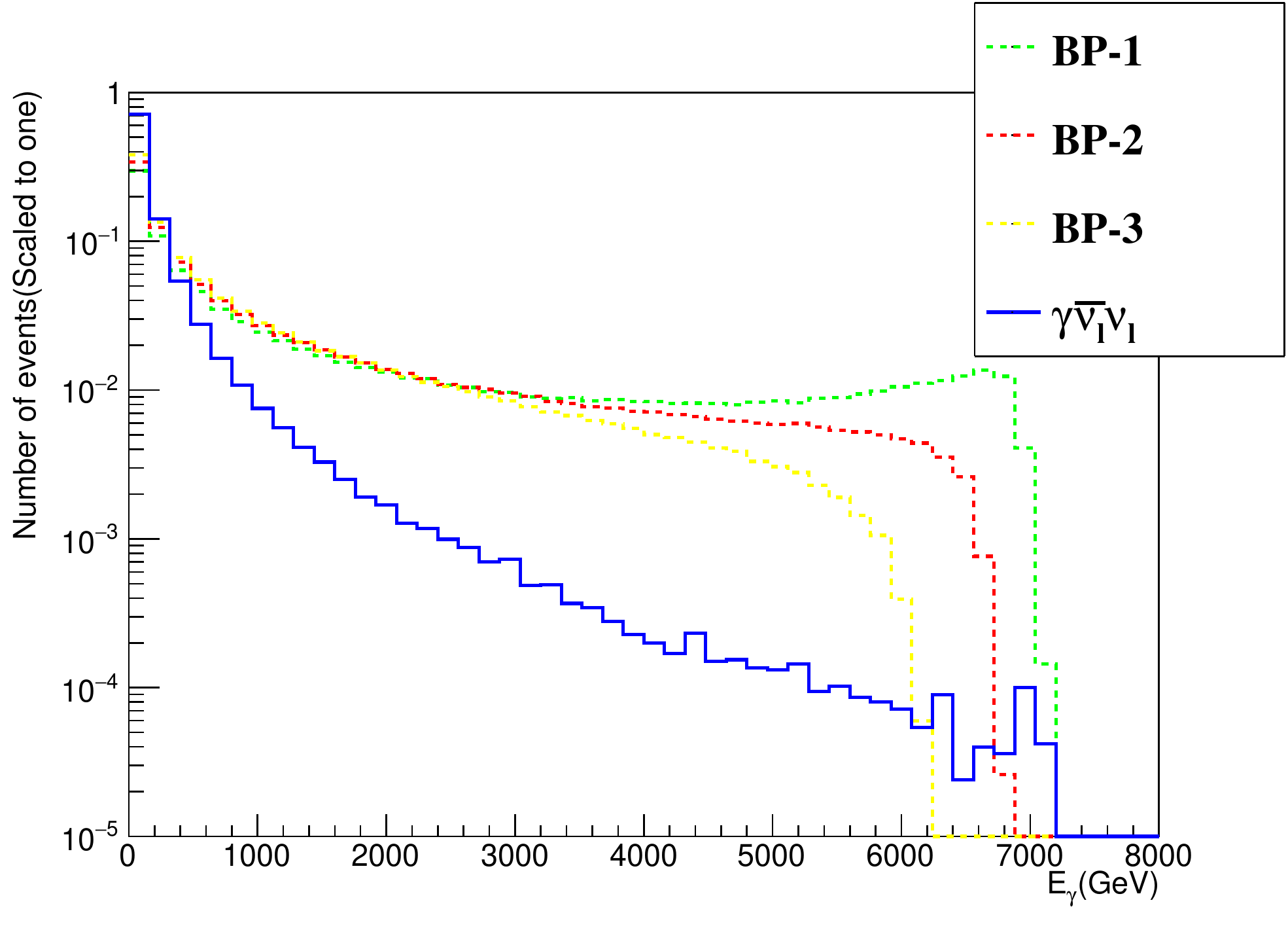}
	\includegraphics[width=0.45\textwidth]{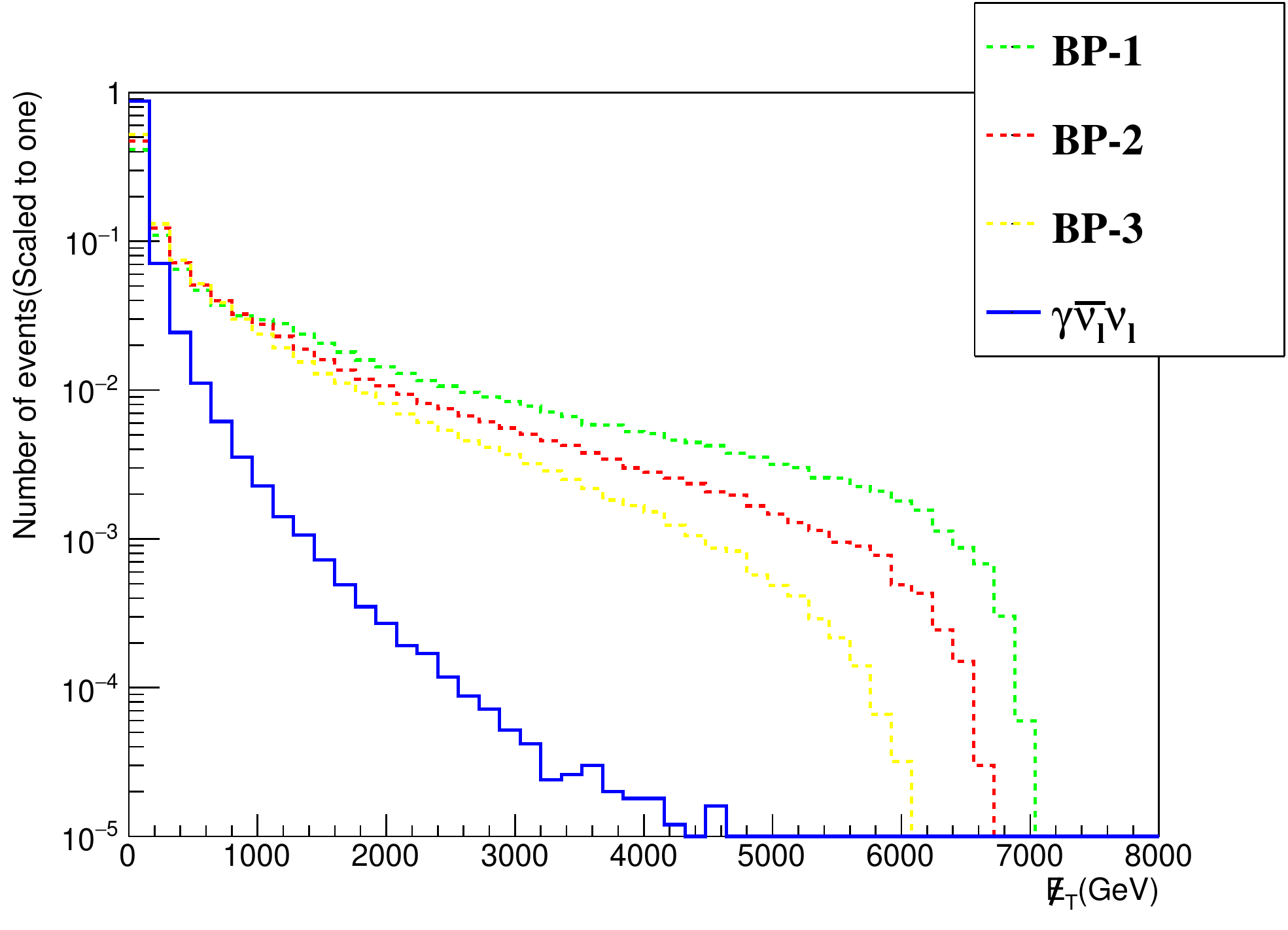}
	\includegraphics[width=0.45\textwidth]{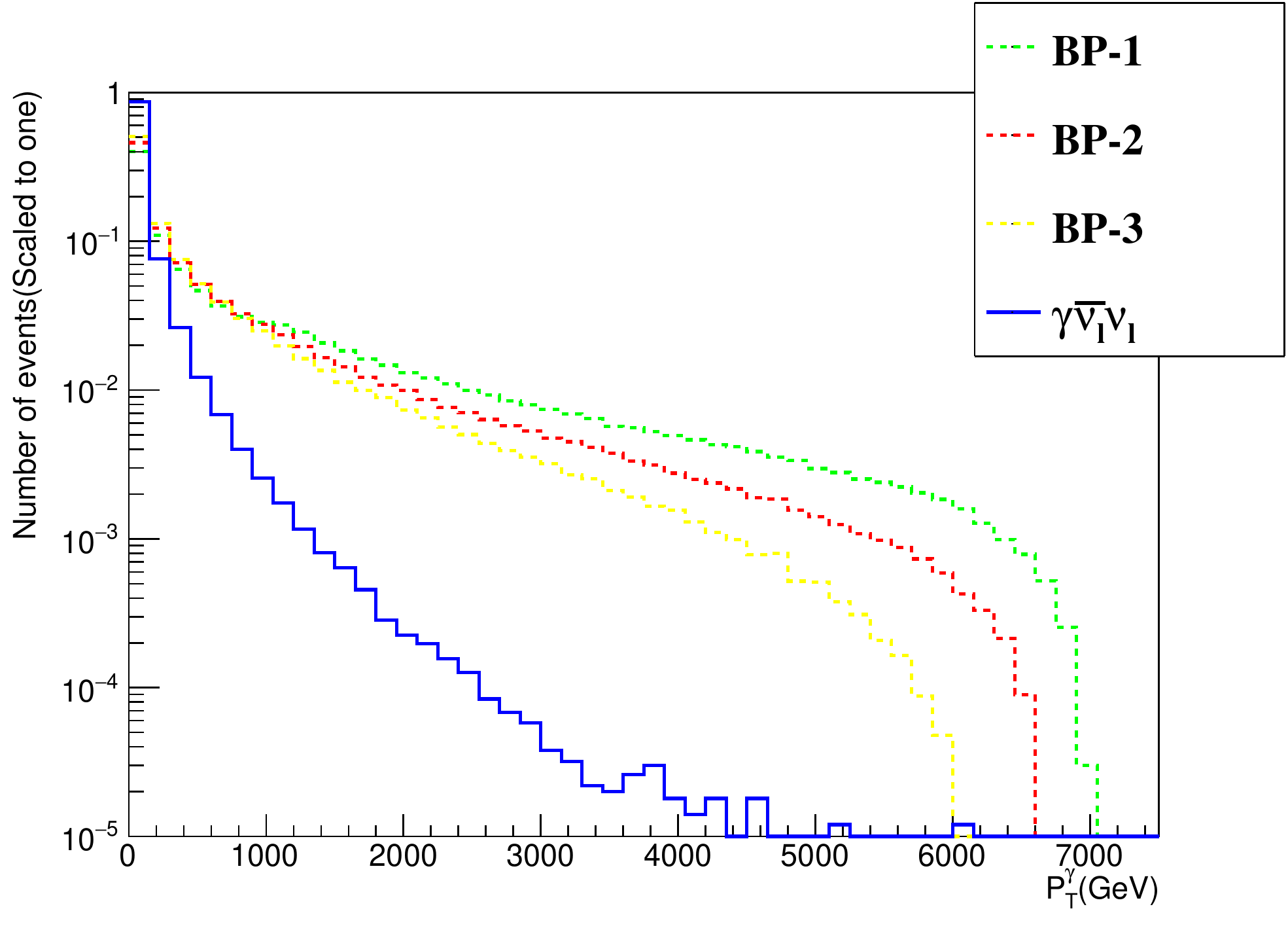}
	\includegraphics[width=0.45\textwidth]{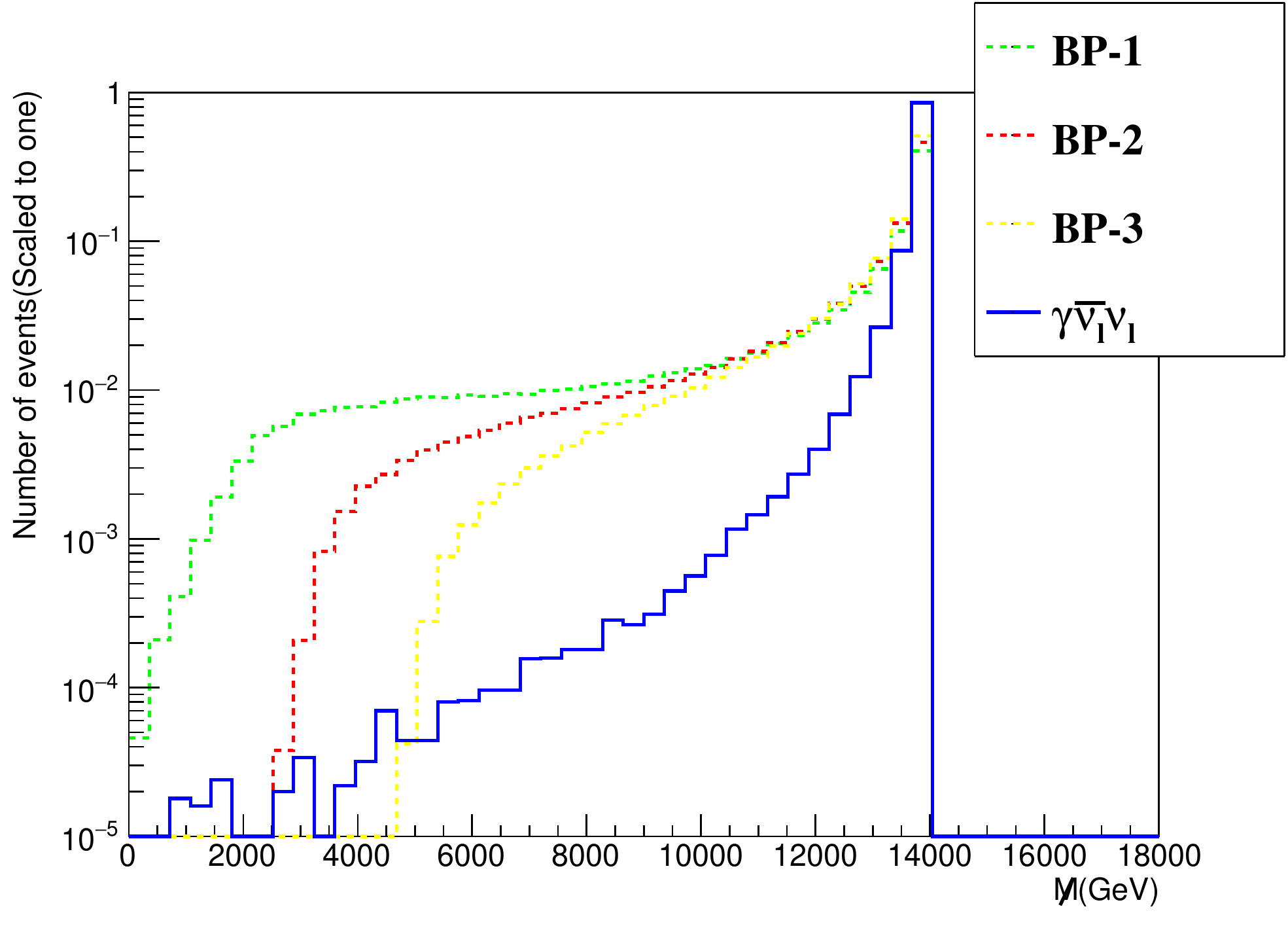}
	\caption{Normalized distributions of the $\theta$ angle of photon $\theta_\gamma$ (up-left panel), energy of the photon $E_\gamma$ (up-right panel), missing transverse energy $\cancel{E}_T$ (middle-left panel), transverse momentum of the photon $P_T^\gamma$ (middle-right panel), and missing mass $\cancel{M}$ (down panel).
		\label{FIG:DISMP}} 	
\end{figure}

The initial cross-section of the benchmark points are typically at the order of $\mathcal{O}(10)$~fb, while the cross-section of background is two orders of magnitudes higher.
In Figure \ref{FIG:DISMP}, the normalized distribution of relevant parameters are shown. The missing mass $\cancel{M}$ is defined as \cite{Han:2020uak}
\begin{equation}\label{Eq:Mmiss}
	\cancel{M}=\sqrt{(p_{\mu^+}+p_{\mu^-}-\sum_i p_i)^2},
\end{equation}
where $p_{\mu^\pm}$ is the momenta of initial muon and $p_i$ is the momenta of the $i$-th observed particles in the final states. From the distribution of missing mass $\cancel{M}$ in Figure \ref{FIG:DISMP}, it is obvious that the signals predict $\cancel{M}>2M_1$. However, applying the cut $\cancel{M}>2M_1$ only rejects very small amount of the background. So we do not apply cut on $\cancel{M}$.

Following Ref.~\cite{Black:2022qlg}, we first select events with
\begin{equation}
	0.69<\theta_\gamma<2.44.
\end{equation}
At this level of cut, the background $\gamma \bar{\nu}_\ell\nu_\ell$ is still at the order of $\mathcal{O}(10^3)$ fb. From the distribution of $E_\gamma$, it is clear the signals have relatively larger values than the background. The same is true for the distributions of $\cancel{E}_T$ and $P_T^\gamma$. We then apply the following cuts
\begin{equation}\label{Eq:EGamma}
	E_\gamma>500~\text{GeV},\cancel{E}_T>1~\text{TeV},P_T^\gamma>2~\text{TeV}.
\end{equation}
The above cuts are much tighter than those in Ref.~\cite{Black:2022qlg}, but they are efficient to suppress the background. After the cuts in Equation \eqref{Eq:EGamma}, the cross-section of the signal and background are comparable. 

The cut flow for the mono-photon signature and background are summarized in Table~\ref{Tab:MonoPhoton}. The final cross-sections of signals are 0.27 fb, 1.07 fb, and 0.84 fb for BP-1, BP-2, and BP-3, respectively. The final cross-section of background is 2.23 fb. Provided an integrated luminosity of 200 fb$^{-1}$, the expected significance will reach 2.41 for BP-1, 8.33 for BP-2, and 6.78 for BP-3. To reach the $5\sigma$ discovery limit, the required luminosity is 857 fb$^{-1}$ for BP-1,  72 fb$^{-1}$ for BP-2, and 109 fb$^{-1}$ for BP-3. Comparing with the dilepton signature, this mono-photon signature is less promising.

\begin{figure} 
	\centering
	\includegraphics[width=0.45\textwidth]{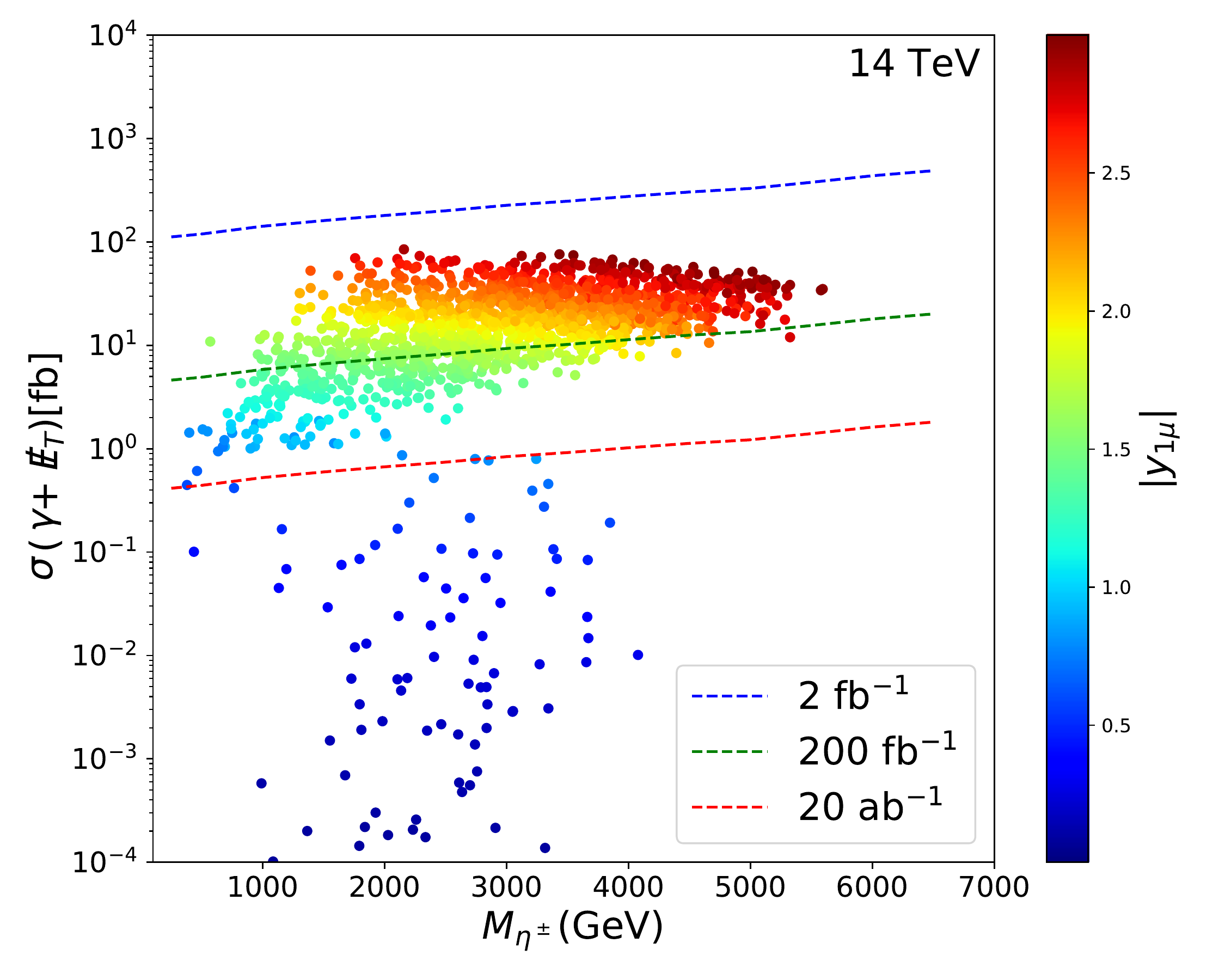}
	\includegraphics[width=0.45\textwidth]{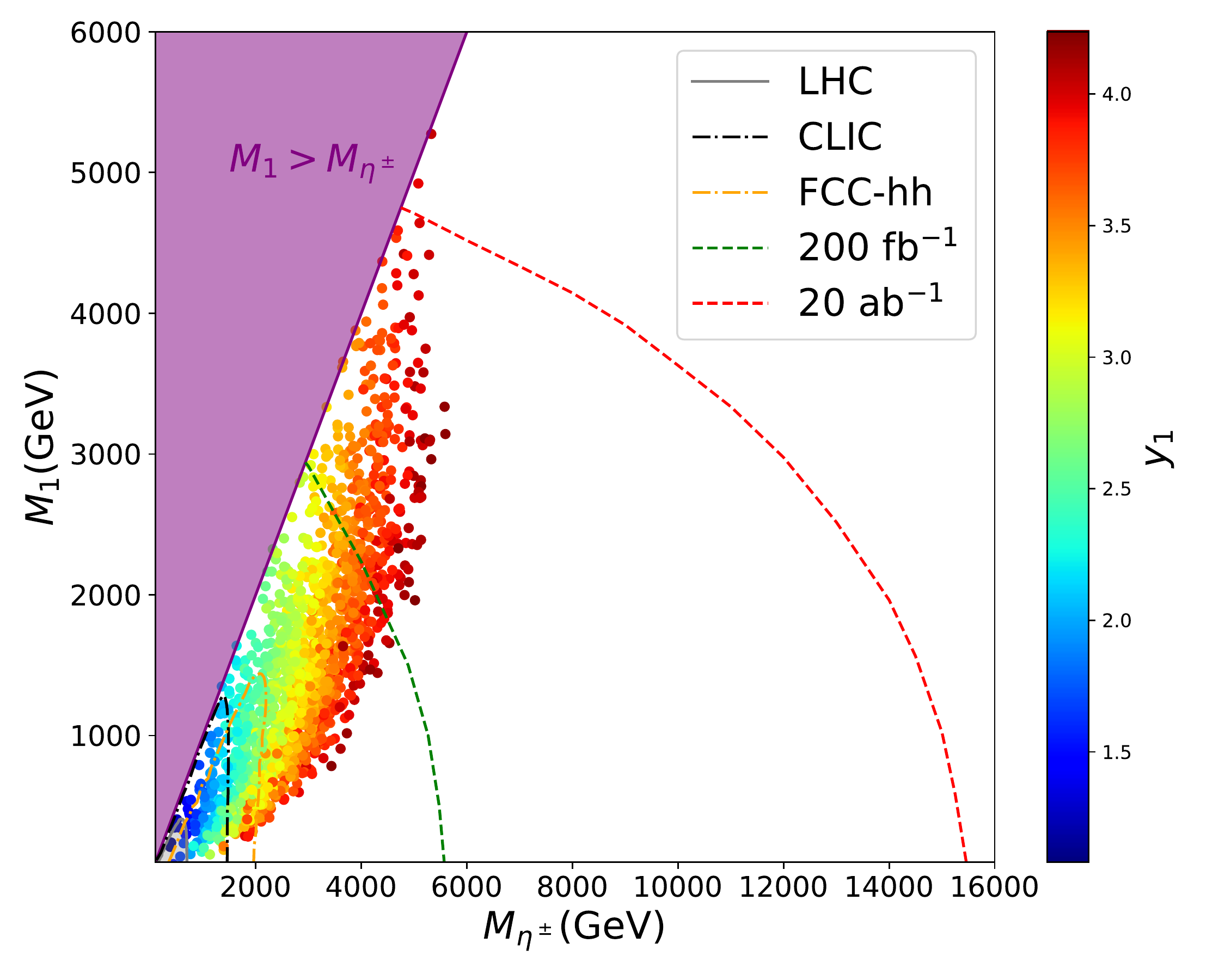}
	\caption{Same as Fig.~\ref{FIG:SG1}, but for the mono-photon signature. 
		\label{FIG:SGMP}} 	
\end{figure}

In Figure \ref{FIG:SGMP}, the $5\sigma$ discovery reaches of the two specific scenarios for the mono-photon signature are shown. In the left panel, the theoretical mono-photon cross-section is calculated by {\bf MadGraph5\_aMC@NLO}~\cite{Alwall:2014hca}. The predicted mono-photon cross-section is less than 80 fb. With the cuts shown in Table \ref{Tab:MonoPhoton}, the cut efficiency for the scenario $M_1=M_{\eta^\pm}/2$ decreases from 0.12 to 0.03 when $M_{\eta^\pm}$ increases. According to our simulation, no samples are within the reach of 2 fb$^{-1}$ data. Provided an integrated luminosity of 200 fb$^{-1}$, the 14 TeV MuC could discover the samples with $\sigma(\gamma+\cancel{E}_T)\gtrsim 10$ fb, which corresponds to $|y_{1\mu}|\gtrsim1.5$. The final 20 ab$^{-1}$ data would push the $5\sigma$ discovery limit down to about 1~fb, but is still hard to probe the $|y_{1\mu}|\lesssim1$ region in this mono-photon channel.

Fixing  $y_{1e}=0.02,y_{1\mu}=y_{1\tau}=2$, the $5\sigma$ discovery limits on the $M_{\eta^\pm}-M_1$ plane are shown in the right panel of Figure ~\ref{FIG:SGMP}. We do not find the corresponding $5\sigma$ discovery limit for 2 fb$^{-1}$ in this analysis. The region with $M_1\lesssim3$ TeV and $M_{\eta^\pm}\lesssim5.5$ TeV is unrevealed by 200 fb$^{-1}$ data.
Increasing the luminosity to 20 ab$^{-1}$ is able to cover almost all samples. Notably, the mono-photon signature is also sensitive to the compressed mass region $M_1\simeq M_{\eta^\pm}$.
For $M_{\eta^\pm}$ above 7 TeV, the mono-photon signature can also reach $5\sigma$ when $M_1\lesssim4$ TeV with 20 ab$^{-1}$ data. So this signal is expected powerful to probe the heavy $\eta^\pm$ region at lower collision energy, e.g., a 6 TeV MuC.

\subsection{Mass Measurement}\label{SEC:Mass}

\begin{figure} 
	\centering
	\includegraphics[width=0.45\textwidth]{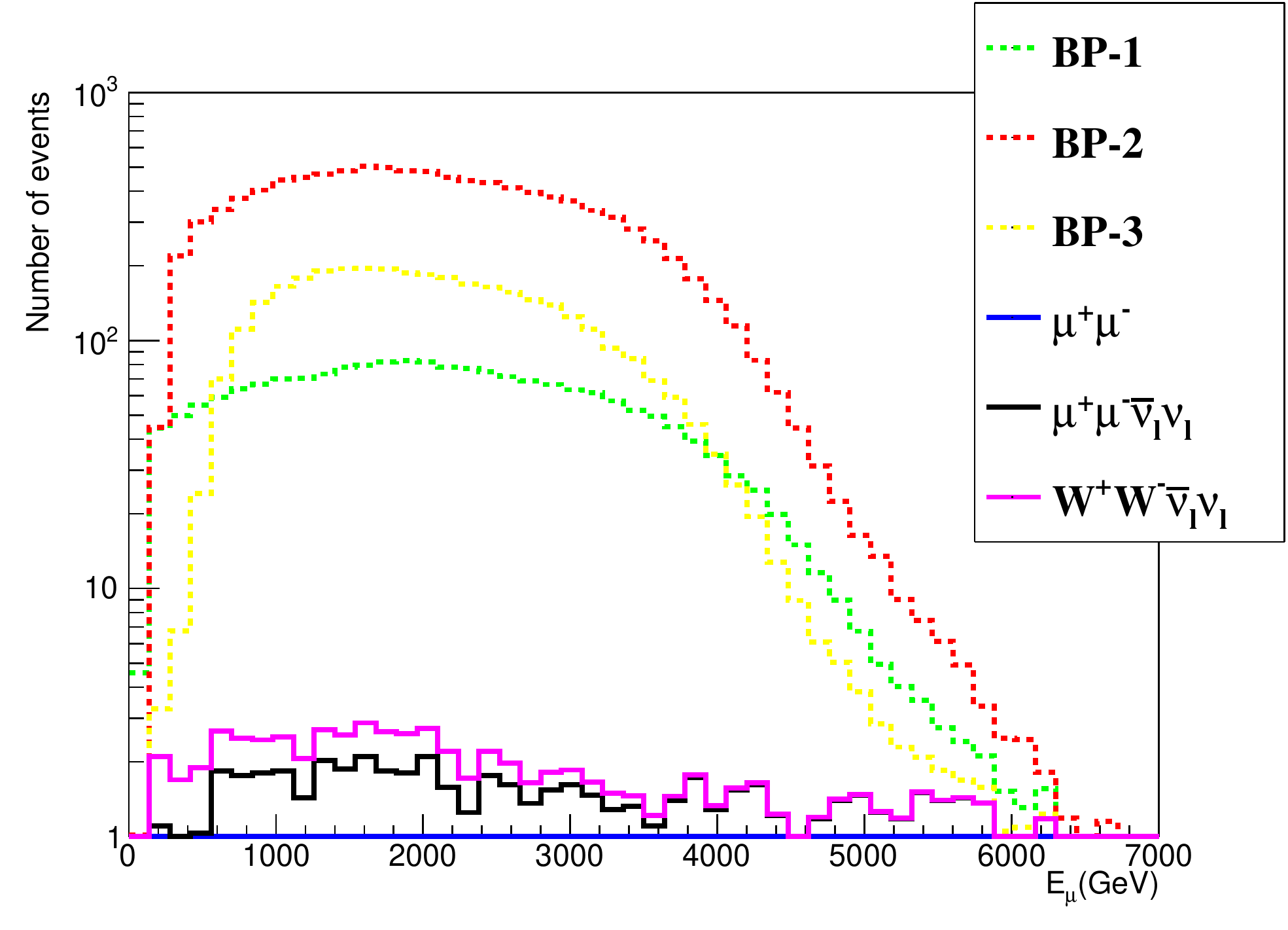}
	\includegraphics[width=0.45\textwidth]{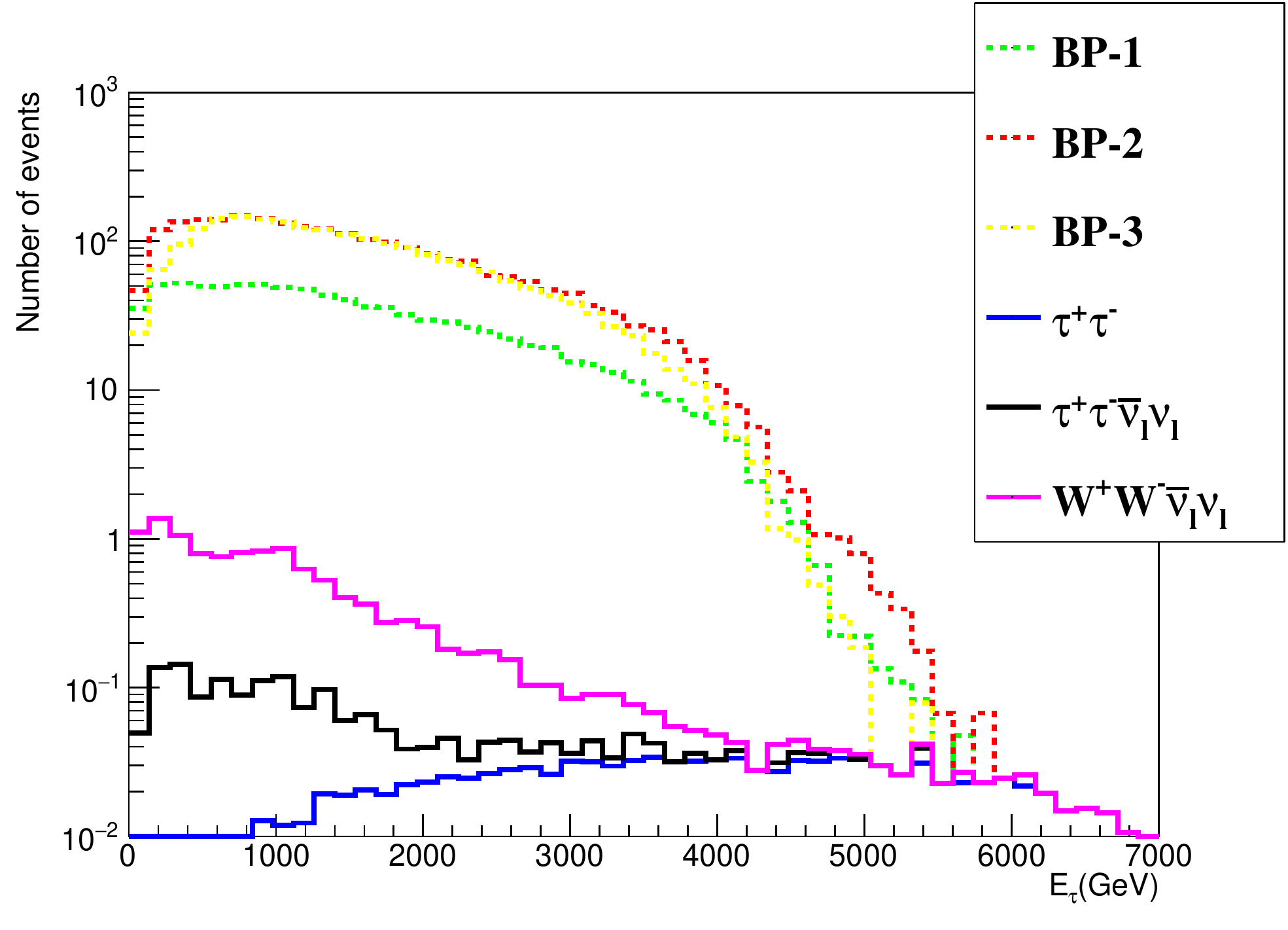}
	\caption{Distribution of the lepton energy $E_\mu$ (left) and $E_\tau$ (right) after applying all cuts.  Here, we assume an integrated luminosity of $200~\text{fb}^{-1}$.
		\label{FIG:SGE}} 	
\end{figure}

At MuC, masses of charged scalar $\eta^\pm$ and dark matter $N_1$ can be extracted from the endpoints of the lepton energy distribution \cite{Feng:1993sd,Homiller:2022iax}
\begin{equation}
	M_{\eta^\pm}=\sqrt{s}\frac{\sqrt{E_L E_H}}{E_L+E_H},
	~M_1=M_{\eta^\pm}\left(1-\frac{2(E_H+E_L)}{\sqrt{s}}\right)^{1/2},
\end{equation}
with the higher and lower endpoints $E_{H,L}$ given by
\begin{equation}
	E_{H,L}=\frac{\sqrt{s}}{4}\left(1-\frac{M_1^2}{M_{\eta^\pm}^2}\right)\left(1\pm\sqrt{1-4\frac{M_{\eta^\pm}^2}{s}}\right).
\end{equation}
For example, BP-2 predicts the higher and lower endpoints to be $E_H=4342$ GeV and $E_L=167$ GeV. 

\begin{figure} 
	\centering
	\includegraphics[width=0.45\textwidth]{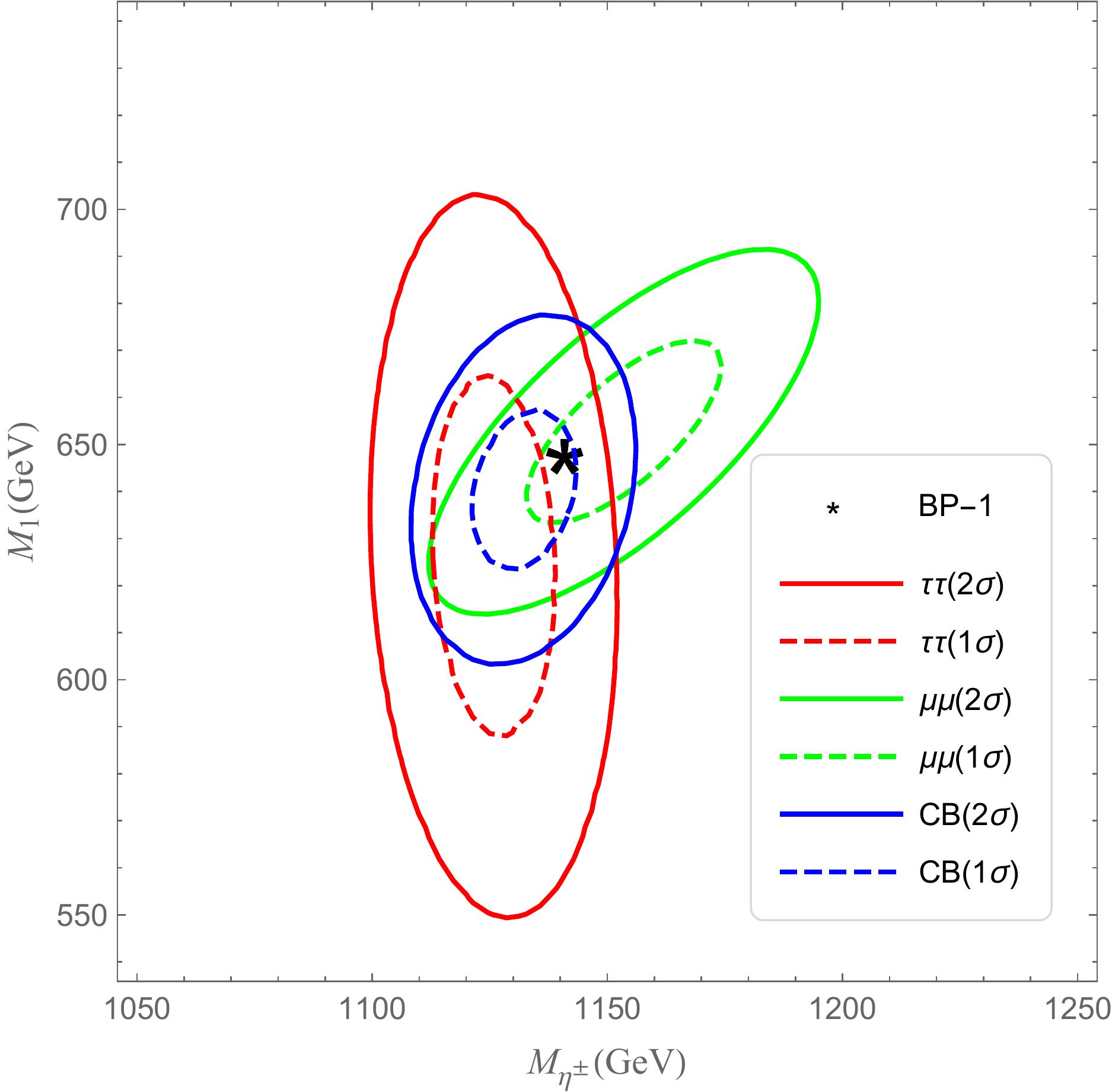}
	\includegraphics[width=0.45\textwidth]{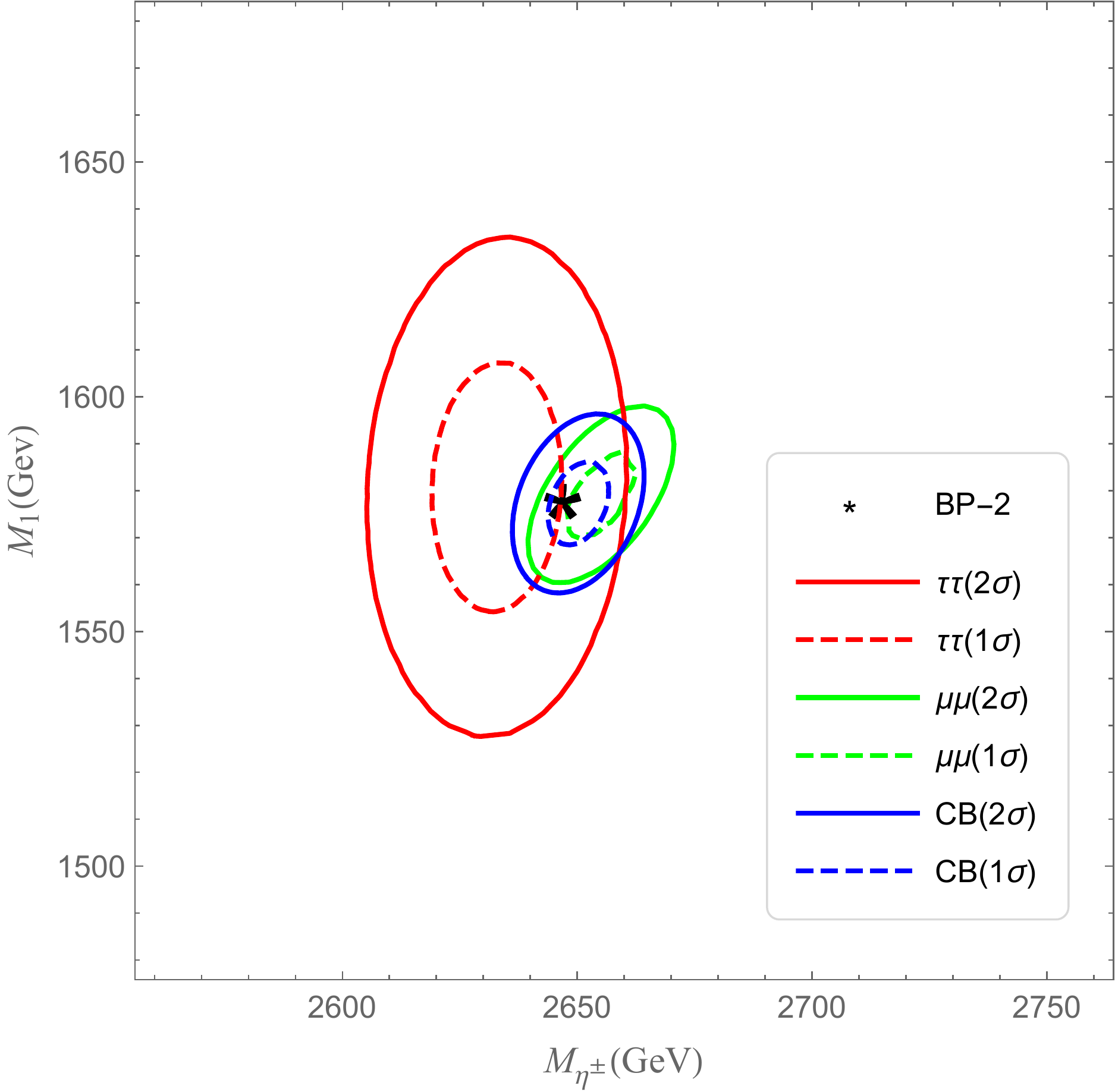}
	\includegraphics[width=0.45\textwidth]{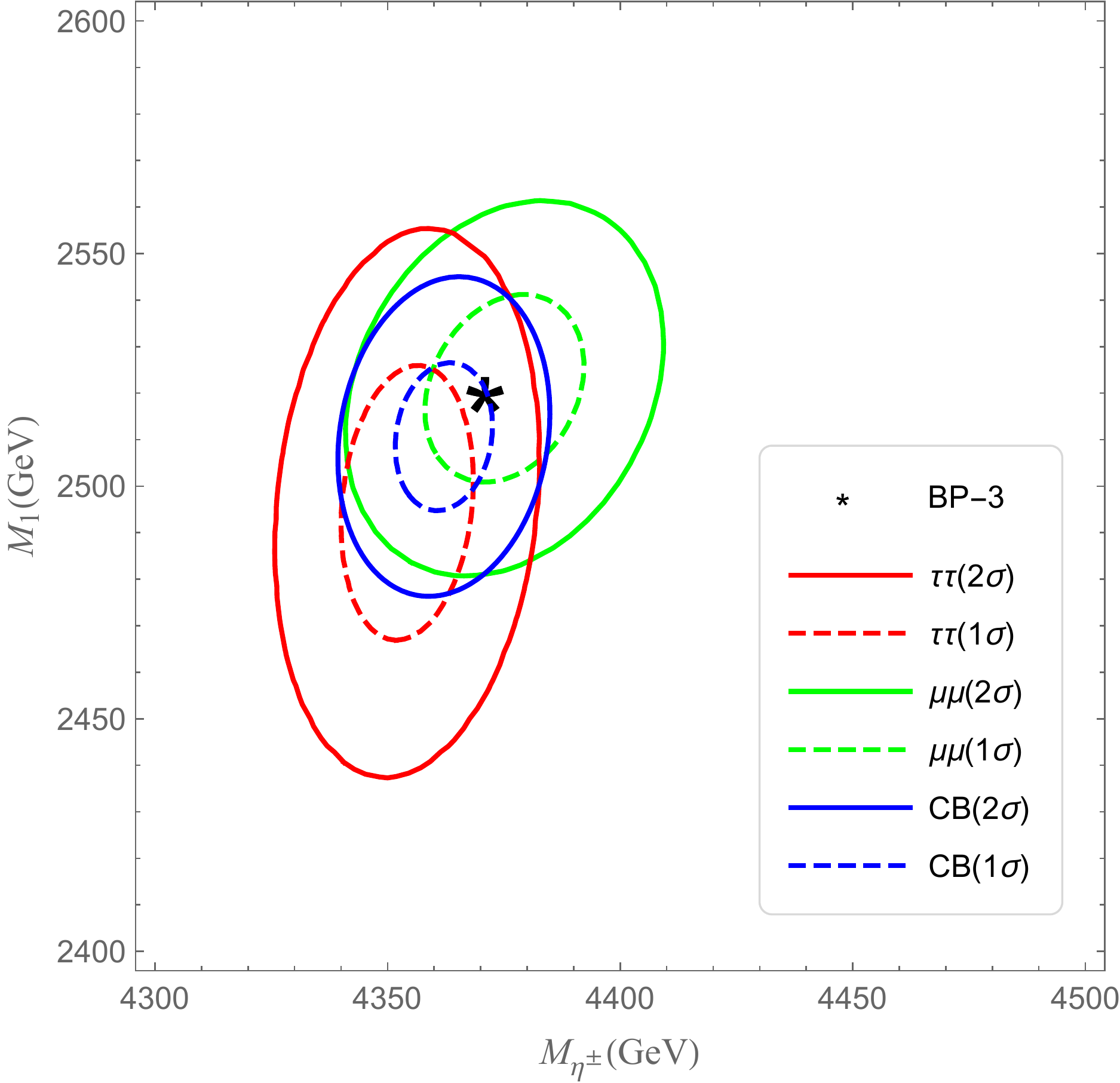}
	\caption{The fitted results of the benchmark points in the $M_{\eta^\pm}-M_1$ plane for both $\mu^+\mu^-+\cancel{E}_T$ and $\tau^+\tau^-+\cancel{E}_T$ channel. Here, we assume an integrated luminosity of $200~\text{fb}^{-1}$. The red and green lines are the results of $\tau^+\tau^-+\cancel{E}_T$ and $\mu^+\mu^-+\cancel{E}_T$ channel. The blue lines are the combined (denoted as CB) results of the dimuon and ditau channel. The dashed and solid lines correspond to the $1\sigma$ and $2\sigma$ range. The stars $\star$ are the actual values for the benchmark points.
		\label{FIG:SGM}} 	
\end{figure}

The backgrounds should be suppressed to tiny level in order to observe these two endpoints. In Figure~\ref{FIG:SGE}, we show the distribution of $E_\mu$ and $E_\tau$ after applying all cuts. Compared with the signal, the backgrounds are small enough. However, the endpoints of the signals are not always so clear to obtain. The lower endpoints in the distribution of $E_\mu$ are clear, but are smeared in the distribution of $E_\tau$. Meanwhile, the higher endpoints in the distributions actually correspond to knee structures. 

To further determine the masses of charged scalar $\eta^\pm$ and DM $N_1$, we then perform a binned likelihood fit. The logarithm of the likelihood is defined as \cite{Feng:1993sd}
\begin{equation}
	\log \mathcal{L}(M_{\eta^\pm},M_1)=\sum_i^{n_b} A_i(M_{\eta^\pm},M_1)\log B_i - B_i,
\end{equation}
where $n_b$ is the number of bins, $A_i(M_{\eta^\pm},M_1)$ is the expected number of events in the $i$-th bin with masses $M_{\eta^\pm}$ and $M_1$, $B_i$ is the measured number of events in the $i$-th bin. During our simulation, we set $B_i$ to be the event number predicted by the benchmark points and corresponding backgrounds.

The fitted results are shown in Figure \ref{FIG:SGM}. Generally speaking, the actual values of benchmark points are within the $2\sigma$ range of the dimuon and ditau results. With a larger cross-section after all cuts, the dimuon channel usually has a better mass resolution than the ditau channel. We also find that the best fit values for $M_{\eta^\pm}$ of the dimuon channel are always larger than those of the ditau channel. There are strong correlations between $M_{\eta^\pm}$ and $M_1$ in the dimuon channel for all three benchmark points. After combining the results of the dimuon and ditau channel, the true values of benchmark points are within the $1\sigma$ range of the fitted results. 

In Table~\ref{Tab:Mass}, we summarize the fitted results for the dimuon, ditau  and combined channel. For BP-1 and BP-3, the actual values are in the $1\sigma$ range of the dimuon channel, but are out the $1\sigma$ range of the ditau channel. For BP-2, the dimuon channel is the most precise one due to the largest cross-section after all cuts. The actual value of BP-2 is on the edges of $1\sigma$ range for both dimuon and ditau channel. The combined results indicate that masses of charged scalar $\eta^\pm$ and dark matter $N_1$ can be measured at the level of $\mathcal{O}(10)$ GeV with $200~\text{fb}^{-1}$ data.

\begin{table}
	\resizebox{\linewidth}{!}{
	\renewcommand\arraystretch{1.25}
		\begin{tabular}{c| c |c| c | c  | c | c } 
			\hline
			& \multicolumn{2}{c|}{BP-1} & \multicolumn{2}{c|}{BP-2} & \multicolumn{2}{c}{BP-3} 
			\\ \cline{2-7}
			& $M_{\eta^\pm}$ & $M_1$ & $M_{\eta^\pm}$  & $M_1$  & $M_{\eta^\pm}$ & $M_1$ 
			\\ \hline
			Actual Value  & 1141 & 648.7 & 2647  & 1579  & 4371  & 2521
			\\ \hline
			$\mu^+\mu^-\!+\!\cancel{E}_T$  & $1153\pm20$  & ~$653\pm19$  & $2655\pm7$~~ & $1579\pm9$~~ & $4375\pm16$ & $2521\pm19$
			\\ \hline
			$\tau^+\tau^-\!+\!\cancel{E}_T$  & $1126\pm13$ & ~$626\pm38$  & $2633\pm14$ & $1581\pm26$ & $4354\pm14$ & $2496\pm29$
			\\ \hline
			Combined  & $1132\pm11$  & ~$640\pm17$  & $2650\pm7$~~  & $1577\pm9$~~  &  $4362\pm10$ & $2511\pm16$ 
			\\ \hline
		\end{tabular}
		} 
		\caption{ The actual values of benchmark points and corresponding fitted results for dimuon, ditau and combined channel. All the masses are in the unit of GeV.
		\label{Tab:Mass}}
\end{table} 

Besides the lepton energy distribution, masses of the dark sector can be reconstructed by other variables separately. One is the missing mass $\cancel{M}$ defined in Equation \eqref{Eq:Mmiss}, which in theory predicts $\cancel{M}>2M_1$ for the signal.
Another one is the $M_{T2}$ variable defined as \cite{Lester:1999tx,Lester:2014yga},
\begin{equation}
	M_{T2}  = \underset{\textbf{q}_{T,1} + \textbf{q}_{T,2} =~ \cancel{\textbf{E}}_T}{\text{min}} \left\{\text{max}\left[M_{T} (\textbf{P}_{T}^{l_{1}},\textbf{q}_{T,1}),M_{T}(\textbf{P}_{T}^{l_{2}},\textbf{q}_{T,2})\right]\right\},
\end{equation}
where $\textbf{P}_{T}^{l_{1}}$ and $\textbf{P}_{T}^{l_{2}}$ are the transverse momentum vectors of the two leptons, $\textbf{q}_{T,1}$ and $\textbf{q}_{T,2}$ are all possible combinations of two transverse momentum vectors that satisfy $\textbf{q}_{T,1} + \textbf{q}_{T,2} =\cancel{\textbf{E}}_T$.  The $M_{T2}$ variable predicts $M_{T2}<M_{\eta^\pm}$ for the signal.

\begin{figure} 
	\centering
	\includegraphics[width=0.45\textwidth]{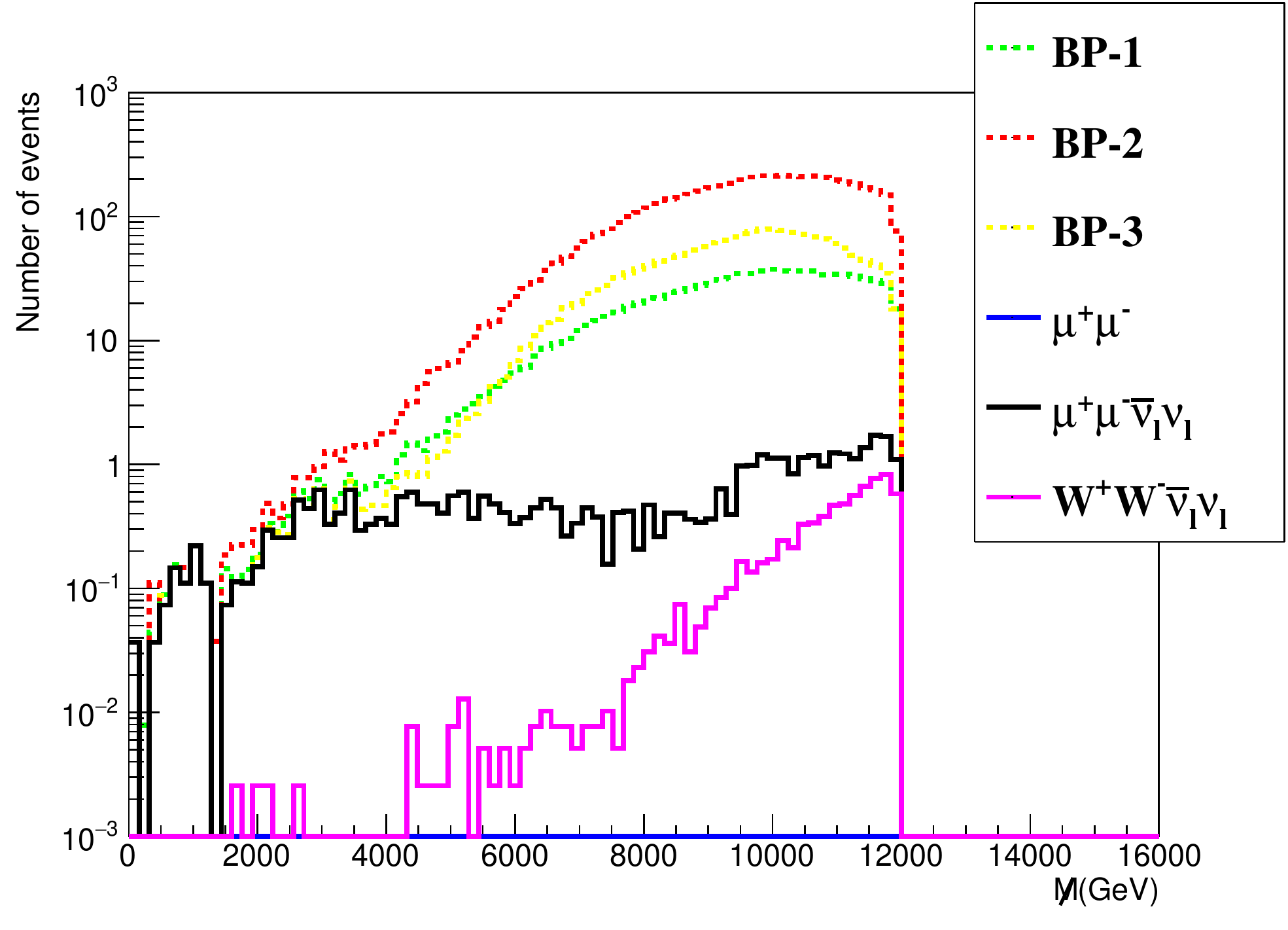}
	\includegraphics[width=0.45\textwidth]{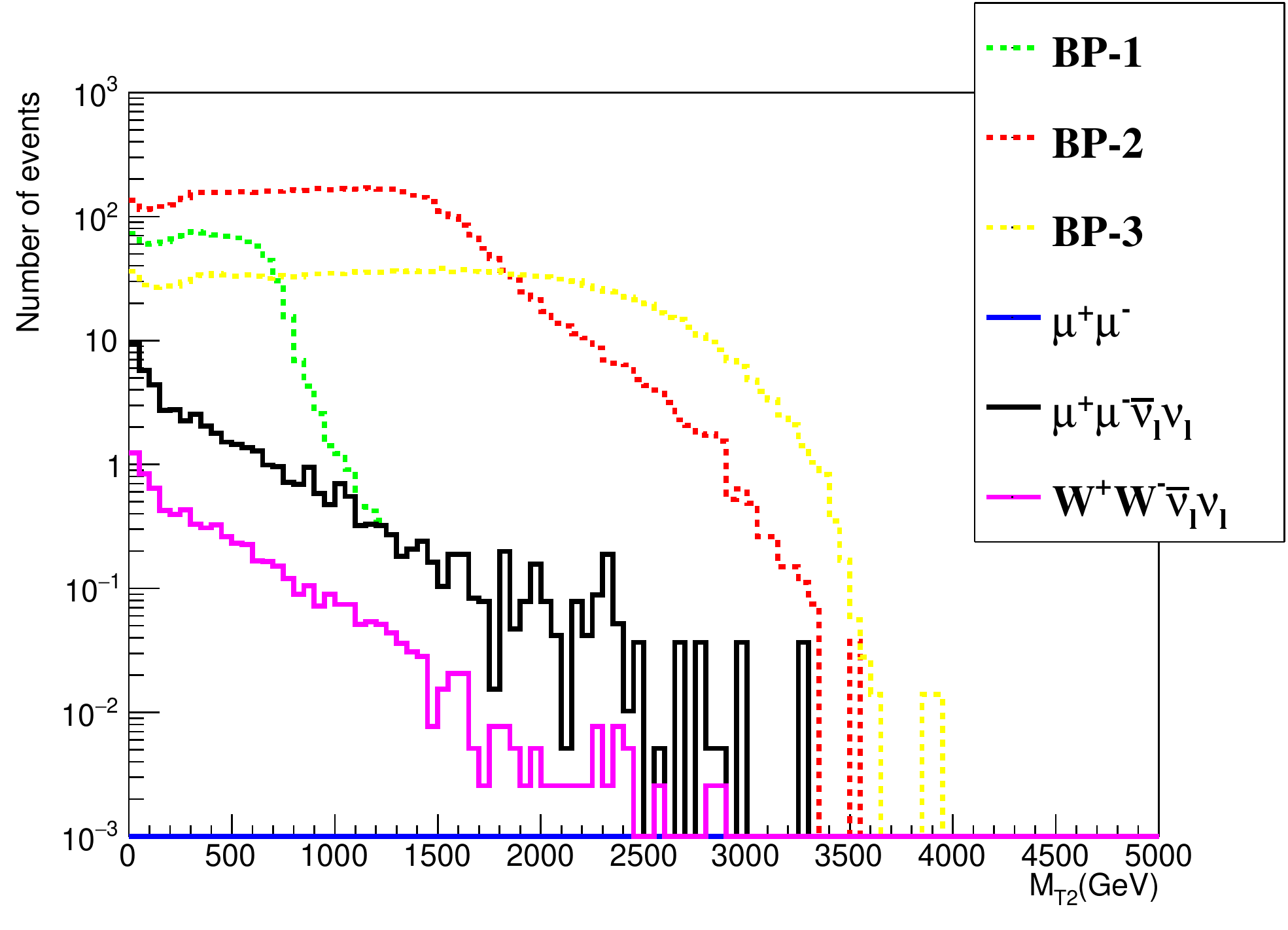}
	\includegraphics[width=0.45\textwidth]{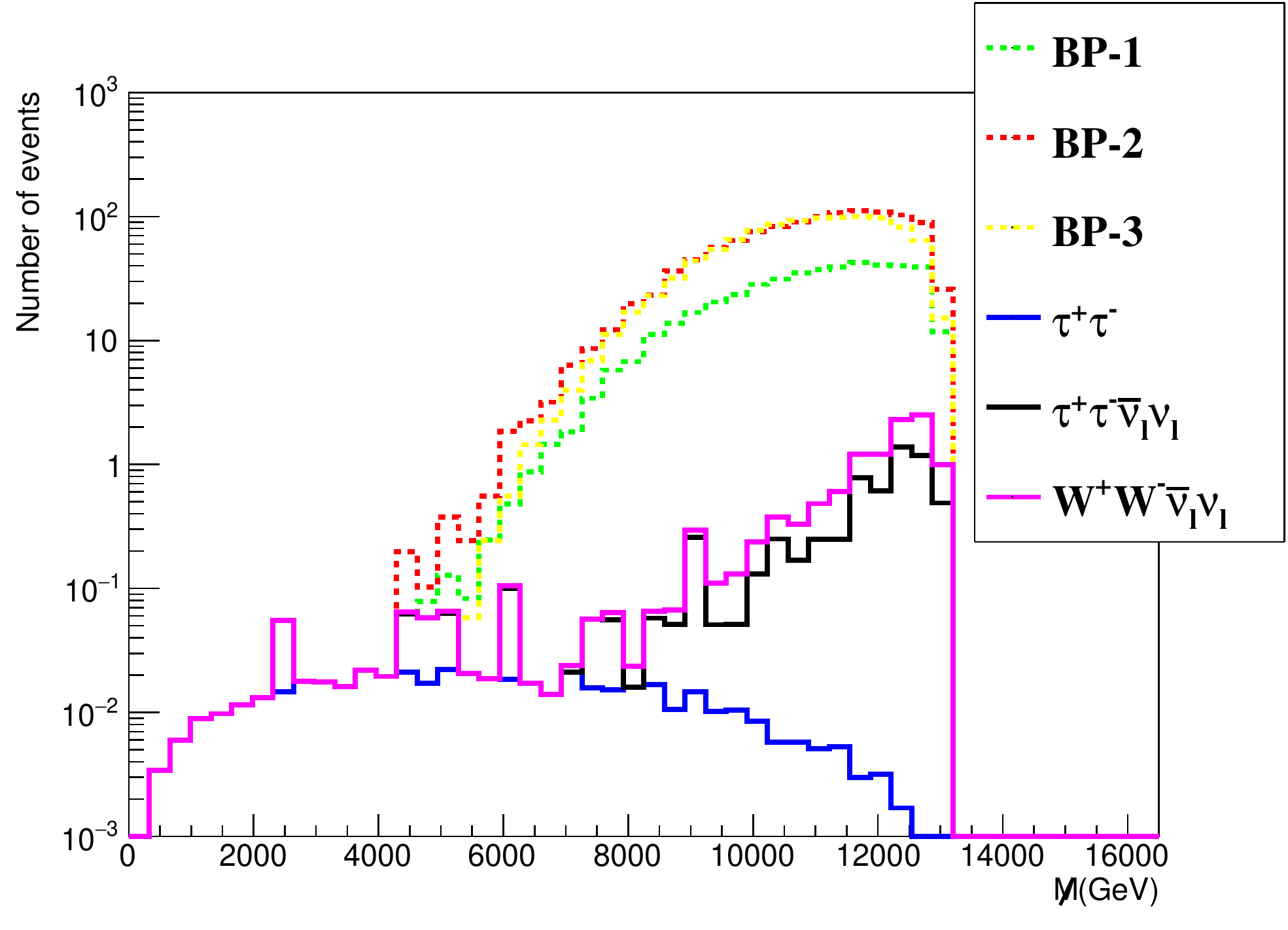}
	\includegraphics[width=0.45\textwidth]{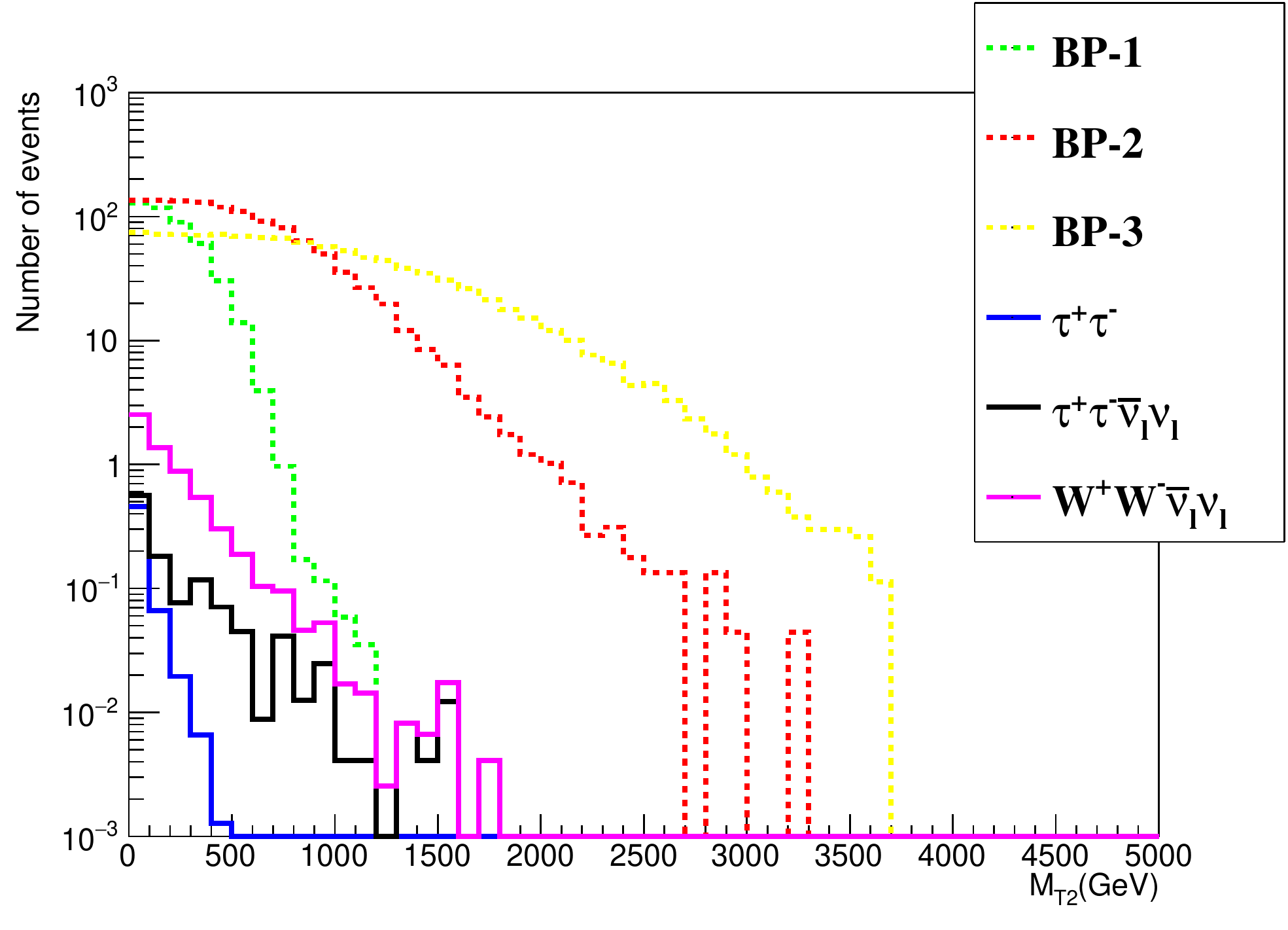}
	\includegraphics[width=0.45\textwidth]{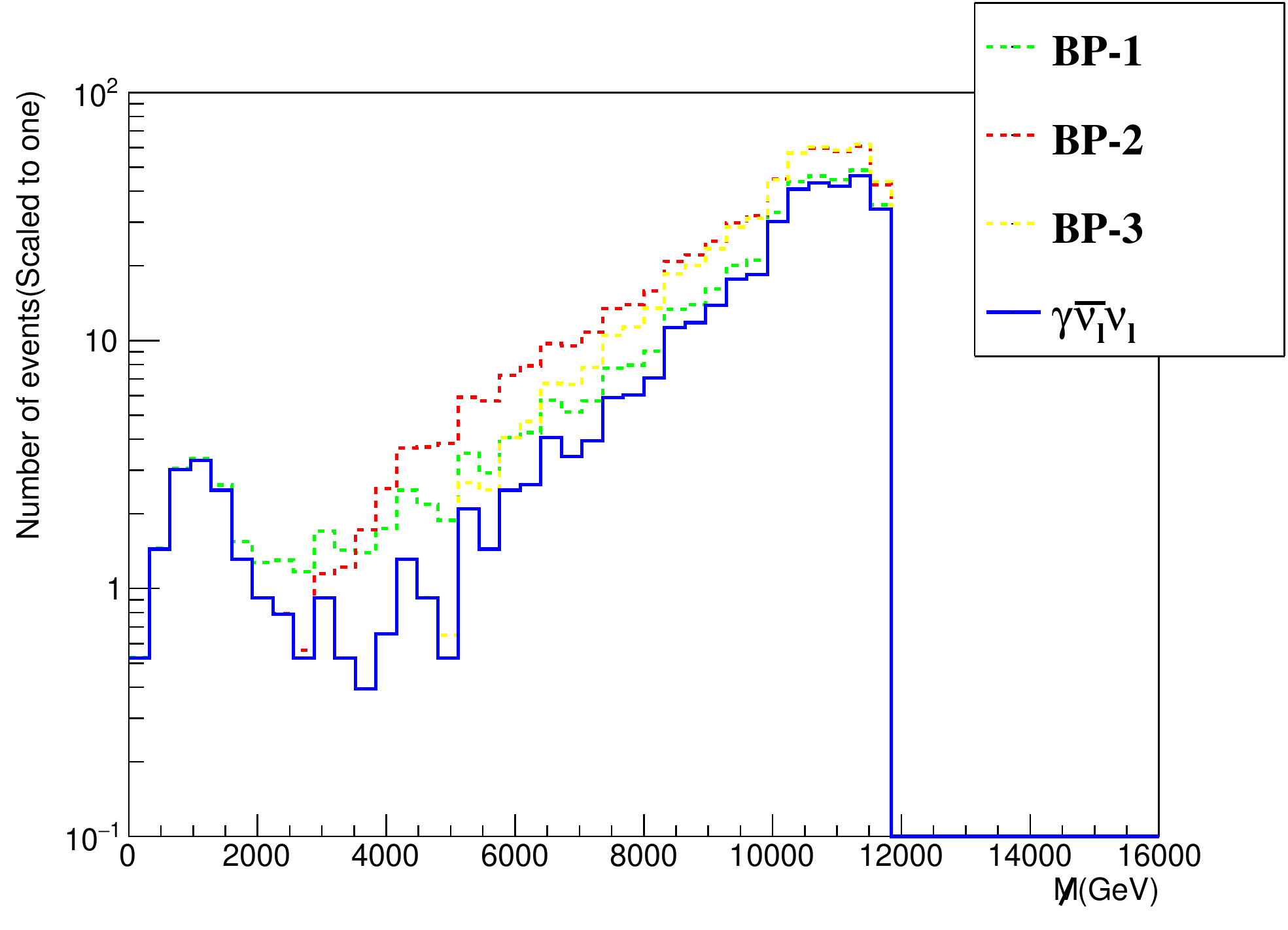}
	\caption{Events number of $\cancel{M}$ and $M_{T2}$ in the dilepton and mono-photon signature after applying all cuts with 200 fb$^{-1}$ data.
		\label{FIG:MT2}} 	
\end{figure}

In Figure \ref{FIG:MT2}, we show the events number of $\cancel{M}$ and $M_{T2}$ after applying all selection cuts in the dilepton and mono-photon signature. In the dimuon and ditau signature, the distribution of $\cancel{M}$ is affected by the applied cuts. In the mono-photon signature, although the lower edges of benchmark points are clear with the signal only, the relatively large background makes it hard to directly observe them. As for the variable $M_{T2}$, the upper edges of benchmark points are obvious. Therefore, we may extract the mass of  dark matter $N_1$  from $\cancel{M}$ in the mono-photon signal and mass of charged scalar $\eta^\pm$ from $M_{T2}$ in the dilepton signal separately by performing binned likelihood fits.
 
\section{Conclusion} \label{Sec:CL}
The Scotogenic model is an appealing way to explain the origin of tiny neutrino mass and dark matter. Three singlet fermion $N_i$ and one inert doublet scalar $\eta$ are introduced in this model. In this paper, we consider the lightest singlet fermion $N_1$ as dark matter. Under the tight constraints from  LFV and relic density, a hierarchical Yukawa structure $|y_{1e}|\ll|y_{1\mu}|\sim|y_{1\tau}|\sim\mathcal{O}(1)$ is usually favored. With large $\mu$-related Yukawa coupling, the cross-section of charged scalar $\eta^\pm$ at MuC is greatly enhanced, which leads to the MuC as an ideal machine to probe the Scotogenic model. In this paper, we investigate the dilepton and mono-photon signature of the Scotogenic model at a 14 TeV MuC.

The most promising channel is the dimuon signature produced via $\mu^+\mu^-\to \eta^+\eta^-\to \mu^+N_1 + \mu^- N_1 \to \mu^+\mu^-+\cancel{E}_T$.  According to our simulation, only $1~\text{fb}^{-1}$ data is enough to reach the $5\sigma$ discovery reach for BP-2 ($M_{\eta^\pm}=2647~\text{GeV},M_1=1579~\text{GeV},|y_{1\mu}|=1.99$). With $200~\text{fb}^{-1}$ data, the 14 TeV MuC is able to unravel most viable samples, except those with small Yukawa coupling $|y_{1\mu}|$ or with compressed mass spectrum $M_1\simeq M_{\eta^\pm}$. The compressed mass region is even hard to probe by the cuts in our analysis even with $20~\text{ab}^{-1}$ data. For the ditau signature produced via $\mu^+\mu^-\to \eta^+\eta^-\to \tau^+N_1 + \tau^- N_1 \to \tau^+\tau^-+\cancel{E}_T$, it is usually less promising than the dimuon channel. To obtain $5\sigma$ discovery, we need about $4.3~\text{fb}^{-1}$ data for BP-2. But when $|y_{1\mu}|$ is too small, the ditau channel becomes more promising. Almost all the samples are within the reach of $20~\text{ab}^{-1}$ data for the ditau channel. The mono-photon signature is produced via $\mu^+\mu^-\to \gamma N_1 N_1\to \gamma +\cancel{E}_T$ by the $t$-channel exchange of $\eta^\pm$. The mono-photon signal is promising with $|y_{1\mu}|\gtrsim 1$. As for the compressed mass region, the mono-photon signature is also able to reach $5\sigma$ significance with 20 ab$^{-1}$ for $M_1\lesssim5$ TeV. At MuC, masses of the charged scalar $\eta^\pm$ and dark matter $N_1$ can be further measured by fitting of the distribution of lepton energy. For instance, the fitted  results of BP-2 are  $M_{\eta^\pm}=2655\pm7$ GeV with $M_1=1579\pm9$ GeV in the dimuon channel and $M_{\eta^\pm}=2633\pm14$~GeV with $M_1=1581\pm26$ GeV in the ditau channel. The combined result is  $M_{\eta^\pm}=2650\pm7$ GeV with $M_1=1577\pm9$ GeV allowing the actual value within the $1\sigma$ range.

Besides the dilepton and mono-photon signature discussed in this paper, we highlight some other interesting signals. With a compressed mass spectrum $M_1\simeq M_{\eta^\pm}$, the calculation of dark matter relic density should be improved by including the coannihilation effect \cite{Vicente:2014wga}. As already shown in Figure~\ref{FIG:SG1} and \ref{FIG:SG2}, the leptons from $\eta^\pm\to \ell^\pm N_1$ decays in this scenario are relatively soft, thus are hard to detect with the cuts in this paper. Searches for the compressed mass region need the {\it LowPt} criterion \cite{ATLAS:2019lng}, which is beyond the scope of this work. In this paper, we only consider the same flavor dilepton signature. The different flavor dilepton signatures as $\mu^\pm\tau^\mp+\cancel{E}_T$, $\tau^\pm e^\mp+\cancel{E}_T$ and $\mu^\pm e^\mp+\cancel{E}_T$ are also possible.  Meanwhile, if the heavier singlet fermions $N_j(j=2,3)$ are light enough, they can be singly and doubly produced at MuC via  $\mu^+\mu^-\to N_j N_1, N_j N_j$. Then the cascade decay $N_j\to \ell^\pm \eta^\mp \to \ell^\pm\ell^\mp N_1$ will lead to dilepton and tetraleton signature. Studies of these signatures will further enlighten the nature of Scotogenic model.

%%%%%%%%%%%%%%%%%%%
\section*{Acknowledgments}
This work is supported by the National Natural Science Foundation of China under Grant No. 11805081 and  11635009, Natural Science Foundation of Shandong Province under Grant No. ZR2019QA021, the Open Project of Guangxi Key Laboratory of Nuclear Physics and Nuclear Technology under Grant No. NLK2021-07.

\bibliographystyle{JHEP}
\bibliography{SG}

\providecommand{\href}[2]{#2}\begingroup\raggedright\begin{thebibliography}{100}

\bibitem{Krauss:2002px}
L.~M. Krauss, S.~Nasri and M.~Trodden, \emph{{A Model for neutrino masses and
  dark matter}}, \href{https://doi.org/10.1103/PhysRevD.67.085002}{\emph{Phys.
  Rev. D} {\bfseries 67} (2003) 085002}
  [\href{https://arxiv.org/abs/hep-ph/0210389}{{\ttfamily hep-ph/0210389}}].

\bibitem{Asaka:2005an}
T.~Asaka, S.~Blanchet and M.~Shaposhnikov, \emph{{The nuMSM, dark matter and
  neutrino masses}},
  \href{https://doi.org/10.1016/j.physletb.2005.09.070}{\emph{Phys. Lett. B}
  {\bfseries 631} (2005) 151}
  [\href{https://arxiv.org/abs/hep-ph/0503065}{{\ttfamily hep-ph/0503065}}].

\bibitem{Ma:2006km}
E.~Ma, \emph{{Verifiable radiative seesaw mechanism of neutrino mass and dark
  matter}}, \href{https://doi.org/10.1103/PhysRevD.73.077301}{\emph{Phys. Rev.
  D} {\bfseries 73} (2006) 077301}
  [\href{https://arxiv.org/abs/hep-ph/0601225}{{\ttfamily hep-ph/0601225}}].

\bibitem{Aoki:2008av}
M.~Aoki, S.~Kanemura and O.~Seto, \emph{{Neutrino mass, Dark Matter and Baryon
  Asymmetry via TeV-Scale Physics without Fine-Tuning}},
  \href{https://doi.org/10.1103/PhysRevLett.102.051805}{\emph{Phys. Rev. Lett.}
  {\bfseries 102} (2009) 051805}
  [\href{https://arxiv.org/abs/0807.0361}{{\ttfamily 0807.0361}}].

\bibitem{Restrepo:2013aga}
D.~Restrepo, O.~Zapata and C.~E. Yaguna, \emph{{Models with radiative neutrino
  masses and viable dark matter candidates}},
  \href{https://doi.org/10.1007/JHEP11(2013)011}{\emph{JHEP} {\bfseries 11}
  (2013) 011} [\href{https://arxiv.org/abs/1308.3655}{{\ttfamily 1308.3655}}].

\bibitem{Escudero:2016ksa}
M.~Escudero, N.~Rius and V.~Sanz, \emph{{Sterile Neutrino portal to Dark Matter
  II: Exact Dark symmetry}},
  \href{https://doi.org/10.1140/epjc/s10052-017-4963-x}{\emph{Eur. Phys. J. C}
  {\bfseries 77} (2017) 397}
  [\href{https://arxiv.org/abs/1607.02373}{{\ttfamily 1607.02373}}].

\bibitem{Cai:2017jrq}
Y.~Cai, J.~Herrero-Garc\'\i{}a, M.~A. Schmidt, A.~Vicente and R.~R. Volkas,
  \emph{{From the trees to the forest: a review of radiative neutrino mass
  models}}, \href{https://doi.org/10.3389/fphy.2017.00063}{\emph{Front. in
  Phys.} {\bfseries 5} (2017) 63}
  [\href{https://arxiv.org/abs/1706.08524}{{\ttfamily 1706.08524}}].

\bibitem{Cacciapaglia:2020psm}
G.~Cacciapaglia and M.~Rosenlyst, \emph{{Loop-generated neutrino masses in
  composite Higgs models}},
  \href{https://doi.org/10.1007/JHEP09(2021)167}{\emph{JHEP} {\bfseries 09}
  (2021) 167} [\href{https://arxiv.org/abs/2010.01437}{{\ttfamily
  2010.01437}}].

\bibitem{Kubo:2006yx}
J.~Kubo, E.~Ma and D.~Suematsu, \emph{{Cold Dark Matter, Radiative Neutrino
  Mass, $\mu \to e\gamma$, and Neutrinoless Double Beta Decay}},
  \href{https://doi.org/10.1016/j.physletb.2006.08.085}{\emph{Phys. Lett. B}
  {\bfseries 642} (2006) 18}
  [\href{https://arxiv.org/abs/hep-ph/0604114}{{\ttfamily hep-ph/0604114}}].

\bibitem{AristizabalSierra:2008cnr}
D.~Aristizabal~Sierra, J.~Kubo, D.~Restrepo, D.~Suematsu and O.~Zapata,
  \emph{{Radiative seesaw: Warm dark matter, collider and lepton flavour
  violating signals}},
  \href{https://doi.org/10.1103/PhysRevD.79.013011}{\emph{Phys. Rev. D}
  {\bfseries 79} (2009) 013011}
  [\href{https://arxiv.org/abs/0808.3340}{{\ttfamily 0808.3340}}].

\bibitem{Suematsu:2009ww}
D.~Suematsu, T.~Toma and T.~Yoshida, \emph{{Reconciliation of CDM abundance and
  mu ---\ensuremath{>} e gamma in a radiative seesaw model}},
  \href{https://doi.org/10.1103/PhysRevD.79.093004}{\emph{Phys. Rev. D}
  {\bfseries 79} (2009) 093004}
  [\href{https://arxiv.org/abs/0903.0287}{{\ttfamily 0903.0287}}].

\bibitem{Kanemura:2011vm}
S.~Kanemura, O.~Seto and T.~Shimomura, \emph{{Masses of dark matter and
  neutrino from TeV scale spontaneous $U(1)_{B-L}$ breaking}},
  \href{https://doi.org/10.1103/PhysRevD.84.016004}{\emph{Phys. Rev. D}
  {\bfseries 84} (2011) 016004}
  [\href{https://arxiv.org/abs/1101.5713}{{\ttfamily 1101.5713}}].

\bibitem{Schmidt:2012yg}
D.~Schmidt, T.~Schwetz and T.~Toma, \emph{{Direct Detection of Leptophilic Dark
  Matter in a Model with Radiative Neutrino Masses}},
  \href{https://doi.org/10.1103/PhysRevD.85.073009}{\emph{Phys. Rev. D}
  {\bfseries 85} (2012) 073009}
  [\href{https://arxiv.org/abs/1201.0906}{{\ttfamily 1201.0906}}].

\bibitem{Baek:2015mna}
S.~Baek, H.~Okada and K.~Yagyu, \emph{{Flavour Dependent Gauged Radiative
  Neutrino Mass Model}},
  \href{https://doi.org/10.1007/JHEP04(2015)049}{\emph{JHEP} {\bfseries 04}
  (2015) 049} [\href{https://arxiv.org/abs/1501.01530}{{\ttfamily
  1501.01530}}].

\bibitem{Merle:2015gea}
A.~Merle and M.~Platscher, \emph{{Parity Problem of the Scotogenic Neutrino
  Model}}, \href{https://doi.org/10.1103/PhysRevD.92.095002}{\emph{Phys. Rev.
  D} {\bfseries 92} (2015) 095002}
  [\href{https://arxiv.org/abs/1502.03098}{{\ttfamily 1502.03098}}].

\bibitem{Ahriche:2017iar}
A.~Ahriche, A.~Jueid and S.~Nasri, \emph{{Radiative neutrino mass and Majorana
  dark matter within an inert Higgs doublet model}},
  \href{https://doi.org/10.1103/PhysRevD.97.095012}{\emph{Phys. Rev. D}
  {\bfseries 97} (2018) 095012}
  [\href{https://arxiv.org/abs/1710.03824}{{\ttfamily 1710.03824}}].

\bibitem{Hugle:2018qbw}
T.~Hugle, M.~Platscher and K.~Schmitz, \emph{{Low-Scale Leptogenesis in the
  Scotogenic Neutrino Mass Model}},
  \href{https://doi.org/10.1103/PhysRevD.98.023020}{\emph{Phys. Rev. D}
  {\bfseries 98} (2018) 023020}
  [\href{https://arxiv.org/abs/1804.09660}{{\ttfamily 1804.09660}}].

\bibitem{Baumholzer:2018sfb}
S.~Baumholzer, V.~Brdar and P.~Schwaller, \emph{{The New $\nu$MSM
  ($\nu\nu$MSM): Radiative Neutrino Masses, keV-Scale Dark Matter and Viable
  Leptogenesis with sub-TeV New Physics}},
  \href{https://doi.org/10.1007/JHEP08(2018)067}{\emph{JHEP} {\bfseries 08}
  (2018) 067} [\href{https://arxiv.org/abs/1806.06864}{{\ttfamily
  1806.06864}}].

\bibitem{Ahriche:2018ger}
A.~Ahriche, A.~Arhrib, A.~Jueid, S.~Nasri and A.~de~La~Puente,
  \emph{{Mono-Higgs Signature in the Scotogenic Model with Majorana Dark
  Matter}}, \href{https://doi.org/10.1103/PhysRevD.101.035038}{\emph{Phys. Rev.
  D} {\bfseries 101} (2020) 035038}
  [\href{https://arxiv.org/abs/1811.00490}{{\ttfamily 1811.00490}}].

\bibitem{Han:2019lux}
Z.-L. Han and W.~Wang, \emph{{Predictive Scotogenic Model with Flavor Dependent
  Symmetry}}, \href{https://doi.org/10.1140/epjc/s10052-019-7033-8}{\emph{Eur.
  Phys. J. C} {\bfseries 79} (2019) 522}
  [\href{https://arxiv.org/abs/1901.07798}{{\ttfamily 1901.07798}}].

\bibitem{Wang:2019byi}
W.~Wang and Z.-L. Han, \emph{{$U(1)_{B-3L_{\alpha}}$ extended scotogenic models
  and single-zero textures of neutrino mass matrices}},
  \href{https://doi.org/10.1103/PhysRevD.101.115040}{\emph{Phys. Rev. D}
  {\bfseries 101} (2020) 115040}
  [\href{https://arxiv.org/abs/1911.00819}{{\ttfamily 1911.00819}}].

\bibitem{Borah:2020wut}
D.~Borah, A.~Dasgupta, K.~Fujikura, S.~K. Kang and D.~Mahanta,
  \emph{{Observable Gravitational Waves in Minimal Scotogenic Model}},
  \href{https://doi.org/10.1088/1475-7516/2020/08/046}{\emph{JCAP} {\bfseries
  08} (2020) 046} [\href{https://arxiv.org/abs/2003.02276}{{\ttfamily
  2003.02276}}].

\bibitem{Liao:2022cwh}
Y.~Liao and X.-D. Ma, \emph{{One-loop Matching of Scotogenic Model onto
  Standard Model Effective Field Theory up to Dimension 7}},
  \href{https://arxiv.org/abs/2210.04270}{{\ttfamily 2210.04270}}.

\bibitem{Dolle:2009fn}
E.~M. Dolle and S.~Su, \emph{{The Inert Dark Matter}},
  \href{https://doi.org/10.1103/PhysRevD.80.055012}{\emph{Phys. Rev. D}
  {\bfseries 80} (2009) 055012}
  [\href{https://arxiv.org/abs/0906.1609}{{\ttfamily 0906.1609}}].

\bibitem{Arhrib:2013ela}
A.~Arhrib, Y.-L.~S. Tsai, Q.~Yuan and T.-C. Yuan, \emph{{An Updated Analysis of
  Inert Higgs Doublet Model in light of the Recent Results from LUX, PLANCK,
  AMS-02 and LHC}},
  \href{https://doi.org/10.1088/1475-7516/2014/06/030}{\emph{JCAP} {\bfseries
  06} (2014) 030} [\href{https://arxiv.org/abs/1310.0358}{{\ttfamily
  1310.0358}}].

\bibitem{Belyaev:2016lok}
A.~Belyaev, G.~Cacciapaglia, I.~P. Ivanov, F.~Rojas-Abatte and M.~Thomas,
  \emph{{Anatomy of the Inert Two Higgs Doublet Model in the light of the LHC
  and non-LHC Dark Matter Searches}},
  \href{https://doi.org/10.1103/PhysRevD.97.035011}{\emph{Phys. Rev. D}
  {\bfseries 97} (2018) 035011}
  [\href{https://arxiv.org/abs/1612.00511}{{\ttfamily 1612.00511}}].

\bibitem{Queiroz:2015utg}
F.~S. Queiroz and C.~E. Yaguna, \emph{{The CTA aims at the Inert Doublet
  Model}}, \href{https://doi.org/10.1088/1475-7516/2016/02/038}{\emph{JCAP}
  {\bfseries 02} (2016) 038}
  [\href{https://arxiv.org/abs/1511.05967}{{\ttfamily 1511.05967}}].

\bibitem{Garcia-Cely:2015khw}
C.~Garcia-Cely, M.~Gustafsson and A.~Ibarra, \emph{{Probing the Inert Doublet
  Dark Matter Model with Cherenkov Telescopes}},
  \href{https://doi.org/10.1088/1475-7516/2016/02/043}{\emph{JCAP} {\bfseries
  02} (2016) 043} [\href{https://arxiv.org/abs/1512.02801}{{\ttfamily
  1512.02801}}].

\bibitem{Eiteneuer:2017hoh}
B.~Eiteneuer, A.~Goudelis and J.~Heisig, \emph{{The inert doublet model in the
  light of Fermi-LAT gamma-ray data: a global fit analysis}},
  \href{https://doi.org/10.1140/epjc/s10052-017-5166-1}{\emph{Eur. Phys. J. C}
  {\bfseries 77} (2017) 624}
  [\href{https://arxiv.org/abs/1705.01458}{{\ttfamily 1705.01458}}].

\bibitem{Dolle:2009ft}
E.~Dolle, X.~Miao, S.~Su and B.~Thomas, \emph{{Dilepton Signals in the Inert
  Doublet Model}},
  \href{https://doi.org/10.1103/PhysRevD.81.035003}{\emph{Phys. Rev. D}
  {\bfseries 81} (2010) 035003}
  [\href{https://arxiv.org/abs/0909.3094}{{\ttfamily 0909.3094}}].

\bibitem{Miao:2010rg}
X.~Miao, S.~Su and B.~Thomas, \emph{{Trilepton Signals in the Inert Doublet
  Model}}, \href{https://doi.org/10.1103/PhysRevD.82.035009}{\emph{Phys. Rev.
  D} {\bfseries 82} (2010) 035009}
  [\href{https://arxiv.org/abs/1005.0090}{{\ttfamily 1005.0090}}].

\bibitem{Kalinowski:2018kdn}
J.~Kalinowski, W.~Kotlarski, T.~Robens, D.~Sokolowska and A.~F. Zarnecki,
  \emph{{Exploring Inert Scalars at CLIC}},
  \href{https://doi.org/10.1007/JHEP07(2019)053}{\emph{JHEP} {\bfseries 07}
  (2019) 053} [\href{https://arxiv.org/abs/1811.06952}{{\ttfamily
  1811.06952}}].

\bibitem{Yang:2021hcu}
F.-X. Yang, Z.-L. Han and Y.~Jin, \emph{{Same-Sign Dilepton Signature in the
  Inert Doublet Model}},
  \href{https://doi.org/10.1088/1674-1137/abf828}{\emph{Chin. Phys. C}
  {\bfseries 45} (2021) 073114}
  [\href{https://arxiv.org/abs/2101.06862}{{\ttfamily 2101.06862}}].

\bibitem{Fan:2022dck}
Y.-Z. Fan, T.-P. Tang, Y.-L.~S. Tsai and L.~Wu, \emph{{Inert Higgs Dark Matter
  for CDF II W-Boson Mass and Detection Prospects}},
  \href{https://doi.org/10.1103/PhysRevLett.129.091802}{\emph{Phys. Rev. Lett.}
  {\bfseries 129} (2022) 091802}
  [\href{https://arxiv.org/abs/2204.03693}{{\ttfamily 2204.03693}}].

\bibitem{Ibarra:2016dlb}
A.~Ibarra, C.~E. Yaguna and O.~Zapata, \emph{{Direct Detection of Fermion Dark
  Matter in the Radiative Seesaw Model}},
  \href{https://doi.org/10.1103/PhysRevD.93.035012}{\emph{Phys. Rev. D}
  {\bfseries 93} (2016) 035012}
  [\href{https://arxiv.org/abs/1601.01163}{{\ttfamily 1601.01163}}].

\bibitem{Toma:2013zsa}
T.~Toma and A.~Vicente, \emph{{Lepton Flavor Violation in the Scotogenic
  Model}}, \href{https://doi.org/10.1007/JHEP01(2014)160}{\emph{JHEP}
  {\bfseries 01} (2014) 160} [\href{https://arxiv.org/abs/1312.2840}{{\ttfamily
  1312.2840}}].

\bibitem{Vicente:2014wga}
A.~Vicente and C.~E. Yaguna, \emph{{Probing the scotogenic model with lepton
  flavor violating processes}},
  \href{https://doi.org/10.1007/JHEP02(2015)144}{\emph{JHEP} {\bfseries 02}
  (2015) 144} [\href{https://arxiv.org/abs/1412.2545}{{\ttfamily 1412.2545}}].

\bibitem{Baumholzer:2019twf}
S.~Baumholzer, V.~Brdar, P.~Schwaller and A.~Segner, \emph{{Shining Light on
  the Scotogenic Model: Interplay of Colliders and Cosmology}},
  \href{https://doi.org/10.1007/JHEP09(2020)136}{\emph{JHEP} {\bfseries 09}
  (2020) 136} [\href{https://arxiv.org/abs/1912.08215}{{\ttfamily
  1912.08215}}].

\bibitem{ATLAS:2019lff}
{\scshape ATLAS} collaboration, G.~Aad et~al., \emph{{Search for electroweak
  production of charginos and sleptons decaying into final states with two
  leptons and missing transverse momentum in $\sqrt{s}=13$ TeV $pp$ collisions
  using the ATLAS detector}},
  \href{https://doi.org/10.1140/epjc/s10052-019-7594-6}{\emph{Eur. Phys. J. C}
  {\bfseries 80} (2020) 123}
  [\href{https://arxiv.org/abs/1908.08215}{{\ttfamily 1908.08215}}].

\bibitem{ATLAS:2019lng}
{\scshape ATLAS} collaboration, G.~Aad et~al., \emph{{Searches for electroweak
  production of supersymmetric particles with compressed mass spectra in
  $\sqrt{s}=$ 13 TeV $pp$ collisions with the ATLAS detector}},
  \href{https://doi.org/10.1103/PhysRevD.101.052005}{\emph{Phys. Rev. D}
  {\bfseries 101} (2020) 052005}
  [\href{https://arxiv.org/abs/1911.12606}{{\ttfamily 1911.12606}}].

\bibitem{CMS:2020bfa}
{\scshape CMS} collaboration, A.~M. Sirunyan et~al., \emph{{Search for
  supersymmetry in final states with two oppositely charged same-flavor leptons
  and missing transverse momentum in proton-proton collisions at $\sqrt{s} =$
  13 TeV}}, \href{https://doi.org/10.1007/JHEP04(2021)123}{\emph{JHEP}
  {\bfseries 04} (2021) 123}
  [\href{https://arxiv.org/abs/2012.08600}{{\ttfamily 2012.08600}}].

\bibitem{Delahaye:2019omf}
J.~P. Delahaye, M.~Diemoz, K.~Long, B.~Mansouli\'e, N.~Pastrone, L.~Rivkin
  et~al., \emph{{Muon Colliders}},
  \href{https://arxiv.org/abs/1901.06150}{{\ttfamily 1901.06150}}.

\bibitem{Long:2020wfp}
K.~Long, D.~Lucchesi, M.~Palmer, N.~Pastrone, D.~Schulte and V.~Shiltsev,
  \emph{{Muon colliders to expand frontiers of particle physics}},
  \href{https://doi.org/10.1038/s41567-020-01130-x}{\emph{Nature Phys.}
  {\bfseries 17} (2021) 289}
  [\href{https://arxiv.org/abs/2007.15684}{{\ttfamily 2007.15684}}].

\bibitem{Han:2020pif}
T.~Han, D.~Liu, I.~Low and X.~Wang, \emph{{Electroweak couplings of the Higgs
  boson at a multi-TeV muon collider}},
  \href{https://doi.org/10.1103/PhysRevD.103.013002}{\emph{Phys. Rev. D}
  {\bfseries 103} (2021) 013002}
  [\href{https://arxiv.org/abs/2008.12204}{{\ttfamily 2008.12204}}].

\bibitem{Han:2020uak}
T.~Han, Z.~Liu, L.-T. Wang and X.~Wang, \emph{{WIMPs at High Energy Muon
  Colliders}}, \href{https://doi.org/10.1103/PhysRevD.103.075004}{\emph{Phys.
  Rev. D} {\bfseries 103} (2021) 075004}
  [\href{https://arxiv.org/abs/2009.11287}{{\ttfamily 2009.11287}}].

\bibitem{Liu:2021jyc}
W.~Liu and K.-P. Xie, \emph{{Probing electroweak phase transition with
  multi-TeV muon colliders and gravitational waves}},
  \href{https://doi.org/10.1007/JHEP04(2021)015}{\emph{JHEP} {\bfseries 04}
  (2021) 015} [\href{https://arxiv.org/abs/2101.10469}{{\ttfamily
  2101.10469}}].

\bibitem{Han:2021udl}
T.~Han, S.~Li, S.~Su, W.~Su and Y.~Wu, \emph{{Heavy Higgs bosons in 2HDM at a
  muon collider}},
  \href{https://doi.org/10.1103/PhysRevD.104.055029}{\emph{Phys. Rev. D}
  {\bfseries 104} (2021) 055029}
  [\href{https://arxiv.org/abs/2102.08386}{{\ttfamily 2102.08386}}].

\bibitem{AlAli:2021let}
H.~Al~Ali et~al., \emph{{The muon Smasher\textquoteright{}s guide}},
  \href{https://doi.org/10.1088/1361-6633/ac6678}{\emph{Rept. Prog. Phys.}
  {\bfseries 85} (2022) 084201}
  [\href{https://arxiv.org/abs/2103.14043}{{\ttfamily 2103.14043}}].

\bibitem{Franceschini:2021aqd}
R.~Franceschini and M.~Greco, \emph{{Higgs and BSM Physics at the Future Muon
  Collider}}, \href{https://doi.org/10.3390/sym13050851}{\emph{Symmetry}
  {\bfseries 13} (2021) 851}
  [\href{https://arxiv.org/abs/2104.05770}{{\ttfamily 2104.05770}}].

\bibitem{Bai:2021ony}
X.-H. Bai, Z.-L. Han, Y.~Jin, H.-L. Li and Z.-X. Meng, \emph{{Same-sign
  tetralepton signature in type-II seesaw at lepton colliders *}},
  \href{https://doi.org/10.1088/1674-1137/ac2ed1}{\emph{Chin. Phys. C}
  {\bfseries 46} (2022) 012001}
  [\href{https://arxiv.org/abs/2105.02474}{{\ttfamily 2105.02474}}].

\bibitem{Bottaro:2021snn}
S.~Bottaro, D.~Buttazzo, M.~Costa, R.~Franceschini, P.~Panci, D.~Redigolo
  et~al., \emph{{Closing the window on WIMP Dark Matter}},
  \href{https://doi.org/10.1140/epjc/s10052-021-09917-9}{\emph{Eur. Phys. J. C}
  {\bfseries 82} (2022) 31} [\href{https://arxiv.org/abs/2107.09688}{{\ttfamily
  2107.09688}}].

\bibitem{Chen:2021pqi}
J.~Chen, T.~Li, C.-T. Lu, Y.~Wu and C.-Y. Yao, \emph{{Measurement of Higgs
  boson self-couplings through 2\textrightarrow{}3 vector bosons scattering in
  future muon colliders}},
  \href{https://doi.org/10.1103/PhysRevD.105.053009}{\emph{Phys. Rev. D}
  {\bfseries 105} (2022) 053009}
  [\href{https://arxiv.org/abs/2112.12507}{{\ttfamily 2112.12507}}].

\bibitem{Aime:2022flm}
C.~Aim\`e et~al., \emph{{Muon Collider Physics Summary}},  in \emph{{2022
  Snowmass Summer Study}}, 3, 2022,
  \href{https://arxiv.org/abs/2203.07256}{{\ttfamily 2203.07256}}.

\bibitem{Forslund:2022xjq}
M.~Forslund and P.~Meade, \emph{{High Precision Higgs from High Energy Muon
  Colliders}},  \href{https://arxiv.org/abs/2203.09425}{{\ttfamily
  2203.09425}}.

\bibitem{Yang:2020rjt}
J.-C. Yang, Z.-B. Qing, X.-Y. Han, Y.-C. Guo and T.~Li, \emph{{Tri-photon at
  muon collider: a new process to probe the anomalous quartic gauge
  couplings}}, \href{https://doi.org/10.1007/JHEP07(2022)053}{\emph{JHEP}
  {\bfseries 07} (2022) 053}
  [\href{https://arxiv.org/abs/2204.08195}{{\ttfamily 2204.08195}}].

\bibitem{Yang:2022fhw}
J.-C. Yang, X.-Y. Han, Z.-B. Qin, T.~Li and Y.-C. Guo, \emph{{Measuring the
  anomalous quartic gauge couplings in the $W^+W^-\to W^+W^-$ process at muon
  collider using artificial neural networks}},
  \href{https://doi.org/10.1007/JHEP09(2022)074}{\emph{JHEP} {\bfseries 09}
  (2022) 074} [\href{https://arxiv.org/abs/2204.10034}{{\ttfamily
  2204.10034}}].

\bibitem{Senol:2022snc}
A.~Senol, S.~Spor, E.~Gurkanli, V.~Cetinkaya, H.~Denizli and M.~K\"oksal,
  \emph{{Model-independent study on the anomalous $ZZ\gamma$ and
  $Z\gamma\gamma$ couplings at the future muon collider}},
  \href{https://arxiv.org/abs/2205.02912}{{\ttfamily 2205.02912}}.

\bibitem{Li:2022kkc}
T.~Li, H.~Qin, C.-Y. Yao and M.~Yuan, \emph{{Probing heavy triplet leptons of
  the type-III seesaw mechanism at future muon colliders}},
  \href{https://doi.org/10.1103/PhysRevD.106.035021}{\emph{Phys. Rev. D}
  {\bfseries 106} (2022) 035021}
  [\href{https://arxiv.org/abs/2205.04214}{{\ttfamily 2205.04214}}].

\bibitem{Bottaro:2022one}
S.~Bottaro, D.~Buttazzo, M.~Costa, R.~Franceschini, P.~Panci, D.~Redigolo
  et~al., \emph{{The last Complex WIMPs standing}},
  \href{https://arxiv.org/abs/2205.04486}{{\ttfamily 2205.04486}}.

\bibitem{Chakraborty:2022pcc}
I.~Chakraborty, H.~Roy and T.~Srivastava, \emph{{Searches for heavy neutrinos
  at multi-TeV muon collider : a resonant leptogenesis perspective}},
  \href{https://arxiv.org/abs/2206.07037}{{\ttfamily 2206.07037}}.

\bibitem{Capdevilla:2020qel}
R.~Capdevilla, D.~Curtin, Y.~Kahn and G.~Krnjaic, \emph{{Discovering the
  physics of $(g-2)_\mu$ at future muon colliders}},
  \href{https://doi.org/10.1103/PhysRevD.103.075028}{\emph{Phys. Rev. D}
  {\bfseries 103} (2021) 075028}
  [\href{https://arxiv.org/abs/2006.16277}{{\ttfamily 2006.16277}}].

\bibitem{Buttazzo:2020ibd}
D.~Buttazzo and P.~Paradisi, \emph{{Probing the muon $g-2$ anomaly with the
  Higgs boson at a muon collider}},
  \href{https://doi.org/10.1103/PhysRevD.104.075021}{\emph{Phys. Rev. D}
  {\bfseries 104} (2021) 075021}
  [\href{https://arxiv.org/abs/2012.02769}{{\ttfamily 2012.02769}}].

\bibitem{Yin:2020afe}
W.~Yin and M.~Yamaguchi, \emph{{Muon $g-2$ at multi-TeV muon collider}},
  \href{https://arxiv.org/abs/2012.03928}{{\ttfamily 2012.03928}}.

\bibitem{Huang:2021nkl}
G.-y. Huang, F.~S. Queiroz and W.~Rodejohann, \emph{{Gauged
  $L^{}_{\mu}{-}L^{}_{\tau}$ at a muon collider}},
  \href{https://doi.org/10.1103/PhysRevD.103.095005}{\emph{Phys. Rev. D}
  {\bfseries 103} (2021) 095005}
  [\href{https://arxiv.org/abs/2101.04956}{{\ttfamily 2101.04956}}].

\bibitem{Capdevilla:2021rwo}
R.~Capdevilla, D.~Curtin, Y.~Kahn and G.~Krnjaic, \emph{{No-lose theorem for
  discovering the new physics of (g-2)\ensuremath{\mu} at muon colliders}},
  \href{https://doi.org/10.1103/PhysRevD.105.015028}{\emph{Phys. Rev. D}
  {\bfseries 105} (2022) 015028}
  [\href{https://arxiv.org/abs/2101.10334}{{\ttfamily 2101.10334}}].

\bibitem{Li:2021lnz}
T.~Li, M.~A. Schmidt, C.-Y. Yao and M.~Yuan, \emph{{Charged lepton flavor
  violation in light of the muon magnetic moment anomaly and colliders}},
  \href{https://doi.org/10.1140/epjc/s10052-021-09569-9}{\emph{Eur. Phys. J. C}
  {\bfseries 81} (2021) 811}
  [\href{https://arxiv.org/abs/2104.04494}{{\ttfamily 2104.04494}}].

\bibitem{Huang:2021biu}
G.-y. Huang, S.~Jana, F.~S. Queiroz and W.~Rodejohann, \emph{{Probing the RK(*)
  anomaly at a muon collider}},
  \href{https://doi.org/10.1103/PhysRevD.105.015013}{\emph{Phys. Rev. D}
  {\bfseries 105} (2022) 015013}
  [\href{https://arxiv.org/abs/2103.01617}{{\ttfamily 2103.01617}}].

\bibitem{Asadi:2021gah}
P.~Asadi, R.~Capdevilla, C.~Cesarotti and S.~Homiller, \emph{{Searching for
  leptoquarks at future muon colliders}},
  \href{https://doi.org/10.1007/JHEP10(2021)182}{\emph{JHEP} {\bfseries 10}
  (2021) 182} [\href{https://arxiv.org/abs/2104.05720}{{\ttfamily
  2104.05720}}].

\bibitem{Qian:2021ihf}
S.~Qian, C.~Li, Q.~Li, F.~Meng, J.~Xiao, T.~Yang et~al., \emph{{Searching for
  heavy leptoquarks at a muon collider}},
  \href{https://doi.org/10.1007/JHEP12(2021)047}{\emph{JHEP} {\bfseries 12}
  (2021) 047} [\href{https://arxiv.org/abs/2109.01265}{{\ttfamily
  2109.01265}}].

\bibitem{Altmannshofer:2022xri}
W.~Altmannshofer, S.~A. Gadam and S.~Profumo, \emph{{Snowmass White Paper:
  Probing New Physics with $\mu^+ \mu^- \to bs$ at a Muon Collider}},  in
  \emph{{2022 Snowmass Summer Study}}, 3, 2022,
  \href{https://arxiv.org/abs/2203.07495}{{\ttfamily 2203.07495}}.

\bibitem{Azatov:2022itm}
A.~Azatov, F.~Garosi, A.~Greljo, D.~Marzocca, J.~Salko and S.~Trifinopoulos,
  \emph{{New Physics in $b \to s \mu \mu$: FCC-hh or a Muon Collider?}},
  \href{https://arxiv.org/abs/2205.13552}{{\ttfamily 2205.13552}}.

\bibitem{Battaglia:2013bha}
M.~Battaglia, J.-J. Blaising, J.~S. Marshall, S.~Poss, A.~Sailer, M.~Thomson
  et~al., \emph{{Physics performance for scalar electron, scalar muon and
  scalar neutrino searches at $\sqrt{s} =$ 3 TeV and 1.4 TeV at CLIC}},
  \href{https://doi.org/10.1007/JHEP09(2013)001}{\emph{JHEP} {\bfseries 09}
  (2013) 001} [\href{https://arxiv.org/abs/1304.2825}{{\ttfamily 1304.2825}}].

\bibitem{Homiller:2022iax}
S.~Homiller, Q.~Lu and M.~Reece, \emph{{Complementary Signals of Lepton Flavor
  Violation at a High-Energy Muon Collider}},  in \emph{{2022 Snowmass Summer
  Study}}, 3, 2022, \href{https://arxiv.org/abs/2203.08825}{{\ttfamily
  2203.08825}}.

\bibitem{Alvarez:2021otp}
A.~Alvarez, R.~Cepedello, M.~Hirsch and W.~Porod, \emph{{Temperature effects on
  the Z2 symmetry breaking in the scotogenic model}},
  \href{https://doi.org/10.1103/PhysRevD.105.035013}{\emph{Phys. Rev. D}
  {\bfseries 105} (2022) 035013}
  [\href{https://arxiv.org/abs/2110.04311}{{\ttfamily 2110.04311}}].

\bibitem{Ginzburg:2010wa}
I.~F. Ginzburg, K.~A. Kanishev, M.~Krawczyk and D.~Sokolowska, \emph{{Evolution
  of Universe to the present inert phase}},
  \href{https://doi.org/10.1103/PhysRevD.82.123533}{\emph{Phys. Rev. D}
  {\bfseries 82} (2010) 123533}
  [\href{https://arxiv.org/abs/1009.4593}{{\ttfamily 1009.4593}}].

\bibitem{Branco:2011iw}
G.~C. Branco, P.~M. Ferreira, L.~Lavoura, M.~N. Rebelo, M.~Sher and J.~P.
  Silva, \emph{{Theory and phenomenology of two-Higgs-doublet models}},
  \href{https://doi.org/10.1016/j.physrep.2012.02.002}{\emph{Phys. Rept.}
  {\bfseries 516} (2012) 1} [\href{https://arxiv.org/abs/1106.0034}{{\ttfamily
  1106.0034}}].

\bibitem{Barbieri:2006dq}
R.~Barbieri, L.~J. Hall and V.~S. Rychkov, \emph{{Improved naturalness with a
  heavy Higgs: An Alternative road to LHC physics}},
  \href{https://doi.org/10.1103/PhysRevD.74.015007}{\emph{Phys. Rev. D}
  {\bfseries 74} (2006) 015007}
  [\href{https://arxiv.org/abs/hep-ph/0603188}{{\ttfamily hep-ph/0603188}}].

\bibitem{Casas:2001sr}
J.~A. Casas and A.~Ibarra, \emph{{Oscillating neutrinos and $\mu \to e,
  \gamma$}}, \href{https://doi.org/10.1016/S0550-3213(01)00475-8}{\emph{Nucl.
  Phys. B} {\bfseries 618} (2001) 171}
  [\href{https://arxiv.org/abs/hep-ph/0103065}{{\ttfamily hep-ph/0103065}}].

\bibitem{deSalas:2020pgw}
P.~F. de~Salas, D.~V. Forero, S.~Gariazzo, P.~Mart\'\i{}nez-Mirav\'e, O.~Mena,
  C.~A. Ternes et~al., \emph{{2020 global reassessment of the neutrino
  oscillation picture}},
  \href{https://doi.org/10.1007/JHEP02(2021)071}{\emph{JHEP} {\bfseries 02}
  (2021) 071} [\href{https://arxiv.org/abs/2006.11237}{{\ttfamily
  2006.11237}}].

\bibitem{MEG:2016leq}
{\scshape MEG} collaboration, A.~M. Baldini et~al., \emph{{Search for the
  lepton flavour violating decay $\mu ^+ \rightarrow \mathrm {e}^+ \gamma $
  with the full dataset of the MEG experiment}},
  \href{https://doi.org/10.1140/epjc/s10052-016-4271-x}{\emph{Eur. Phys. J. C}
  {\bfseries 76} (2016) 434}
  [\href{https://arxiv.org/abs/1605.05081}{{\ttfamily 1605.05081}}].

\bibitem{BaBar:2009hkt}
{\scshape BaBar} collaboration, B.~Aubert et~al., \emph{{Searches for Lepton
  Flavor Violation in the Decays tau+- ---\ensuremath{>} e+- gamma and tau+-
  ---\ensuremath{>} mu+- gamma}},
  \href{https://doi.org/10.1103/PhysRevLett.104.021802}{\emph{Phys. Rev. Lett.}
  {\bfseries 104} (2010) 021802}
  [\href{https://arxiv.org/abs/0908.2381}{{\ttfamily 0908.2381}}].

\bibitem{SINDRUM:1987nra}
{\scshape SINDRUM} collaboration, U.~Bellgardt et~al., \emph{{Search for the
  Decay mu+ ---\ensuremath{>} e+ e+ e-}},
  \href{https://doi.org/10.1016/0550-3213(88)90462-2}{\emph{Nucl. Phys. B}
  {\bfseries 299} (1988) 1}.

\bibitem{Hayasaka:2010np}
K.~Hayasaka et~al., \emph{{Search for Lepton Flavor Violating Tau Decays into
  Three Leptons with 719 Million Produced Tau+Tau- Pairs}},
  \href{https://doi.org/10.1016/j.physletb.2010.03.037}{\emph{Phys. Lett. B}
  {\bfseries 687} (2010) 139}
  [\href{https://arxiv.org/abs/1001.3221}{{\ttfamily 1001.3221}}].

\bibitem{Guo:2020qin}
S.-Y. Guo and Z.-L. Han, \emph{{Observable Signatures of Scotogenic Dirac
  Model}}, \href{https://doi.org/10.1007/JHEP12(2020)062}{\emph{JHEP}
  {\bfseries 12} (2020) 062}
  [\href{https://arxiv.org/abs/2005.08287}{{\ttfamily 2005.08287}}].

\bibitem{SINDRUMII:1993gxf}
{\scshape SINDRUM II} collaboration, C.~Dohmen et~al., \emph{{Test of lepton
  flavor conservation in mu ---\ensuremath{>} e conversion on titanium}},
  \href{https://doi.org/10.1016/0370-2693(93)91383-X}{\emph{Phys. Lett. B}
  {\bfseries 317} (1993) 631}.

\bibitem{SINDRUMII:2006dvw}
{\scshape SINDRUM II} collaboration, W.~H. Bertl et~al., \emph{{A Search for
  muon to electron conversion in muonic gold}},
  \href{https://doi.org/10.1140/epjc/s2006-02582-x}{\emph{Eur. Phys. J. C}
  {\bfseries 47} (2006) 337}.

\bibitem{Kitano:2002mt}
R.~Kitano, M.~Koike and Y.~Okada, \emph{{Detailed calculation of lepton flavor
  violating muon electron conversion rate for various nuclei}},
  \href{https://doi.org/10.1103/PhysRevD.76.059902}{\emph{Phys. Rev. D}
  {\bfseries 66} (2002) 096002}
  [\href{https://arxiv.org/abs/hep-ph/0203110}{{\ttfamily hep-ph/0203110}}].

\bibitem{Arganda:2007jw}
E.~Arganda, M.~J. Herrero and A.~M. Teixeira, \emph{{mu-e conversion in nuclei
  within the CMSSM seesaw: Universality versus non-universality}},
  \href{https://doi.org/10.1088/1126-6708/2007/10/104}{\emph{JHEP} {\bfseries
  10} (2007) 104} [\href{https://arxiv.org/abs/0707.2955}{{\ttfamily
  0707.2955}}].

\bibitem{Borah:2016zbd}
D.~Borah and A.~Dasgupta, \emph{{Common Origin of Neutrino Mass, Dark Matter
  and Dirac Leptogenesis}},
  \href{https://doi.org/10.1088/1475-7516/2016/12/034}{\emph{JCAP} {\bfseries
  12} (2016) 034} [\href{https://arxiv.org/abs/1608.03872}{{\ttfamily
  1608.03872}}].

\bibitem{ACME:2013pal}
{\scshape ACME} collaboration, J.~Baron et~al., \emph{{Order of Magnitude
  Smaller Limit on the Electric Dipole Moment of the Electron}},
  \href{https://doi.org/10.1126/science.1248213}{\emph{Science} {\bfseries 343}
  (2014) 269} [\href{https://arxiv.org/abs/1310.7534}{{\ttfamily 1310.7534}}].

\bibitem{XENON:2018voc}
{\scshape XENON} collaboration, E.~Aprile et~al., \emph{{Dark Matter Search
  Results from a One Ton-Year Exposure of XENON1T}},
  \href{https://doi.org/10.1103/PhysRevLett.121.111302}{\emph{Phys. Rev. Lett.}
  {\bfseries 121} (2018) 111302}
  [\href{https://arxiv.org/abs/1805.12562}{{\ttfamily 1805.12562}}].

\bibitem{PandaX-4T:2021bab}
{\scshape PandaX-4T} collaboration, Y.~Meng et~al., \emph{{Dark Matter Search
  Results from the PandaX-4T Commissioning Run}},
  \href{https://doi.org/10.1103/PhysRevLett.127.261802}{\emph{Phys. Rev. Lett.}
  {\bfseries 127} (2021) 261802}
  [\href{https://arxiv.org/abs/2107.13438}{{\ttfamily 2107.13438}}].

\bibitem{Planck:2018vyg}
{\scshape Planck} collaboration, N.~Aghanim et~al., \emph{{Planck 2018 results.
  VI. Cosmological parameters}},
  \href{https://doi.org/10.1051/0004-6361/201833910}{\emph{Astron. Astrophys.}
  {\bfseries 641} (2020) A6}
  [\href{https://arxiv.org/abs/1807.06209}{{\ttfamily 1807.06209}}].

\bibitem{Djouadi:2001yk}
A.~Djouadi, M.~Drees and J.~L. Kneur, \emph{{Constraints on the minimal
  supergravity model and prospects for SUSY particle production at future
  linear $e^{+} e^{-}$ colliders}},
  \href{https://doi.org/10.1088/1126-6708/2001/08/055}{\emph{JHEP} {\bfseries
  08} (2001) 055} [\href{https://arxiv.org/abs/hep-ph/0107316}{{\ttfamily
  hep-ph/0107316}}].

\bibitem{Costantini:2020stv}
A.~Costantini, F.~De~Lillo, F.~Maltoni, L.~Mantani, O.~Mattelaer, R.~Ruiz
  et~al., \emph{{Vector boson fusion at multi-TeV muon colliders}},
  \href{https://doi.org/10.1007/JHEP09(2020)080}{\emph{JHEP} {\bfseries 09}
  (2020) 080} [\href{https://arxiv.org/abs/2005.10289}{{\ttfamily
  2005.10289}}].

\bibitem{Han:2020uid}
T.~Han, Y.~Ma and K.~Xie, \emph{{High energy leptonic collisions and
  electroweak parton distribution functions}},
  \href{https://doi.org/10.1103/PhysRevD.103.L031301}{\emph{Phys. Rev. D}
  {\bfseries 103} (2021) L031301}
  [\href{https://arxiv.org/abs/2007.14300}{{\ttfamily 2007.14300}}].

\bibitem{Ruiz:2021tdt}
R.~Ruiz, A.~Costantini, F.~Maltoni and O.~Mattelaer, \emph{{The Effective
  Vector Boson Approximation in high-energy muon collisions}},
  \href{https://doi.org/10.1007/JHEP06(2022)114}{\emph{JHEP} {\bfseries 06}
  (2022) 114} [\href{https://arxiv.org/abs/2111.02442}{{\ttfamily
  2111.02442}}].

\bibitem{Alwall:2014hca}
J.~Alwall, R.~Frederix, S.~Frixione, V.~Hirschi, F.~Maltoni, O.~Mattelaer
  et~al., \emph{{The automated computation of tree-level and next-to-leading
  order differential cross sections, and their matching to parton shower
  simulations}}, \href{https://doi.org/10.1007/JHEP07(2014)079}{\emph{JHEP}
  {\bfseries 07} (2014) 079} [\href{https://arxiv.org/abs/1405.0301}{{\ttfamily
  1405.0301}}].

\bibitem{deFavereau:2013fsa}
{\scshape DELPHES 3} collaboration, J.~de~Favereau, C.~Delaere, P.~Demin,
  A.~Giammanco, V.~Lema\^\i{}tre, A.~Mertens et~al., \emph{{DELPHES 3, A
  modular framework for fast simulation of a generic collider experiment}},
  \href{https://doi.org/10.1007/JHEP02(2014)057}{\emph{JHEP} {\bfseries 02}
  (2014) 057} [\href{https://arxiv.org/abs/1307.6346}{{\ttfamily 1307.6346}}].

\bibitem{CMS:2018jrd}
{\scshape CMS} collaboration, A.~M. Sirunyan et~al., \emph{{Performance of
  reconstruction and identification of $\tau$ leptons decaying to hadrons and
  $\nu_\tau$ in pp collisions at $\sqrt{s}=$ 13 TeV}},
  \href{https://doi.org/10.1088/1748-0221/13/10/P10005}{\emph{JINST} {\bfseries
  13} (2018) P10005} [\href{https://arxiv.org/abs/1809.02816}{{\ttfamily
  1809.02816}}].

\bibitem{Casarsa:2021rud}
M.~Casarsa, M.~Fabbrichesi and E.~Gabrielli, \emph{{Monochromatic single photon
  events at the muon collider}},
  \href{https://doi.org/10.1103/PhysRevD.105.075008}{\emph{Phys. Rev. D}
  {\bfseries 105} (2022) 075008}
  [\href{https://arxiv.org/abs/2111.13220}{{\ttfamily 2111.13220}}].

\bibitem{Habermehl:2020njb}
M.~Habermehl, M.~Berggren and J.~List, \emph{{WIMP Dark Matter at the
  International Linear Collider}},
  \href{https://doi.org/10.1103/PhysRevD.101.075053}{\emph{Phys. Rev. D}
  {\bfseries 101} (2020) 075053}
  [\href{https://arxiv.org/abs/2001.03011}{{\ttfamily 2001.03011}}].

\bibitem{Black:2022qlg}
K.~Black, T.~Bose, Y.~Chen, S.~Dasu, H.~Jia, D.~Pinna et~al., \emph{{Prospects
  for Heavy WIMP Dark Matter Searches at Muon Colliders}},  in \emph{{2022
  Snowmass Summer Study}}, 5, 2022,
  \href{https://arxiv.org/abs/2205.10404}{{\ttfamily 2205.10404}}.

\bibitem{Feng:1993sd}
J.~L. Feng and D.~E. Finnell, \emph{{Squark mass determination at the next
  generation of linear $e^{+} e^{-}$ colliders}},
  \href{https://doi.org/10.1103/PhysRevD.49.2369}{\emph{Phys. Rev. D}
  {\bfseries 49} (1994) 2369}
  [\href{https://arxiv.org/abs/hep-ph/9310211}{{\ttfamily hep-ph/9310211}}].

\bibitem{Lester:1999tx}
C.~G. Lester and D.~J. Summers, \emph{{Measuring masses of semiinvisibly
  decaying particles pair produced at hadron colliders}},
  \href{https://doi.org/10.1016/S0370-2693(99)00945-4}{\emph{Phys. Lett. B}
  {\bfseries 463} (1999) 99}
  [\href{https://arxiv.org/abs/hep-ph/9906349}{{\ttfamily hep-ph/9906349}}].

\bibitem{Lester:2014yga}
C.~G. Lester and B.~Nachman, \emph{{Bisection-based asymmetric M$_{T2}$
  computation: a higher precision calculator than existing symmetric methods}},
  \href{https://doi.org/10.1007/JHEP03(2015)100}{\emph{JHEP} {\bfseries 03}
  (2015) 100} [\href{https://arxiv.org/abs/1411.4312}{{\ttfamily 1411.4312}}].

\end{thebibliography}\endgroup

\end{document}